\documentclass[twocolumn,aps,prd,reprint,groupedaddress,showpacs]{revtex4-1}

\usepackage{graphicx}
\usepackage{atlasphysics}
\usepackage{subfigure}
\usepackage{amssymb,amsmath}
\usepackage{hyperref}
\usepackage{placeins}

\def\ptl{\ensuremath{p_{\mathrm{T}}^\ell}}

\def\ptltwo{\ensuremath{p_{\mathrm{T}}^{\ell_2}}}

\usepackage{preprintcover}  
\PreprintCoverPaperTitle{Further search for supersymmetry at $\sqrt{s}
  = 7$~\TeV~in final states with jets, missing transverse momentum and
  isolated leptons with the ATLAS detector}  
\PreprintIdNumber{CERN-PH-EP-2012-204}  
\PreprintCoverAbstract{  This work presents a new inclusive
  search for supersymmetry (SUSY)
  by
  the ATLAS experiment at the LHC
  in proton-proton collisions at a center-of-mass
  energy $\sqrt{s} = 7$~\TeV~ in final states with jets, missing
  transverse momentum and one or more isolated electrons and/or muons.
  The search is based on data
  from the full 2011 data-taking period, corresponding to an integrated
  luminosity of 4.7 \ifb.  Single- and multi-lepton channels are
  treated together in one analysis.  
  An increase in sensitivity is
  obtained by simultaneously fitting 
  the number of events in 
  statistically independent
  signal regions, and the
  shapes of distributions within those regions. A dedicated signal
  region is introduced to be sensitive to decay cascades of SUSY
  particles with
  small mass differences (``compressed SUSY''). Background
  uncertainties are constrained by fitting to the jet multiplicity
  distribution in background control regions.   Observations
  are consistent with Standard Model expectations, and limits are
  set or 
  extended on a number of SUSY models. 
}  
\PreprintJournalName{Physical Review D}

\begin{document}

\title{Further search for supersymmetry at $\sqrt{s} = 7$~\TeV~in final
states with jets, missing transverse momentum and isolated leptons
with the ATLAS detector}

\author{The ATLAS Collaboration}

\begin{abstract}
  This work presents a new inclusive
  search for supersymmetry (SUSY)
  by
  the ATLAS experiment at the LHC
  in proton-proton collisions at a center-of-mass
  energy $\sqrt{s} = 7$~\TeV~ in final states with jets, missing
  transverse momentum and one or more isolated electrons and/or muons.
  The search is based on data
  from the full 2011 data-taking period, corresponding to an integrated
  luminosity of 4.7 \ifb.  Single- and multi-lepton channels are
  treated together in one analysis.  
  An increase in sensitivity is
  obtained by simultaneously fitting 
  the number of events in 
  statistically independent
  signal regions, and the
  shapes of distributions within those regions. A dedicated signal
  region is introduced to be sensitive to decay cascades of SUSY
  particles with
  small mass differences (``compressed SUSY''). Background
  uncertainties are constrained by fitting to the jet multiplicity
  distribution in background control regions.   Observations
  are consistent with Standard Model expectations, and limits are
  set or 
  extended on a number of SUSY models. 
\end{abstract}

\maketitle

\section{Introduction}

Supersymmetry (SUSY)
\cite{Miyazawa:1966,Ramond:1971gb,Golfand:1971iw,Neveu:1971rx,Neveu:1971iv,Gervais:1971ji,Volkov:1973ix,Wess:1973kz,Wess:1974tw}
is a candidate for physics beyond the Standard
Model (SM).  If strongly interacting supersymmetric particles are
present at the \TeV~scale, they may be copiously produced
in 7 \TeV~ proton-proton
collisions at the Large Hadron Collider \cite{Evans:2008zzb}.
In the minimal supersymmetric extension of the Standard Model (MSSM)
\cite{Fayet:1976et,Fayet:1977yc,Farrar:1978xj,Fayet:1979sa,Dimopoulos:1981zb}
such particles decay into jets, leptons and
the lightest supersymmetric particle (LSP).  Jets arise in
the decays of squarks and gluinos, while leptons can arise in decays
involving charginos or neutralinos.   A long-lived,
weakly interacting LSP will
escape detection, leading to missing transverse momentum
($\vec{p}_{\rm{T}}^{\rm{~miss}}$ and its magnitude \met) in the
final state.  Significant
\met~can also arise in scenarios where neutrinos are created
somewhere in the SUSY decay cascade.

This paper presents a new inclusive search with the ATLAS detector
for SUSY in final states
containing jets, one or more isolated leptons (electrons or muons) and \met.
Previous searches in these channels have been conducted by both the
ATLAS \cite{ATLAS:2011ad,Aad:2011cwa} and CMS
\cite{Chatrchyan:2011qs,Chatrchyan:2012ye,Chatrchyan:2012th,Chatrchyan:2012te}
collaborations.  In this paper, the
analysis is extended to 4.7~\ifb~and single- and multi-lepton channels
(with jets and \met) are treated simultaneously.  A 
signal region with a soft lepton and soft jets is introduced in order to
probe SUSY decays involving small mass differences between the
particles in the decay chain.  A new,
simultaneous fit to 
the yield in
multiple signal regions and to the shapes of distributions
within those signal regions is employed.
Background uncertainties are constrained
by fitting to the jet multiplicity distribution in background control regions.

\section{The ATLAS Detector}

The ATLAS detector 
\cite{Aad:2008zzm,Aad:2009wy} consists of a
tracking system (inner detector, ID)
surrounded by a
thin superconducting solenoid providing a 2~T magnetic field,
electromagnetic and hadronic
calorimeters and a muon spectrometer (MS).  The ID consists of pixel
and silicon microstrip detectors, surrounded by a straw-tube tracker
with transition radiation detection (transition
radiation tracker, TRT).  The electromagnetic calorimeter is a
lead liquid-argon (LAr) detector.  Hadronic calorimetry is based on two
different detector technologies, with scintillator-tiles or LAr 
as active media, and with either steel, copper, or tungsten as the
absorber material.
The MS is based on
three large superconducting toroid systems arranged with an eight-fold
azimuthal coil symmetry around the calorimeters, and three
stations of chambers for the trigger and  for precise position measurements.
The nominal $pp$ interaction point at the center of the
detector
is defined as the origin
of a right-handed coordinate system.
The positive $x$-axis is defined by the direction
from the interaction point to the center of the LHC ring, with the
positive $y$-axis pointing upwards, while the
beam direction defines the $z$-axis. The azimuthal angle $\phi$ is measured
around the beam axis and the polar angle $\theta$ is the angle from
the $z$-axis. The pseudorapidity is defined as $\eta = -\ln
\tan(\theta/2)$. Transverse coordinates, such as the transverse
momentum, \pt, are defined in the (x--y) plane.

\section{SUSY Signal Modeling and Simulated Event Samples}
\label{sec:MCsamples}

\begin{table*}[ht]
\begin{center}
\begin{tabular}{l|l|c| c c}
\hline\hline
                &          & Cross        & \multicolumn{2}{c}{Calculation} \\
Physics process & Generator& section (pb) & \multicolumn{2}{c}{accuracy} \\
\hline\hline
$t\bar{t}$ & ALPGEN 2.13~\cite{Mangano:2002ea} & 166.8 & NLO+NLL  & \cite{Aliev:2010zk}\\
$W(\rightarrow \ell\nu)$ + jets  & ALPGEN 2.13~\cite{Mangano:2002ea} & 10460 & NNLO &\cite{Melnikov:2006kv}   \\
$W(\rightarrow \ell\nu)$ + $b\overline{b}$ + jets & ALPGEN 2.13~\cite{Mangano:2002ea} & 130 & LO$\times$K &   \\
$W(\rightarrow \ell\nu)$ + $c\overline{c}$ + jets & ALPGEN 2.13~\cite{Mangano:2002ea} & 360 & LO$\times$K &  \\
$W(\rightarrow \ell\nu)$ + $c$ + jets & ALPGEN 2.13~\cite{Mangano:2002ea} & 1100 & LO$\times$K &   \\
$Z/\gamma^{\ast}(\rightarrow \ell \ell)$ + jets ($m_{\ell\ell}>40$~\GeV) &ALPGEN 2.13~\cite{Mangano:2002ea} & 1070
&NNLO&\cite{Melnikov:2006kv} \\
$Z/\gamma^{\ast}(\rightarrow \ell \ell)$ + jets ($10 \GeV <
m_{\ell\ell} < 40$~\GeV) &ALPGEN 2.13~\cite{Mangano:2002ea} & 3970
&NNLO&\cite{Melnikov:2006kv} \\
$Z/\gamma^{\ast}(\rightarrow \ell \ell)$ + $b\overline{b}$ + jets ($m_{\ell\ell}>40$~\GeV) &ALPGEN 2.13~\cite{Mangano:2002ea} & 10.3
&LO& \\
Single-top ($t$-chan) & AcerMC 3.8~\cite{Kersevan:2002dd} & 7.0 & NLO & \\
Single-top ($s$-chan) & MC$@$NLO 4.01~\cite{Frixione:2002ik} & 0.5 & NLO & \\
Single-top ($Wt$-chan) & MC$@$NLO 4.01~\cite{Frixione:2002ik} & 15.7 & NLO & \\
$WW$ & HERWIG 6.5.20 \cite{Corcella:2000bw} & 44.9 & NLO  & \cite{Campbell:2003hd}\\
$WZ/\gamma^{\ast}~(m_{Z/\gamma^{\ast}} > 60~\rm{\GeV})$ & HERWIG 6.5.20 \cite{Corcella:2000bw} & 18.5 & NLO  & \cite{Campbell:2003hd}\\
$Z/\gamma^{\ast}Z/\gamma^{\ast}~(m_{Z/\gamma^{\ast}} > 60~\rm{\GeV})$ & HERWIG 6.5.20 \cite{Corcella:2000bw} & 5.96 & NLO  & \cite{Campbell:2003hd}\\
\ttbar+$W$ & MADGRAPH5 \cite{Alwall:2011uj} & 0.169 & NLO & \cite{Campbell:2012dh} \\
\ttbar+$Z$ & MADGRAPH5 \cite{Alwall:2011uj} & 0.120 & LO$\times$K & \cite{Lazopoulos:2008de}\\

\hline\hline

\end{tabular}
\caption{Simulated background event
  samples used in this analysis, with the corresponding production cross
  sections. The notation LO$\times$K indicates that the process is
  calculated at leading-order and corrected by a factor derived from
  the ratio of NLO to LO cross sections for a closely related process.
  The  \ttbar, $W$+ light-jets and $Z$+ light-jets
  samples are normalized using the inclusive
  cross sections; the values shown for 
  the $W$+ light-jets and $Z$+ light-jets samples
  are for a single
  lepton flavor.  The single-top cross sections are listed for a
  single lepton flavor in the $s$- and $t$-channels.
  Further details are given in the text.}
\label{tab:MC}
\end{center}
\end{table*}

The SUSY models
considered are MSUGRA/CMSSM \cite{msugra,Kane:1993td},  minimal GMSB
\cite{AlvarezGaume:1981wy,Dine:1981za,Dimopoulos:1981au,Nappi:1982hm,Dine:1993yw}
and a number of
simplified models \cite{Alwall:2008ag,Alves:2011wf}.  The
MSUGRA/CMSSM model is characterized by five parameters: the universal
scalar and gaugino mass parameters $m_{0}$ and $m_{1/2}$, a universal
trilinear coupling parameter $A_{0}$, the ratio of the vacuum
expectation values of the two Higgs doublets $\tan\beta$, and the sign
of the Higgsino mass parameter $\mu$.  In this analysis, the values of
$m_{0}$ and $m_{1/2}$ are
scanned, and the other parameters are fixed as follows: $\tan\beta = 10$,
$A_{0} = 0$ and $\mu$ is taken to be positive.
A diagram showing the decay of the associated production of squark and gluino 
is depicted in Fig.~\ref{fig:feynman} (a).
Other diagrams representative for the SUSY models
discussed in the following are shown in Fig.~\ref{fig:feynman} (b-d).

The minimal GMSB model
has six parameters: the SUSY breaking scale $\Lambda$, the mass scale
of the messenger fields $M_{\rm{mes}}$, the number of messenger
fields $N_{5}$, the scale of the gravitino coupling $C_{\rm{grav}}$,
the ratio of the vacuum expectation values of the two Higgs doublets
$\tan\beta$, and the sign of the Higgsino mass parameter $\mu$.
For the minimal GMSB model, the parameters
$\tan\beta$ and $\Lambda$ are scanned and the other parameters are
assigned fixed values: $M_{\rm{mes}}=250$~\TeV, $N_{5}=3$, $C_{\rm{grav}}=1$ and
the sign of $\mu$ is taken to be positive.   The mass scale of the
colored superpartners is set by the parameter $\Lambda$, while the
next-to-lightest SUSY particle (NLSP) is determined by a combination
of $\Lambda$ and $\tan\beta$. At low values of
$\Lambda$, the NLSP is the lightest neutralino
($\tilde{\chi}_{1}^{0}$) while at the higher values of
$\Lambda$ where this search provides new sensitivity, the NLSP is a
stau for $\tan\beta \gtrsim 10$ and a slepton of the first and second
generation otherwise.  The NLSP decays into its SM partner and a
nearly massless gravitino. The gaugino and sfermion masses are
proportional to $N_{5}$ and $\sqrt{N_{5}}$, respectively.  The
parameter $C_{\rm{grav}}$ determines the NLSP lifetime, set here such that
all NLSPs decay promptly.

\begin{figure*}[htbp]
\begin{center}
\includegraphics[height=0.12\textheight]{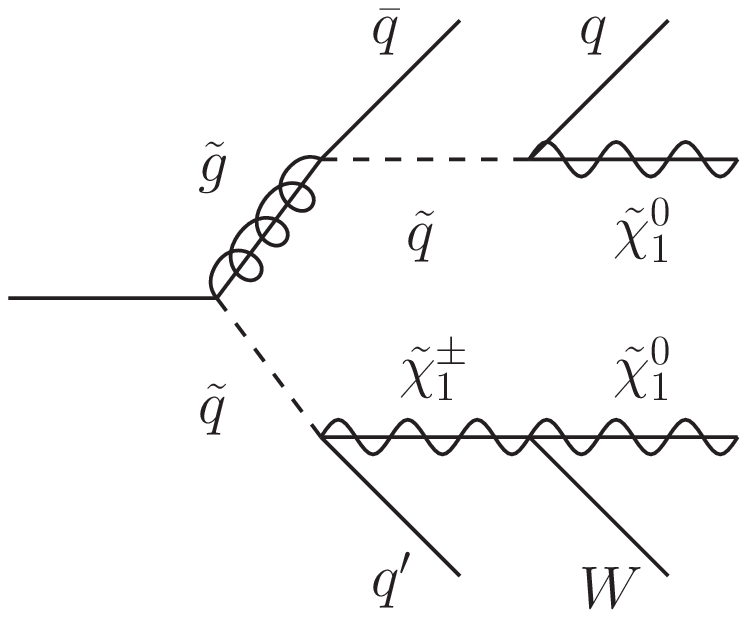}\hfill
\includegraphics[height=0.12\textheight]{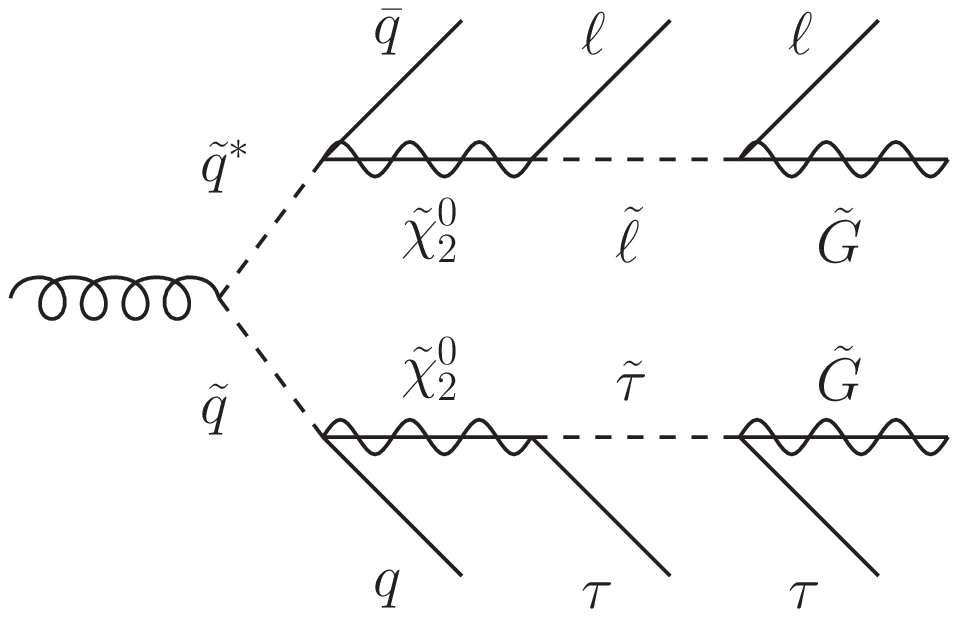}\hfill
\includegraphics[height=0.12\textheight]{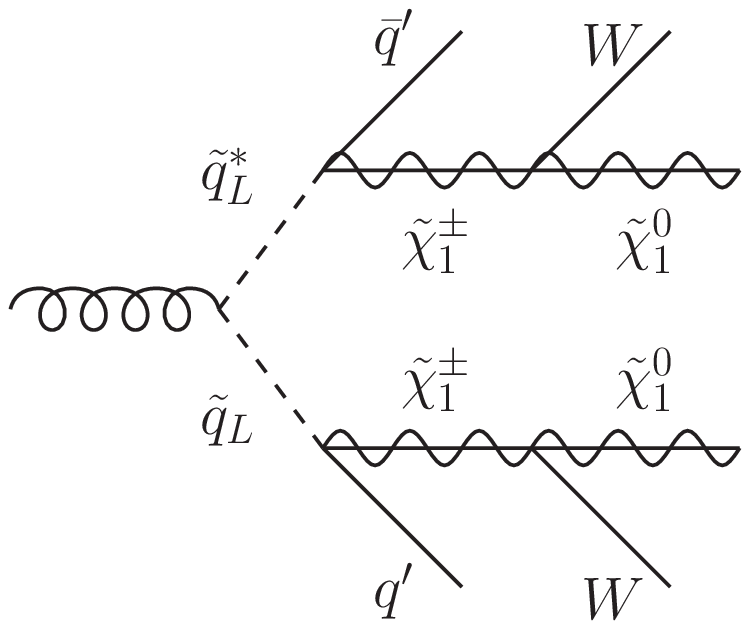}\hfill
\includegraphics[height=0.12\textheight]{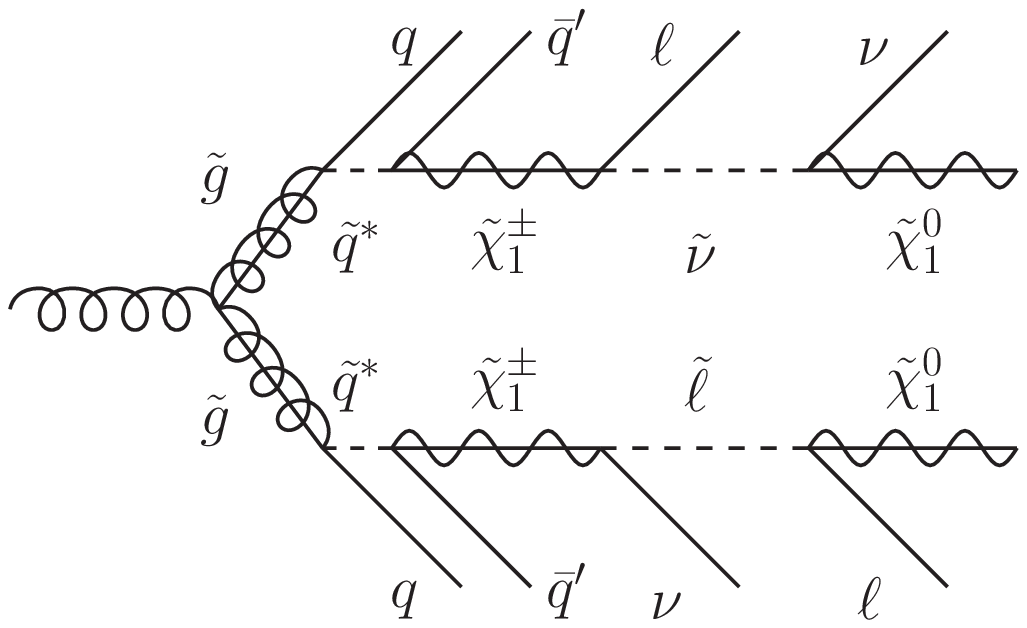}\\[-5mm]
\end{center}
(a) \hspace*{0.25\textwidth} (b) \hspace*{0.25\textwidth} (c) \hspace*{0.25\textwidth} (d)
                                                                 
\caption{Representative diagrams for the different SUSY
  models considered
  in this analysis: (a) MSUGRA/CMSSM model with $pp \rightarrow
  \tilde{q}\tilde{g}$ and subsequent decay of the squark
  via a chargino; (b) GMSB model with $pp \rightarrow
  \tilde{q}\tilde{q}^\ast$ and subsequent decay via
  sleptons and staus; (c) one-step simplified model 
  with $pp \rightarrow \tilde{q}_L\tilde{q}_L^\ast$ and subsequent decay
  via charginos; (d) two-step
  simplified model with $pp \rightarrow
  \tilde{g}\tilde{g}$ and subsequent decays via charginos
  and sleptons or sneutrinos.}
\label{fig:feynman}
\end{figure*}

Several simplified models are 
considered in this paper.
In the ``one-step'' models,
SUSY production proceeds via either $pp
\rightarrow \tilde{g}\tilde{g}$ or $pp \rightarrow
\tilde{q}_{L}\tilde{q}_{L}^{\ast}$, where only left-handed
squarks of the first-
and second-generation are considered.  The gluino decays to
the neutralino LSP via $\tilde{g} \rightarrow
q\overline{q}'\tilde{\chi}_{1}^{\pm} \rightarrow
q\overline{q}'W^{\pm}\tilde{\chi}_{1}^{0}$, and the squark via
$\tilde{q}_{L} \rightarrow
q' \tilde{\chi}_{1}^{\pm} \rightarrow q'W^{\pm}\tilde{\chi}_{1}^{0}$,
where the $W$-boson can be real
or virtual.  The gluino
and LSP masses are varied while the
chargino mass is set to be halfway between them.
In a variant of the one-step model, the LSP mass is held fixed at
60 \GeV~while the gluino (squark) and chargino masses are scanned.

In the ``two-step'' models, SUSY production proceeds via either
$pp \rightarrow \tilde{g}\tilde{g}$ or $pp \rightarrow
\tilde{q}_{L}\tilde{q}_{L}^{\ast}$, again where squarks of the first-
and second-generation are considered.  In the first class of
two-step models all squarks and
gluinos decay via a chargino: $\tilde{g} \rightarrow
q\overline{q}'\tilde{\chi}_{1}^{\pm}$~and $\tilde{q}_{L} \rightarrow
q' \tilde{\chi}_{1}^{\pm}$.  The charginos decay via 
$\tilde{\chi}_{1}^{\pm} \rightarrow \ell
\tilde{\nu}_{L}$ or  $\tilde{\chi}_{1}^{\pm} \rightarrow \nu
\tilde{\ell}_{L}$; in case of third generation sleptons, the decay to
the stau is via $\tilde{\chi}_{1}^{\pm} \rightarrow \nu
\tilde{\tau}_{1}$.  All three
generations of sleptons and sneutrinos are allowed with equal
probability, resulting in equal branching ratio to sleptons and 
to sneutrinos.  In the second class of two-step models, the gluinos or
left-handed squarks decay either via a chargino ($\tilde{g} \rightarrow
q\overline{q}'\tilde{\chi}_{1}^{\pm}$ or $\tilde{q}_{L} \rightarrow
q' \tilde{\chi}_{1}^{\pm}$) or via a neutralino ($\tilde{g} \rightarrow
q\overline{q}\tilde{\chi}_{2}^{0}$ or $\tilde{q}_{L} \rightarrow
q \tilde{\chi}_{2}^{0}$). The events are generated such that one
chargino and one neutralino are always present in the decays of the pair produced
gluinos or left-handed squarks. Neutralino decays proceed via either 
$\tilde{\chi}_{2}^{0} \rightarrow \ell \tilde{\ell}_{L}$
or $\tilde{\chi}_{2}^{0} \rightarrow \nu \tilde{\nu}$.  As in the
first two-step model, all three generations of sleptons and sneutrinos
are allowed with equal probability, resulting in a 50\% branching
ratio to sleptons and to sneutrinos.  Finally, in the third class
of two-step models without intermediate sleptons, the gluino and squark decay via
$\tilde{g} \rightarrow q\overline{q}'\tilde{\chi}_{1}^{\pm}$~ or
$\tilde{q}_{L} \rightarrow q' \tilde{\chi}_{1}^{\pm}$; the decay of the chargino then
proceeds  via $\tilde{\chi}_{1}^{\pm} \rightarrow W^{(\ast)\pm} \tilde{\chi}_{2}^{0} \rightarrow
W^{(\ast)\pm} Z^{(\ast)} \tilde{\chi}_{1}^{0}$.  This signature is
realized in the MSSM in a parameter region where additional decay
modes, not contained in the simplified model, may lead to a
significant reduction of the cross section times branching fraction of
the $WZ$ signature.

In the first two types of two-step models, the chargino and neutralino have
equal masses (again set to be halfway between the gluino/squark and
LSP mass); the slepton and sneutrino masses are set to be equal and
halfway between the chargino/neutralino and LSP masses.  In the third
two-step model, the $\tilde{\chi}_{1}^{\pm}$~mass is set halfway between
the gluino/squark and LSP while
the $\tilde{\chi}_{2}^{0}$~mass is set halfway between
the chargino and LSP.
In all the simplified models, the
superpartners that have not been
mentioned are decoupled by setting their masses to multi-\TeV~ values.

Simulated event samples are used for estimating the signal acceptance,
the detector efficiency, and for estimating many of the backgrounds
(in most cases in association with data-driven techniques).
The MSUGRA/CMSSM and minimal GMSB 
signal samples are generated with Herwig++ 2.5.2 \cite{Bahr:2008pv}
and $\rm{MRST2007LO}^{*}$ \cite{Sherstnev:2007nd}
parton distribution functions (PDFs);
ISAJET 7.80 \cite{Paige:2003mg} is used to generate the physical
particle masses. 
The simplified models are generated with one extra jet in the matrix
element using
MADGRAPH5 \cite{Alwall:2011uj}, interfaced to PYTHIA \cite{Pythia}, with the
CTEQ6L1 \cite{Pumplin:2002vw} PDF set; MLM
matching \cite{Alwall:2007fs} is done with a scale parameter
that is set to one-fourth of the mass of the lightest sparticle
in the hard-scattering
matrix element.
Signal cross sections are calculated in the MSSM at next-to-leading order in the
strong coupling constant, including the resummation of soft gluon
emission at next-to-leading-logarithmic accuracy
(NLO+NLL)~\cite{Beenakker:1996ch,Kulesza:2008jb,Kulesza:2009kq,Beenakker:2009ha,Beenakker:2011fu}.

The simulated event samples for the SM backgrounds
are summarized in Table \ref{tab:MC}. 
The ALPGEN and MADGRAPH samples are produced with the MLM matching
scheme.
The ALPGEN samples are generated with a number of partons
  $0 \le N_{\rm{parton}} \le 5$ in the matrix element, except for $W$
  + light-flavored jets which are generated with up to 6 partons.
  The $Wb\overline{b}$, $Wc\overline{c}$ and
  $Wc$
  cross sections shown are the
  leading-order values from ALPGEN multiplied by a K-factor of 1.2,
  based on the K-factor for light-flavored jets.
  For the final result, measured cross sections are used for the 
  $W/Z$+ heavy-flavor-jets samples~\cite{ATLAS:2012an}.
  The overlap between the heavy-flavored and
  light-flavored $W/Z$+jets samples is removed.  The cross section for
  $Z$+jets with $10 \GeV < m_{\ell\ell} < 40$~\GeV~is obtained by
  assuming the same K-factor as for  $m_{\ell\ell} > 40$~\GeV.
  The single-top cross sections are taken from
  MC$@$NLO; for the $s$- and $t$-channels, they are listed for a
  single lepton flavor.
  
The theoretical cross sections for $W$+jets and $Z$+jets
are calculated with FEWZ \cite{Melnikov:2006kv} with the MSTW2008NNLO
\cite{Martin:2007bv} PDF set.
  For the diboson cross sections,
MCFM \cite{Campbell:2003hd} with the
MSTW2008NLO PDFs is used.  The \ttbar~cross section is calculated
with HATHOR 1.2 \cite{Aliev:2010zk} using MSTW2008NNLO PDFs.  The
\ttbar+$W$ cross section is taken from Ref. \cite{Campbell:2012dh}. The
\ttbar+$Z$ cross section is the leading-order value multiplied by a
K-factor deduced from the NLO calculation at $\sqrt{s}=14$~\TeV~
\cite{Lazopoulos:2008de}.

  Parton shower and fragmentation processes are simulated
for the ALPGEN and MC$@$NLO samples using  HERWIG
\cite{Corcella:2000bw} with
JIMMY \cite{Butterworth:1996zw} for underlying event modeling;
PYTHIA is used for the AcerMC single-top sample and \ttbar+$W/Z$.
The PDFs used in this analysis
are: CTEQ6L1 for the
ALPGEN and MADGRAPH samples,
CT10 \cite{Lai:2010vv} for MC$@$NLO, and MRSTMCal
$(\rm{LO}^{**})$ \cite{Sherstnev:2008dm} for HERWIG.
The underlying event tunes are the ATLAS AUET2B\_LO$^{**}$ tunes \cite{AUET2B}.

The detector simulation
\cite{Aad:2010ah} is performed using GEANT4 \cite{Agostinelli:2002hh}.
All samples are produced with
a range of simulated minimum-bias interactions overlaid on the
hard-scattering event to account for
multiple $pp$ interactions in
the same beam crossing (pile-up). The overlay
also treats the impact of pile-up from beam crossings
other than the one in which the event occurred.
Corrections are applied to the simulated samples to account for
differences between data and simulation for the lepton
trigger and reconstruction efficiencies, momentum scale and
resolution, and for the efficiency and
mis-tag rates for $b$-quark tagging.

\section{Object Reconstruction}

This analysis is based on three broad classes of event selection: 
{\it i)} a hard single-lepton channel that is an extension to higher masses
of the previous search \cite{ATLAS:2011ad}, {\it ii)} a soft
single-lepton channel
geared towards SUSY models with small mass differences in the decay
cascade, and {\it iii)} a multi-lepton channel aimed at decay chains with
higher lepton multiplicities.
The event selection requirements are described in detail
in Sec. \ref{sec:EventSelection}.  Here the final-state
object reconstruction and selection are discussed.

\subsection{Object Preselection}

The primary vertex
\cite{PV} is
required to be consistent with the beam spot envelope and to
have at least five associated tracks;
when more than one such vertex is found, the vertex with
the largest summed $|\pt|^{2}$ of the associated tracks is chosen.

Electrons are reconstructed from energy clusters in the electromagnetic
calorimeter matched to a track in the ID \cite{Aad:2011mk}.
Pre-selected electrons are required to have $|\eta| < 2.47$~and
pass a variant of the
``medium'' selection defined in  Ref. \cite{Aad:2011mk} that differs
mainly in having  a tighter track-cluster matching in $\eta$, stricter
pixel hit requirements, additional requirements in the TRT,  and tighter
shower-shape requirements for $|\eta| > 2.0$.  These requirements provide
background rejection close to the ``tight'' selection of Ref.
\cite{Aad:2011mk} with only a few percent loss in efficiency with
respect to ``medium''.
Pre-selected electrons are further required to pass
a \pt~requirement depending on the analysis channel:
10 \GeV~for the hard-lepton and multi-lepton channels, and 7 \GeV~in
the soft-lepton channel.

Muons are identified either as a combined track in the MS
and ID systems, or as an ID track matched with a  MS segment
\cite{ATLAS-CONF-2011-021,ATLAS-CONF-2011-063}.  Requirements on the
quality of the ID track are identical to those in
Ref. \cite{ATLAS:2011ad}.  Pre-selected muons are required to have
$|\eta| < 2.4$ and a \pt~requirement that depends on the analysis channel:
10 \GeV~for the hard-lepton and multi-lepton channels, and 6 \GeV~in
the soft-lepton channel.

Jets are reconstructed using the anti-$k_{t}$ algorithm
\cite{Cacciari:2008gp, Cacciari:2005hq}
with a radius parameter $R = $ 0.4.   Jets
arising from detector noise, cosmic rays or other non-collision
sources are
rejected \cite{Aad:2011he}.
To account for the differences between the
calorimeter response
to electrons and hadrons, \pt- and $\eta$-dependent factors,
derived from simulated events
and validated with test beam and collision data, 
are applied to each jet to provide an average
energy scale correction \cite{Aad:2011he} back to particle level.
Pre-selected jets are  required to have \pt~$>$~20 \GeV~and $|\eta| <
4.5$.
Since electrons are also reconstructed as jets, pre-selected
jets which overlap with pre-selected electrons within a distance
$\Delta R = \sqrt{(\Delta \eta)^2 +
  (\Delta \phi)^2} = 0.2$ are discarded.

\subsection{Signal Object Selection}

For the final selection of signal events,
``signal'' electrons are required to pass a variant of the ``tight'' selection
of Ref. \cite{Aad:2011mk}, providing 1--2\% gain in efficiency and slightly
better background rejection.  Signal electrons must have
$|\eta| < 2.47$ and a distance to the closest
jet $\Delta R > 0.4$.   
They are also required to satisfy isolation criteria:
the scalar sum
of the \pt~of tracks within a cone of radius
$\Delta R = 0.2$ around the electron
(excluding the electron itself) is required to be
less than 10\% of the electron \pt.  

Muons in the final selection (``signal'' muons) are required to have 
$|\eta| < 2.4$ and $\Delta R > 0.4$
with respect to the closest jet.   Further
isolation criteria are imposed: the scalar sum
of the \pt~of tracks within a cone of radius $\Delta R = 0.2$ around the muon
candidate (excluding the muon itself) is required to be
less than 1.8 \GeV.  The \pt~requirements for signal electrons and
muons 
depend on the signal regions and are described in
Sec. \ref{sec:EventSelection}. 

Signal jets are required to have \pt~$>$ 25 \GeV~and
$|\eta| < 2.5$.   In addition, they are
required to be associated with the hard-scattering process, by
demanding that at least 75\% of the scalar sum of the
\pt~of all tracks associated with the jet come from tracks
associated with the primary vertex of the event. Jets with no associated
tracks are rejected. The above requirements are applied to cope
with the high pile-up conditions of
the 2011 data-taking, in particular the later part of the run.

The missing transverse momentum is
computed as the negative of the vector sum of the \pt~of all
pre-selected electrons, pre-selected muons and pre-selected
jets (after removing those overlapping with pre-selected
electrons),
and all calorimeter clusters with $|\eta| < 4.9$ that are
not associated with any of the above-mentioned objects.  

For approximately 20\% 
of the 2011 data-taking period, an electronics failure created a region in
the electromagnetic calorimeter, located at $0 < \eta < 1.4$ and $-0.8
< \phi < -0.6$, where no signals could be read out.
Events with an electron in
this region are vetoed for the entire dataset,
leading to an acceptance loss of less than
1\% for signal events in the signal region.
For jets, the amount of transverse energy (\ET)
lost in the dead region can be estimated
from the energy deposited in the neighboring calorimeter cells.
If this lost \ET~projected along the \met~direction
amounts to more than 10 \GeV~and constitutes more than 10\% of the
\met, the
event is rejected.   The effect of the electronics failure
is described in the detector simulation,
and the loss of signal acceptance from this requirement is 
negligible.

Jets arising from $b$-quarks  are identified using information
about track impact parameters and reconstructed secondary vertices
\cite{ATLAS-CONF-2011-102}; the  $b$-tagging algorithm is based on
a neural network using the output weights of the JetFitter+IP3D, IP3D,
and SV1 algorithms (defined in Ref. \cite{ATLAS-CONF-2011-102})
as input.  The $b$-tagging requirements are set at
an operating point corresponding to an average efficiency of 60\% for $b$-jets
in simulated \ttbar~events, for which the algorithm provides a
rejection factor of approximately
200--400 for light-quark and gluon jets (depending on the \pt~of the
jet) and a rejection of
approximately 7--10 for charm jets.

\section{Trigger and Data Collection}

The data used in this analysis
were collected from March through October 2011, during which the instantaneous
luminosity of the LHC reached $3.65 \times 10^{33}
\rm{cm}^{-2} \rm{s}^{-1}$.  The average number of 
interactions per beam crossing ranged from approximately 4 to 16
during the run, with
an average of 10.  After the application of
beam, detector, and data-quality requirements, the total
integrated luminosity is 4.7 \ifb.
The uncertainty on the luminosity is
determined to be 3.9\% \cite{Aad:2011dr,ATLAS-CONF-2011-116}.

Three types of triggers were used to collect the data: electron, muon
and \met.  The electron
trigger selects
events containing one or more electron candidates, based on the
presence of a cluster in the electromagnetic calorimeter, with a
shower shape consistent
with that of an electron. The transverse energy threshold at the
trigger level was either 20 \GeV~or 22 \GeV, depending on the
instantaneous luminosity.  For signal electrons satisfying \pt~$>$ 25 \GeV, the
trigger efficiency is in the plateau region and
ranges between 95\% and 97\%.  In order to recover some of
the efficiency for high-\pt~electrons during running periods with
the highest instantaneous
luminosities, events were also collected with an electron trigger
with looser shower shape requirements but 
with a \pt~threshold of 45 \GeV.

The muon trigger selects events containing one or more muon
candidates based on tracks identified in the MS and ID.  The muon
trigger
\pt~threshold was 18 \GeV.  During running periods with the highest
instantaneous luminosities, the trigger requirements on the number of MS
hits were tightened; in order to recover some of the resulting
loss in efficiency, events were also collected with a muon
trigger that maintained the looser requirement on the number of
hits chambers but that required in addition
a jet
with \pt~greater than 10 \GeV.   This jet requirement is
fully efficient for jets with offline calibrated
\pt~greater than approximately 50 \GeV.
The muon triggers reach their efficiency plateaus
below a signal muon \pt~threshold
of 20 \GeV.  The plateau efficiency ranges from about 70\% for
$|\eta|<1.05$  to 88\% for $1.05 < |\eta| < 2.4$.

The \met~trigger bases the bulk of its rejection on
the vector sum of transverse energies
deposited in projective trigger towers (each with a size of approximately
$\Delta\eta \times \Delta\phi \sim 0.1 \times 0.1$~for $|\eta|<2.5$
and larger and less regular in the more forward regions).  A more
refined calculation based on the vector sum of all calorimeter cells
above threshold is made at a later stage in the trigger processing.
The trigger required \met~$>$~60 \GeV, reaching its efficiency
plateau for offline calibrated \met~ $>$ 180 \GeV.
The efficiency on the plateau is close to 100\%.

\section{Event Selection}
\label{sec:EventSelection}

Two variables, derived from the kinematic properties of the
reconstructed objects,  are used in the event selection.
The transverse mass ($m_{\rm{T}}$) computed from the momentum
of the lepton ($\ell$)
and the missing transverse momentum ($\vec{p}_{\rm{T}}^{\rm{~miss}}$),
defined as
    \[
    m_{\mathrm{T}} =
\sqrt{2 p_{\mathrm{T}}^{\ell} \met
  (1-\cos(\Delta\phi(\vec{\ell},\vec{p}_{\rm{T}}^{\rm{~miss}})))},
  \]
is useful in rejecting events containing a single $W$~boson.
The inclusive effective mass ($m_{\rm{eff}}^{\rm{inc}}$) is the scalar sum of
the \pt~of
the leptons, the jets and \met:
    \[
    m_{\mathrm{eff}}^{\mathrm{inc}} =
    \sum_{i=1}^{N_{lep}} p_{\mathrm{T},i}^{\ell} +
    \sum_{j=1}^{N_{jet}} p_{\mathrm{T},j} + \met
    \]
where the index $i$ runs over all the signal leptons
and $j$ runs over all the signal jets in the event.
The inclusive effective
    mass is correlated with the overall mass scale of the
    hard-scattering process and provides good discrimination against
    the SM
    background, without being too sensitive to the details of the SUSY
    decay cascade.    The analysis
    in Ref. \cite{ATLAS:2011ad} used 
    the three or four leading-\pt~jets in the calculation of the
    effective mass; the additional jets used here improve the
    discrimination between signal and background.
    A second definition for the effective
    mass, denoted by $m_{\rm{eff}}$, is based on the sum over the
    2-, 3-, or 4-leading \pt~jets,
    depending on the minimum number of jets required in a given signal
    region. This variable is used to compute the ratio
    \met/$m_{\rm{eff}}$~which reflects the
    fluctuations in  the \met\ as a function of the
    calorimeter activity in the event; the definition used here
    improves the rejection of the
    background from mismeasured jets.

\begin{table*}[tbp]
\begin{center}
\begin{tabular}{lccccc}
\hline\hline
 & \hspace*{25mm} & \hspace*{25mm}
& \hspace*{25mm}& \hspace*{25mm}& \hspace*{25mm}\\[-3mm]
& \multicolumn{3}{c}{\bf single-lepton} & \multicolumn{2}{c}{\bf multi-lepton}\\
 & {\bf 3-jet} & {\bf 4-jet} & {\bf soft-lepton} &
    {\bf 2-jet} & {\bf 4-jet}\\[1mm]
\hline\hline
Trigger   & \multicolumn{2}{c}{Single electron or muon (+jet)} &
Missing $E_{\rm{T}}$ & \multicolumn{2}{c}{Single electron or muon (+jet)}\\
\hline
$N_{\rm{lep}}$ &  1 &  1 &  1 & $\ge 2$ & $\ge 2$ \\
\ptl\ (\GeV)   & $>$ 25 (20) & $>$ 25 (20) & 7 to 25 (6 to 20) & 25 (20)
& 25 (20)\\
\ptltwo (\GeV) & $<$ 10 & $<$ 10 &  $<$ 7 (6) & $>$ 10 & $>$ 10\\
\hline
$N_{\rm{jet}}$  & $\geq$ 3 & $\geq$ 4 & $\geq$ 2 & $\ge 2$ & $\ge 4$ \\
$p_{\rm T}^{\rm{jet}}$ (\GeV) & $>$ 100, 25, 25 & $>$ 80, 80, 80, 80 & $>$ 130,25 & $>$
200,200 & $>$ 50,50,50,50 \\
$p_{\rm T}^{\rm{add. jet}}$ (\GeV) & $<$ 80 & --- & --- & $<$ 50 & --- \\
\hline
\met\ (\GeV) & $>$ 250 & $>$ 250 & $>$ 250 & $>$ 300 & $>$ 100 \\
$m_{\rm{T}}$ (\GeV) & $>$ 100 & $>$ 100 & $>$ 100 & --- & --- \\
\met/$m_{\rm{eff}}$ & $>$ 0.3 & $>$ 0.2 & $>$ 0.3 & --- & 0.2 \\
$m_{\rm{eff}}^{\rm{inc}}$ (\GeV) & $>$ 1200 & $>$ 800 & --- & --- & $>$
650\\
\hline\hline
\end{tabular}
\caption{Overview of the
  selection criteria for the signal regions used in this analysis. The
  \pt~selections for leptons are given for electrons (muons).}
\label{tab:SR}
\end{center}
\end{table*}

This analysis is based on five signal regions, each tailored to
maximize the sensitivity to different SUSY event
topologies: {\it 1,2)} Signal regions requiring a hard lepton plus 3- or 4-jets
are extensions of the previous
analysis \cite{ATLAS:2011ad} to higher SUSY mass scales; these signal
regions have been
optimized for the MSUGRA/CMSSM model as well as for the bulk of the
one-step simplified models with large mass difference ($\Delta m$) between the
gluino and the LSP; {\it 3)}  A soft-lepton signal region targets
the simplified models with small $\Delta m$, where the hard leading
jet
comes from initial-state radiation (ISR); {\it 4)}  A
multi-lepton signal region with $\ge 2$ jets is tailored to
GMSB models; {\it 5)} A multi-lepton signal region with  $\ge 4$
jets is geared towards
the two-step simplified models with intermediate sleptons and
sneutrinos.  These signal regions are described in more detail and
summarized in Table \ref{tab:SR}.

\begin{enumerate}
\item {\it Hard lepton plus three jets}.
  Events are selected with the electron and muon triggers.
  The number of signal leptons with \pt~$>$ 25 (20) \GeV~for electrons
  (muons) is required to be
  exactly one.  Events containing additional signal leptons
  with \pt~$>$ 10~\GeV~are rejected.  
  The number of signal jets is
  required to be $\ge$ 3, with a leading jet satisfying \pt~$>$ 100
  \GeV~and the other jets having \pt~$>$ 25 \GeV.  Events with four or
  more jets are rejected if the fourth jet has \pt~$>$
  80 \GeV; this requirement keeps this signal region disjoint from the
  4-jet signal region.
  In addition, the  following conditions are imposed:
  $m_{\rm{T}} > 100$~\GeV, \met~ $>$ 250 \GeV, \met/$m_{\rm{eff}} >
  0.3$, and $m_{\rm{eff}}^{\rm{inc}} > 1200$~\GeV.
\item {\it Hard lepton plus four jets}.
  The lepton requirements are the same as in the previous signal region.
  The number of signal jets is
  required to be $\ge$~4, with the four leading jets satisfying \pt~$>$ 80
  \GeV.  
  In addition, the  following requirements are applied:
  $m_{\rm{T}} > 100$~\GeV, \met~ $>$ 250~\GeV, \met/$m_{\rm{eff}} >
  0.2$, and $m_{\rm{eff}}^{\rm{inc}} > 800$~\GeV.
\item {\it Soft-lepton selection}.
  Events are selected with the \met~trigger.
  The number of signal leptons (electron or muon) is required to be
  exactly one.  Electrons are required to have 7~\GeV~$<$ \pt~$<$ 25
  \GeV, and muons are required to be in the range 6~\GeV~$<$ \pt~$<$
  20 \GeV.  Events containing an additional signal electron (muon) with
  \pt~$>$ 7~(6)~\GeV~are rejected.
  The number of signal jets is
  required to be $\ge$~2, with the leading jet satisfying \pt~$>$ 130
  \GeV~and the second jet having \pt~$>$ 25~\GeV.
  In addition, the  following conditions are required:
  $m_{\rm{T}} > 100$~\GeV, \met~ $>$ 250 \GeV, and \met/$m_{\rm{eff}} >
  0.3$.  No explicit requirement on $m_{\rm{eff}}^{\rm{inc}}$ is
  applied.
\item {\it Multi-lepton plus two jets}.
  Events are selected with the electron and muon triggers.  Two or
  more signal leptons are required, with a leading electron (muon) with
  \pt~$>$ 25 (20) \GeV~and sub-leading leptons with \pt~$>$ 10 \GeV.
  The two leading leptons must have opposite charge. At
  least two signal jets with \pt~$>$ 200 \GeV~are required. Events with four or
  more signal jets are rejected if the fourth leading jet has \pt~$>$
  50 \GeV; this requirement keeps this signal region disjoint from the
  multi-lepton plus 4-jet signal region.  In addition the \met~is
  required to be $>$ 300 \GeV. No explicit requirements are
  made on \met/$m_{\rm{eff}}$ or $m_{\rm{eff}}^{\rm{inc}}$.
\item {\it Multi-lepton plus four jets}.
  The lepton requirements are the same as in the multi-lepton plus two
  jets signal region.  At least four signal jets with \pt~$>$ 50
  \GeV~are
  required.  In addition, the following requirements are imposed: 
  \met~ $>$ 100 \GeV, \met/$m_{\rm{eff}} >
  0.2$, and $m_{\rm{eff}}^{\rm{inc}} > 650$~\GeV.
\end{enumerate}

In contrast to the previous analysis \cite{ATLAS:2011ad},
no requirement on the azimuthal
angle between the \met~ vector and any of the jets is imposed as
the background from multijet events is already low. This
adds sensitivity to SUSY decay chains where the LSP is boosted along
the jet direction.

\section{Background Estimation}

The dominant sources of background in the single-lepton channels are
the production of semi- and
fully-leptonic \ttbar~events, and $W$+jets where the $W$ decays
leptonically.  For the multi-lepton channels, the main background
sources are $Z$+jets and \ttbar.  Other background processes
which are considered are multijets, single-top, dibosons and
\ttbar~plus vector boson.

The major backgrounds are estimated by
isolating each of them in a dedicated control region, normalizing
the simulation to data in that control region, and then
using the simulation to extrapolate the background expectations
into the signal region.  The
multijet background is determined from the data by a matrix
method described below.
All other (smaller)
backgrounds are estimated entirely from the simulation, using the most
accurate theoretical cross sections available (Table \ref{tab:MC}).
To account for the
cross-contamination of physics processes across control regions, the
final estimate of the background is obtained with a simultaneous,
combined fit to
all control regions, as 
described in Sec.~\ref{sec:bkgfit}.

\begin{table*}[ht]
\begin{center}
\begin{tabular}{lcccccc}
\hline\hline
 & \hspace*{22mm} & \hspace*{22mm}
& \hspace*{22mm}& \hspace*{22mm}& \hspace*{22mm}& \hspace*{22mm}\\[-3mm]
 & \multicolumn{2}{c}{{\bf hard-lepton}} & \multicolumn{2}{c}{{\bf soft-lepton}} & \multicolumn{2}{c}{{\bf multi-lepton}}\\
 & {\bf $W$ CR} & {\bf \ttbar~CR} & {\bf $W$ CR} & {\bf \ttbar~CR} & {\bf $Z$ CR} & {\bf \ttbar~CR}\\[1mm]
\hline\hline
$N_{\rm{jet}}$  & $\geq$ 3 & $\geq$ 3 & $\geq$ 2 & $\geq$ 2 & $\geq$ 2 & $\geq 2$ \\
$p_{\rm{T}}^{\rm{jet}}$ (\GeV) & $>$ 80, 25, 25 & $>$ 80, 25, 25 & $>$ 130,25 & $>$ 130,25 &
$>$ 80,50 or & $>$ 80,50 or \\
            &                &                &        &        &
$>$ 50,50,50,50 & $>$ 50,50,50,50 \\
$N_{\rm{jet}}$ ($b$-tagged) & 0 & $\geq$ 1 & 0 & $\geq$ 1 & --- & $\geq$ 1 \\
\hline
\met\ (\GeV) & [40,150] & [40,150] & [180,250] & [180,250] & $< 50$ & [30,80] \\
$m_{\rm{T}}$ (\GeV) & [40,80] & [40,80] & [40,80] & [40,80] & --- & --- \\
$m_{\rm{eff}}^{\rm{inc}}$ (\GeV) & $>$ 500 & $>$ 500 & --- & --- & --- & ---\\
$m_{\ell\ell}$ (\GeV) & --- & --- & --- & --- & [81,101] & $< 81$ or $>101$ \\
\hline\hline
\end{tabular}
\caption{Overview of the
  selection criteria for the $W$+jets, $Z$+jets and \ttbar~control
  regions (CR). Only the criteria that are different from
the signal selection criteria listed in Table \ref{tab:SR} are shown.}
\label{tab:CR}
\end{center}
\end{table*}

Several correction factors are applied to the simulation.  The
\pt~of the $Z$ boson is reweighted based on a comparison of data with
simulation in an event sample enriched in $Z$+jets events.
The same correction factor is applied to $W$~boson
production and improves the agreement between data and simulation
in the \met~distribution.  Other
correction factors are derived during the combined fit.
The relative normalization of the ALPGEN
samples ($W$+jets, $Z$+jets and \ttbar)
with different numbers of partons ($N_{\rm{parton}}$)
in the matrix element is
adjusted by comparing the jet multiplicity distributions in
data and simulation in all control regions.  A common set of
corrections is obtained for the $W$+jets and $Z$+jets samples, and a
separate set of common corrections is obtained for semi-leptonic and
fully-leptonic \ttbar~decays.  Neither the reweighting based on the
\pt~distribution of the $Z$ boson nor the $N_{\rm{parton}}$~weights are 
applied in Figs. \ref{fig:h1lCR}-\ref{fig:h2lCR} below.

\subsection{$W/Z$+jets and \ttbar~Control Regions}

The $W$+jets and \ttbar~processes are isolated in control regions
defined by the
following requirements.  For the hard single-lepton channel,
$\ge 3$ jets are required, with a leading jet \pt~$>$ 80
\GeV~and the other jets above 25 \GeV.  The lepton requirements are the
same as in the signal region.  The \met~is required to be between 40
and 150 \GeV~while the transverse mass is required to be between 40
and 80 \GeV.  Furthermore,
the $m_{\rm{eff}}^{\rm{inc}}$ requirement is relaxed to be $>$ 500
\GeV.  The $W$+jets and \ttbar~control regions are distinguished by
requirements on the number of $b$-tagged jets.
For the $W$+jets control region, events are rejected
if any of the three highest \pt~jets is $b$-tagged; the rejected events
then define the \ttbar~control region.  Table \ref{tab:CR} summarizes
the control region definitions; Fig. \ref{fig:h1lCR}
shows the composition of the $W$+jets and \ttbar~control
regions as a function of $m_{\rm{eff}}^{\rm{inc}}$ and of the jet
multiplicity.  A discrepancy between simulation and data can be seen
in the $m_{\rm{eff}}^{\rm{inc}}$ distribution and is discussed in Sec.~\ref{sec:reweight}.

\begin{figure*}[htbp]
\center
\includegraphics[width=0.49\textwidth]{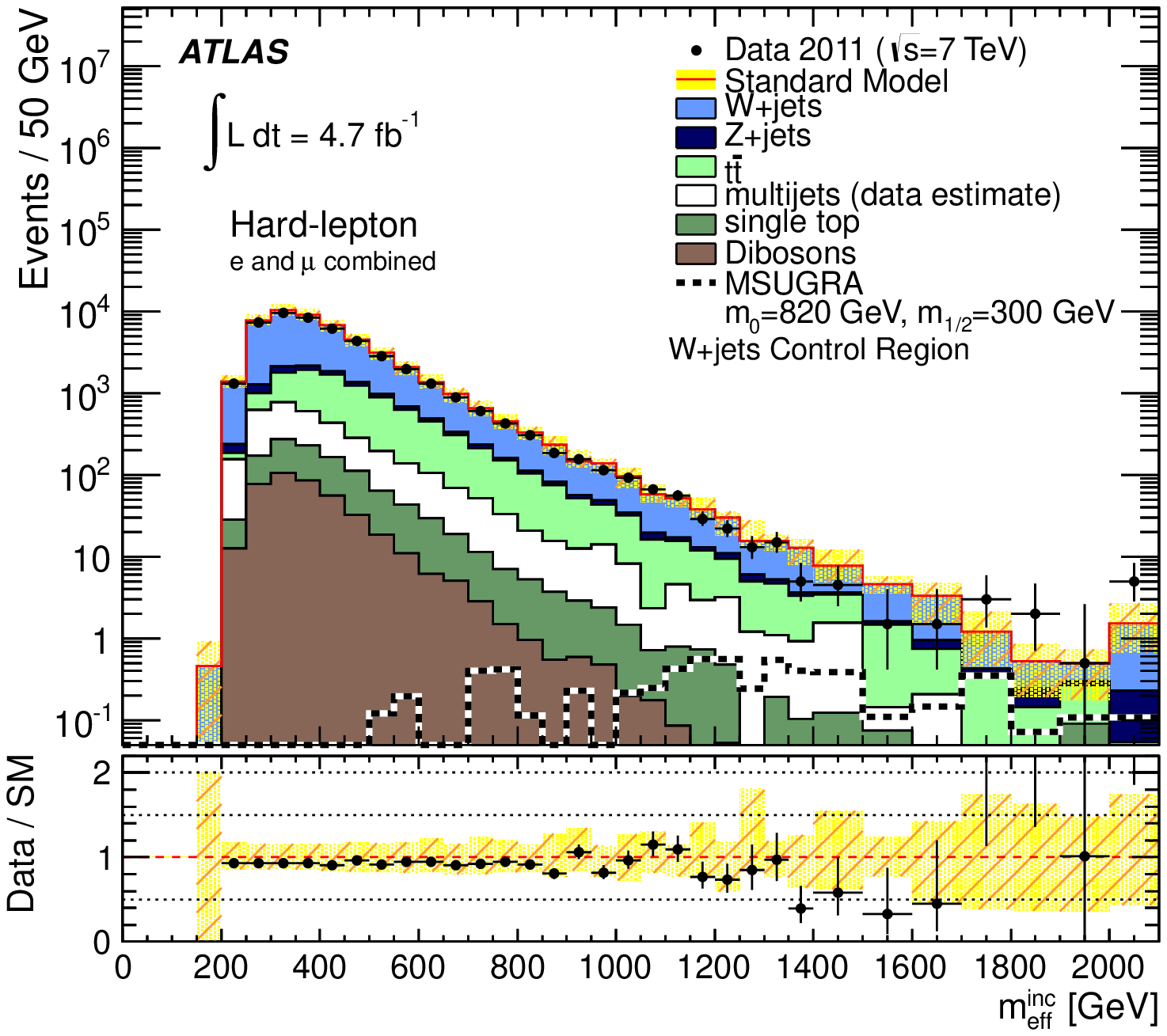}
\includegraphics[width=0.49\textwidth]{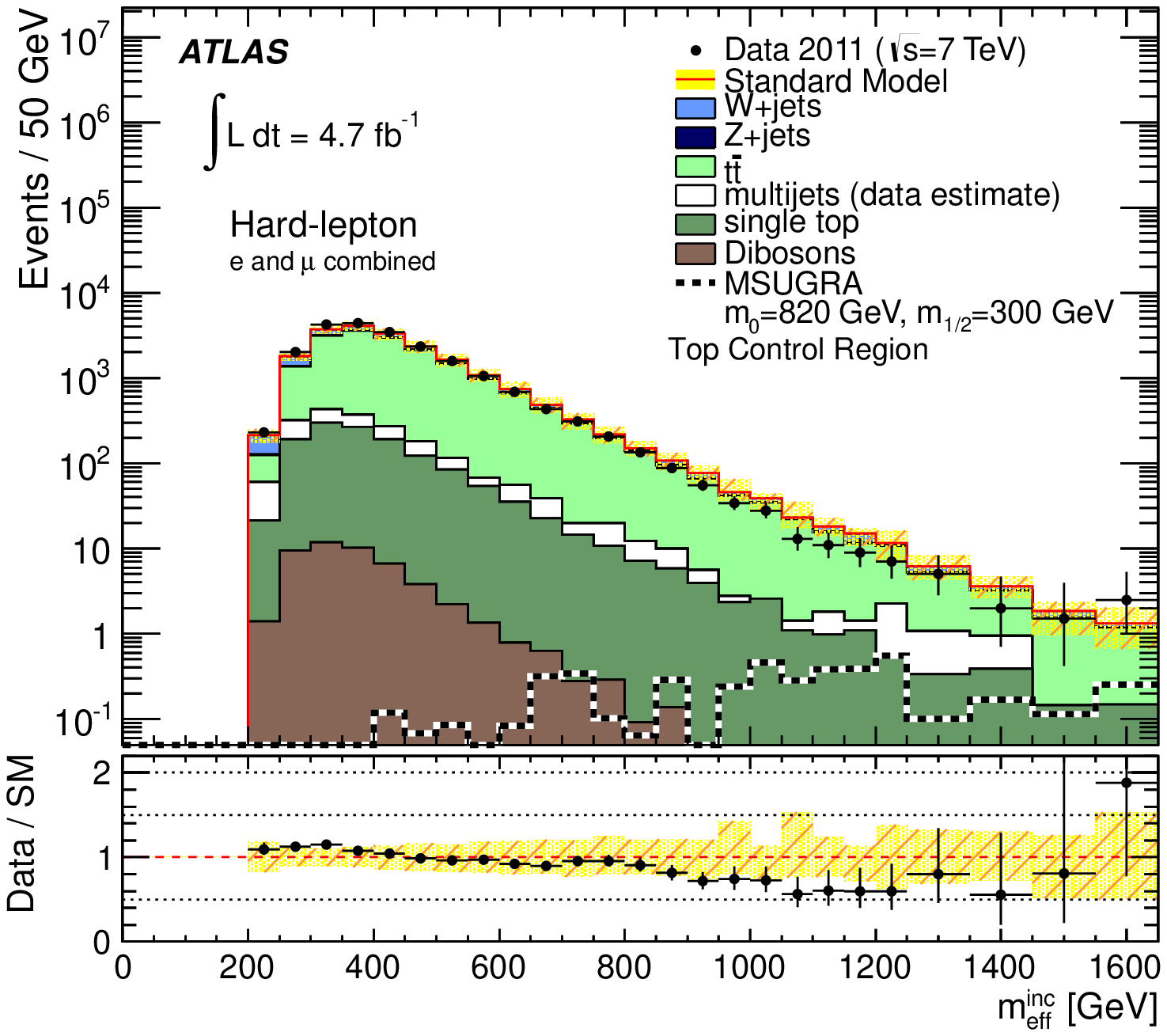}
\newline
\includegraphics[width=0.49\textwidth]{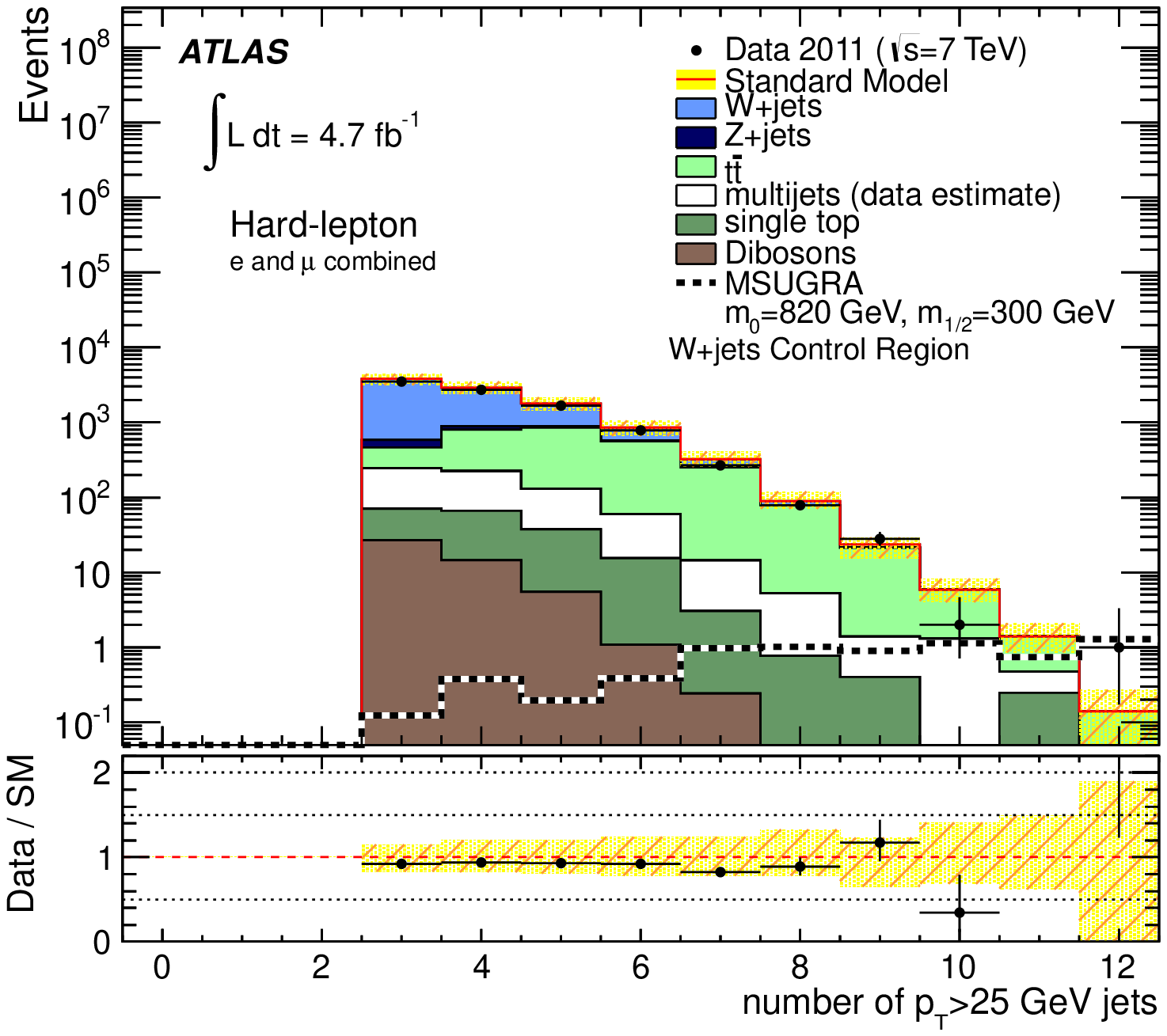}
\includegraphics[width=0.49\textwidth]{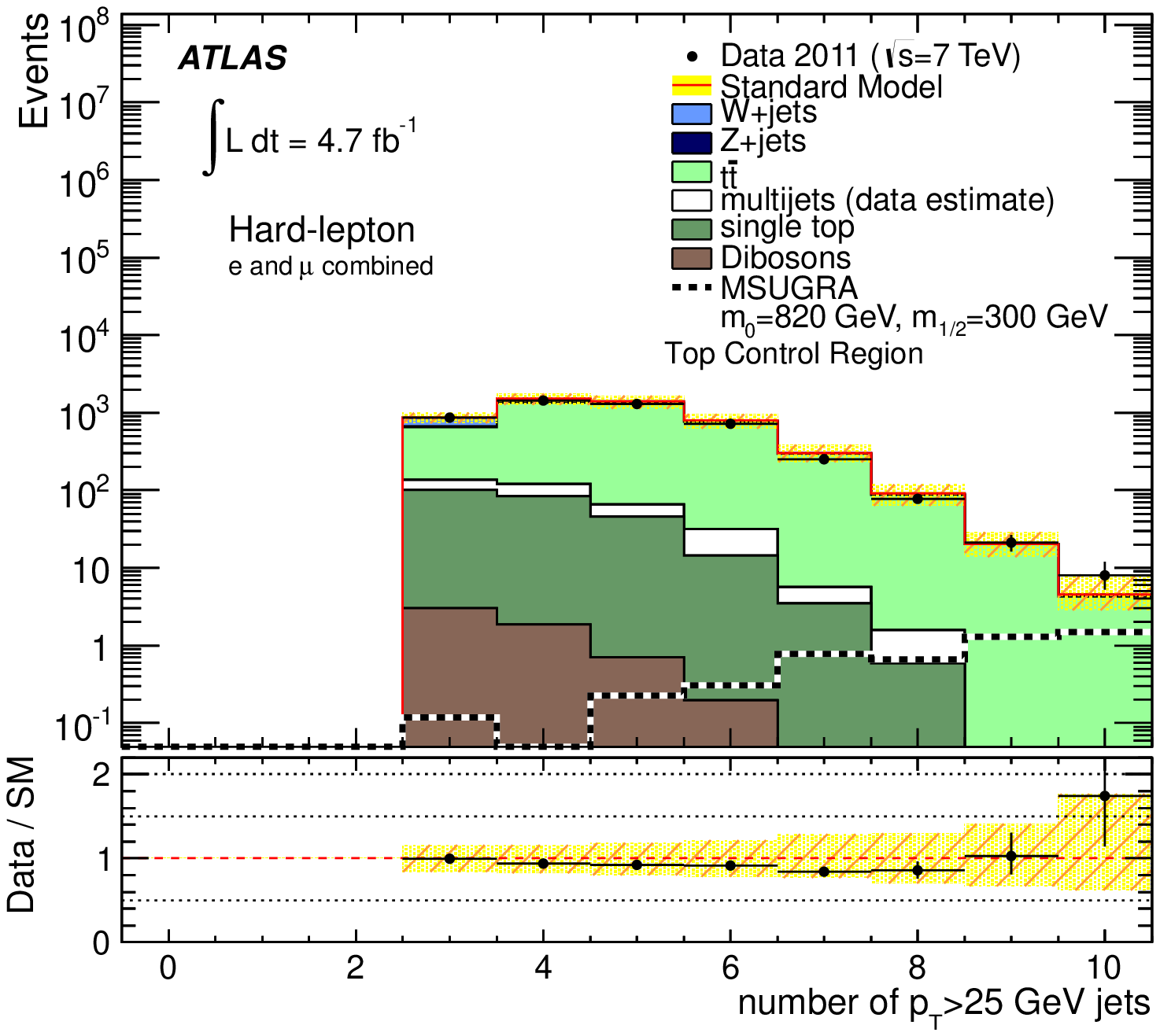}

\caption{ 
  Top: $m_{\rm{eff}}^{\rm{inc}}$ distribution in the
  $W$+jets (left) and \ttbar~(right) control regions
  for data and simulation for the single hard-lepton channels. 
  Bottom: Distribution of the number of jets in the
  $W$+jets (left) and \ttbar~(right) control regions.
  In all plots, the last bin 
  includes all overflows.  The electron and muon channels are
  combined for ease of presentation.  
The ``Data/SM'' plots show the ratio between
data and the total Standard Model expectation.  The expectation for
multijets is derived from the data.  The remaining Standard Model
expectation is entirely derived from simulation, normalized
to the theoretical cross sections. 
The uncertainty band around the Standard Model expectation combines the
statistical uncertainty on the simulated event samples with
the systematic uncertainties on the jet energy scale,
$b$-tagging,
data-driven multijet background, and luminosity. The systematic
uncertainties are largely correlated from bin to bin.  An example of the
distribution for a simulated signal is also shown (not
stacked); the signal point is chosen to be near the exclusion
limit of the analysis in Ref. \cite{ATLAS:2011ad}.}
\label{fig:h1lCR}
\end{figure*}

For the soft-lepton channel, the control region
requirements on the leptons and jets
are the same as in the signal region.  However, the \met~is required to be
between 180 \GeV~and 250 \GeV~and the transverse mass to be
between 40 \GeV~and 80 \GeV. The tighter \met~requirement, compared to the
hard single-lepton control regions, is dictated by the trigger
selection for this channel. Again, the $W$+jets and \ttbar~control regions
are distinguished by the presence of $b$-tagged jets.
For $W$+jets, events are rejected if any of the two highest \pt~jets
is $b$-tagged; the rejected events form the \ttbar~control
region.
Figure \ref{fig:s1lsjCR} shows the
composition of the $W$+jets and \ttbar~control regions for the
soft-lepton channel as a function of
\met/$m_{\rm{eff}}$ and the jet multiplicity.

\begin{figure*}[htbp]
\center
\includegraphics[width=0.49\textwidth]{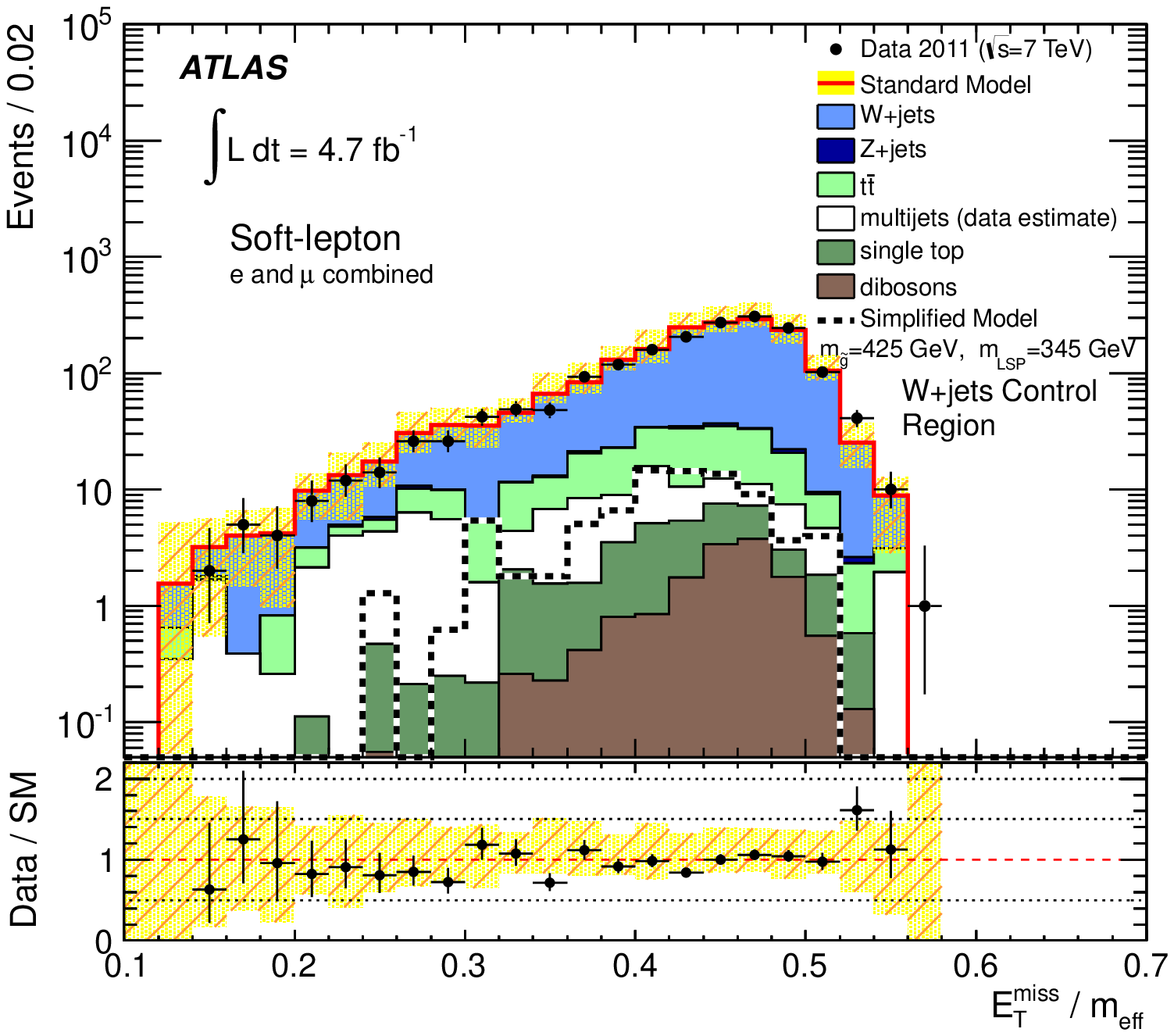}
\includegraphics[width=0.49\textwidth]{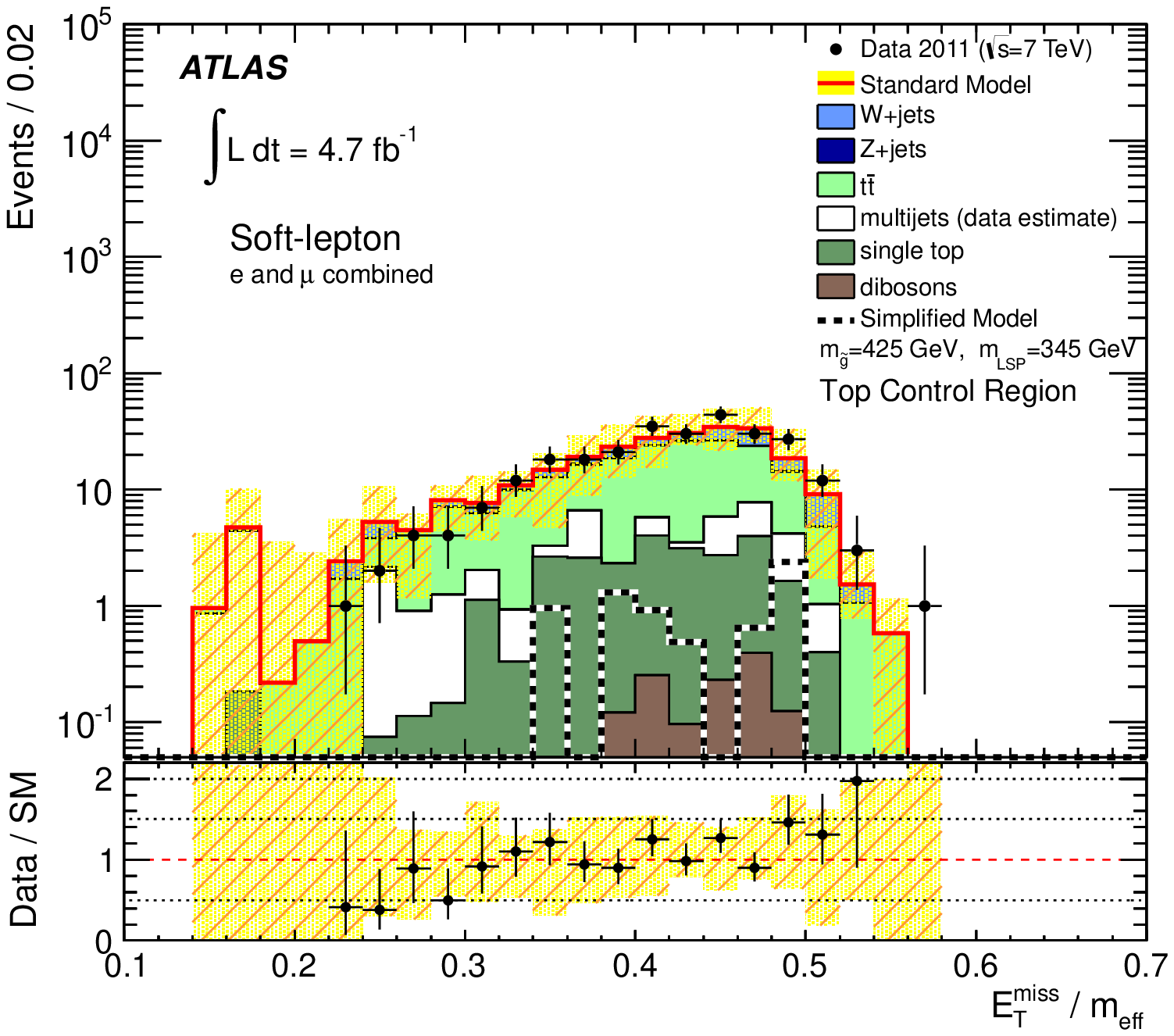}
\includegraphics[width=0.49\textwidth]{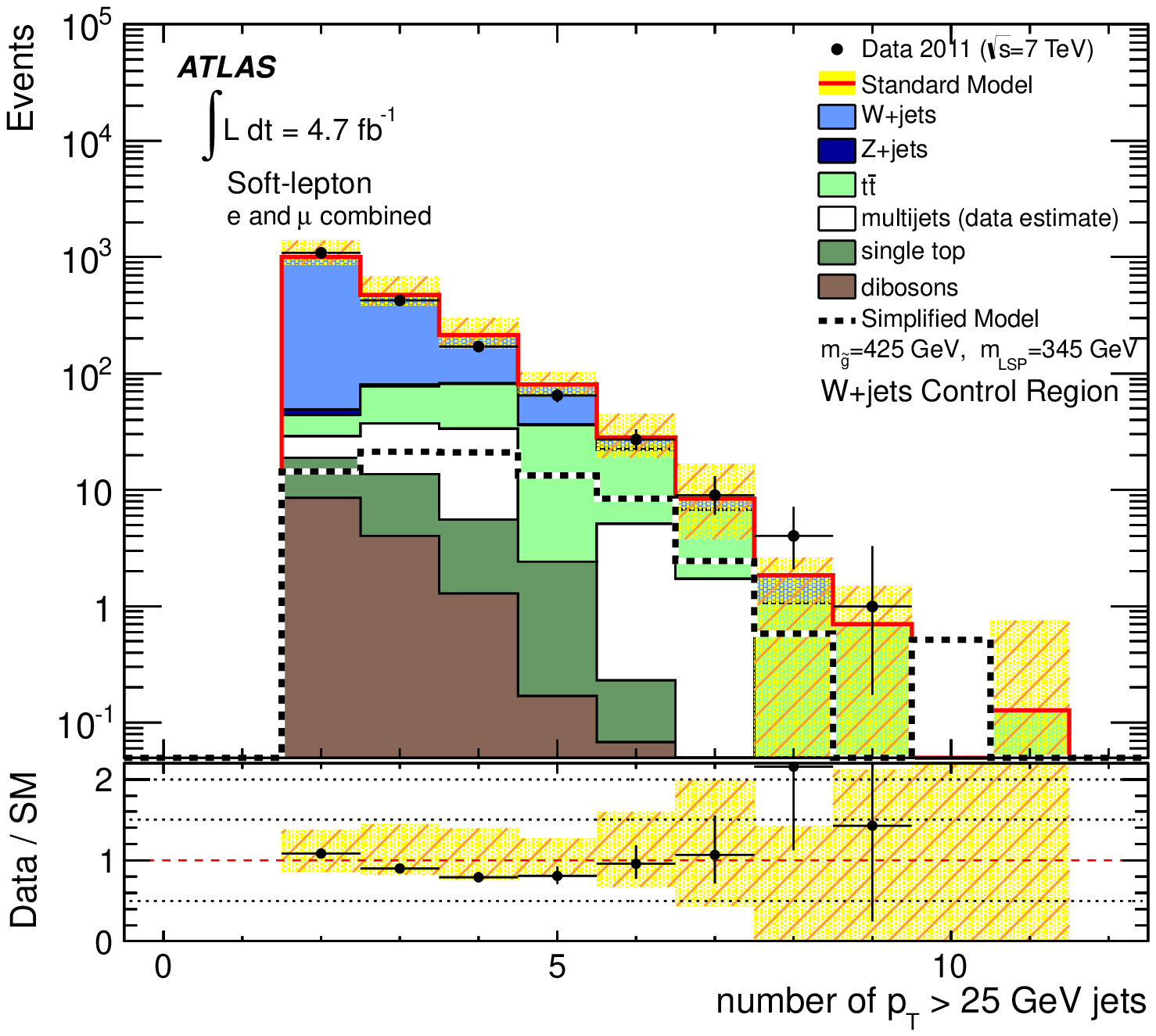}
\includegraphics[width=0.49\textwidth]{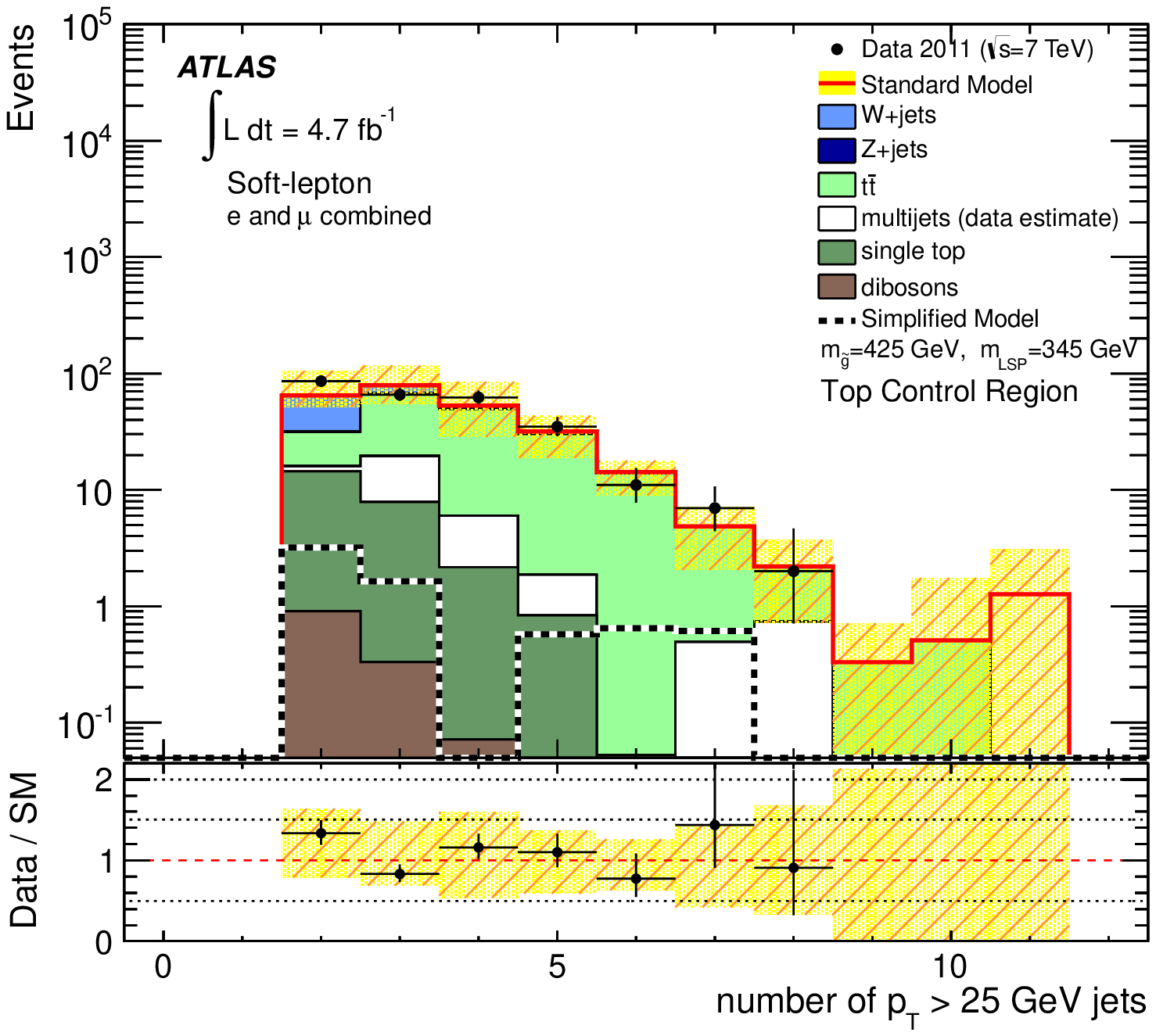}

\caption{
  Top: \met/$m_{\rm{eff}}$ distribution in the
  $W$+jets (left) and \ttbar~(right)
  control regions
  for data and simulation for the soft-lepton channel.
  Bottom: Jet multiplicity distribution
  in the $W$+jets (left) and
  \ttbar~(right) control regions.  In all distributions, electron and
  muon channels are combined.
  The ``Data/SM'' plots show the ratio between
data and the total Standard Model expectation.  The expectation for
multijets is derived from the data.  The remaining Standard Model
expectation is entirely derived from simulation, normalized
to the theoretical cross sections.
The uncertainty band around the Standard Model expectation combines the
statistical uncertainty on the simulated event samples with
the systematic uncertainties on the jet energy scale,
$b$-tagging,
data-driven multijet background, and luminosity. The systematic
uncertainties are largely correlated from bin to bin.  An example of
the distribution for a simulated signal is also shown (not stacked);
the signal point is near the exclusion limit of this analysis.}
\label{fig:s1lsjCR}
\end{figure*}

For the multi-lepton channels, the $Z$+jets control region is defined
by requiring $\ge 2$ jets with the two leading jets having \pt~$>$~80
\GeV~and 50 \GeV, respectively, or with four leading jets having
\pt~$>$~50 \GeV.  In addition, \met~$<$ 50 \GeV~and an
opposite-sign, same-flavor dilepton pair with
invariant mass between 81~\GeV~and 101~\GeV~are required. The lepton selection
requirements are the same as in the signal region.  The
\ttbar~control region is defined with the same jet requirements as the
$Z$+jets control region; at least one jet is required to be
$b$-tagged.   In addition, \met~between 30 \GeV~and 80 \GeV~and a
dilepton invariant mass outside the
window [81,101] \GeV~are required. Figure \ref{fig:h2lCR} (top) shows the
composition of the $Z$+jets and \ttbar~control regions for the
multi-lepton channel as a function of
$m_{\rm{eff}}^{\rm{inc}}$.

\begin{figure*}[htbp]
\center
\includegraphics[width=0.49\textwidth]{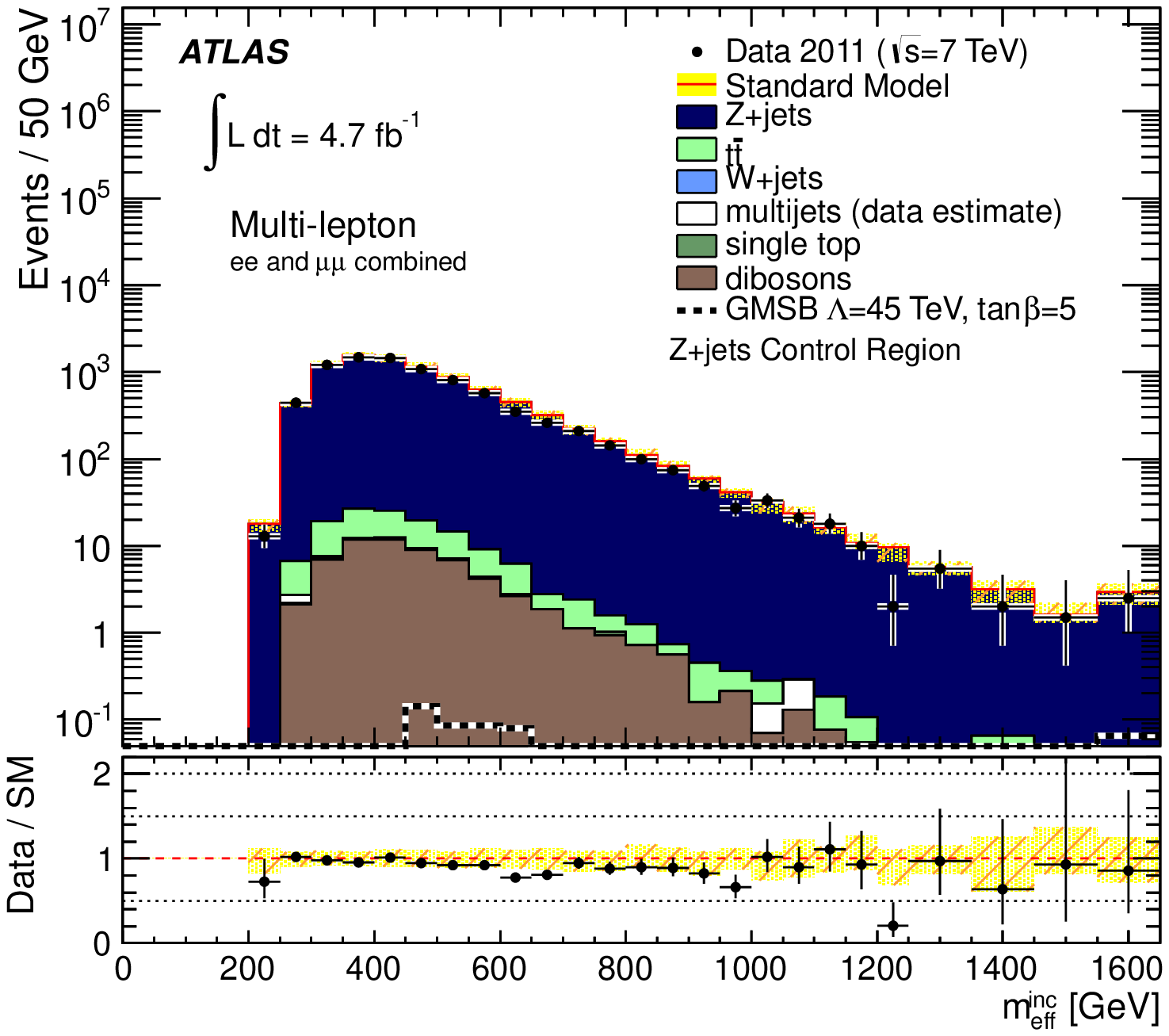}
\includegraphics[width=0.49\textwidth]{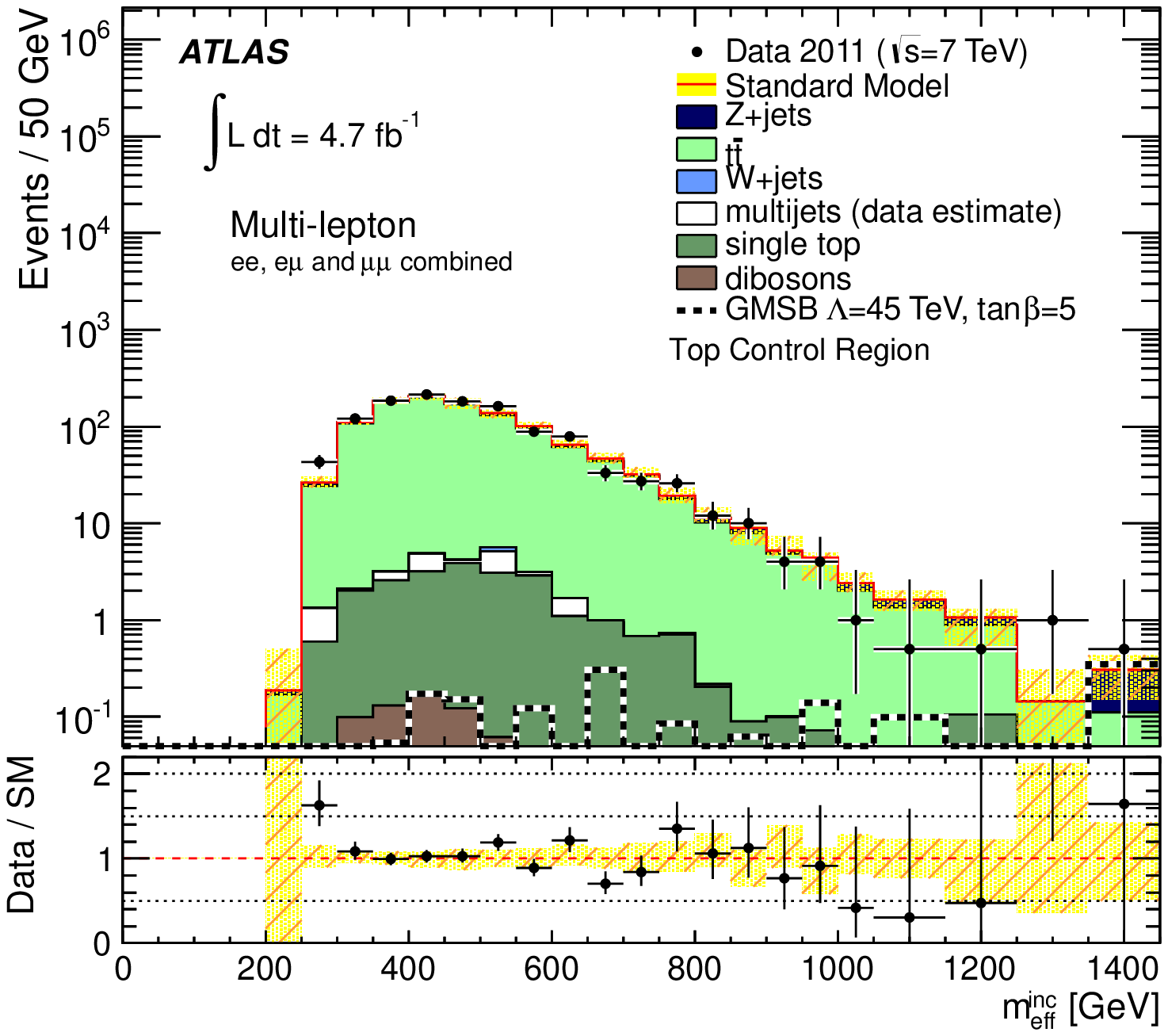}
\newline
\includegraphics[width=0.49\textwidth]{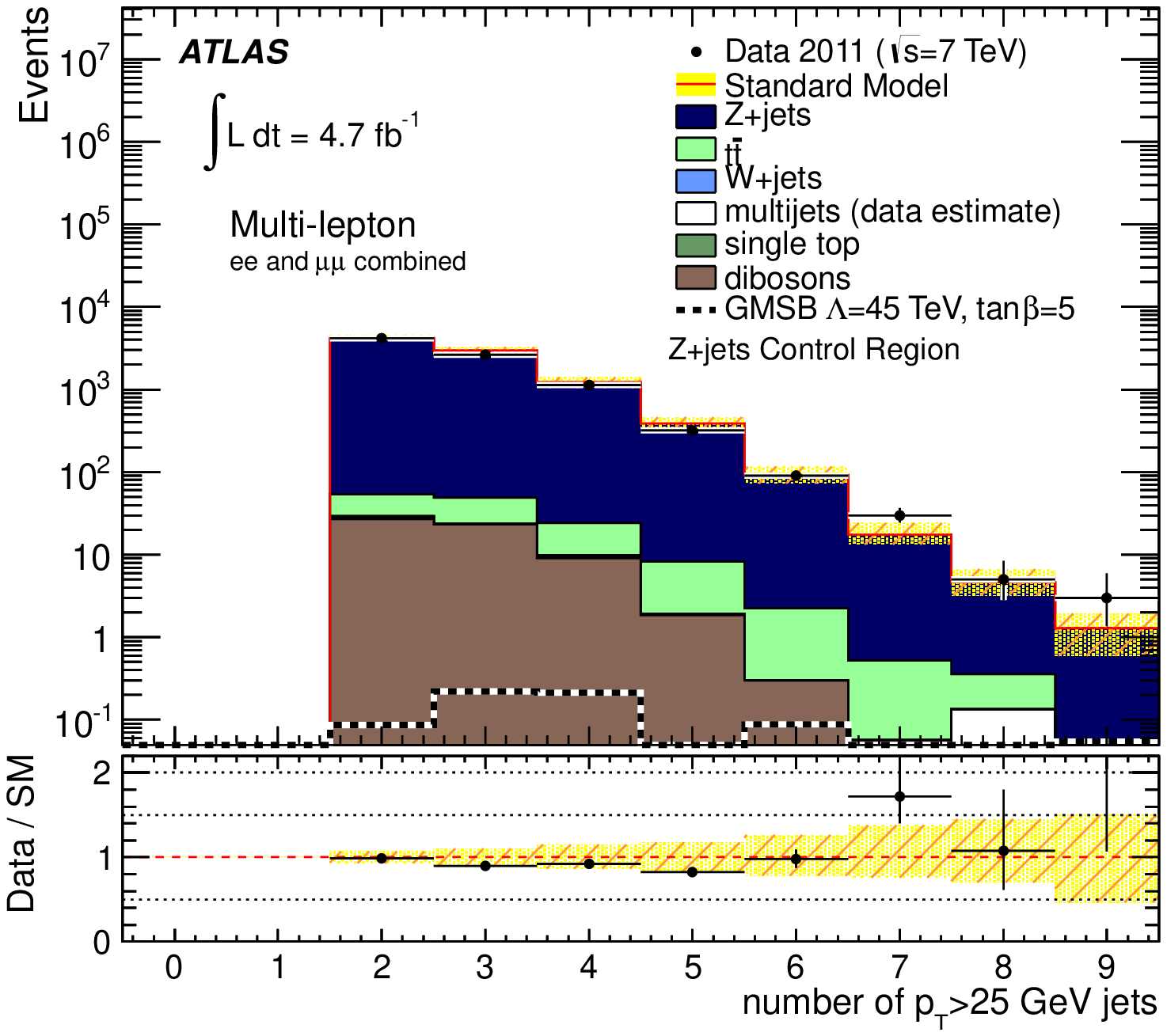}
\includegraphics[width=0.49\textwidth]{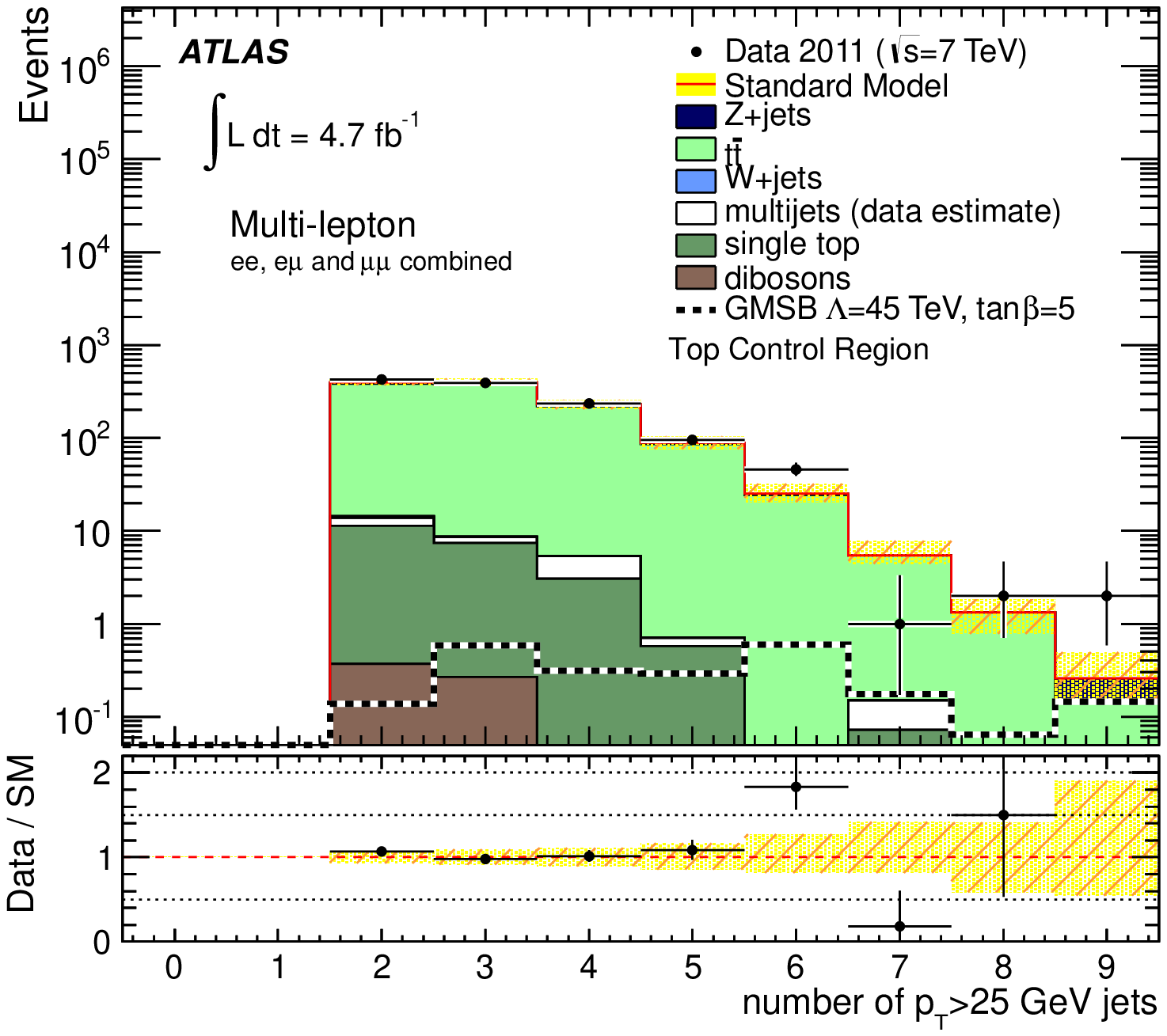}

\caption{ 
  Top: $m_{\rm{eff}}^{\rm{inc}}$ distribution in the
  $Z$+jets (left) and \ttbar~(right) control regions
  for data and simulation for the multi-lepton channels. 
  Bottom: Distribution of the number of jets in the
  $Z$+jets (left) and \ttbar~(right)
  control regions; the last bin 
  includes all overflows.  The $ee$ and  $\mu\mu$ channels are
  combined for $Z$+jets and $ee$, $\mu\mu$ and $e\mu$ channels are
  combined for the \ttbar~distributions for ease of presentation.
The ``Data/SM'' plots show the ratio between
data and the total Standard Model expectation.  The expectation for
multijets is derived from the data.  The remaining Standard Model
expectation is entirely derived from simulation, normalized
to the theoretical cross sections. 
The uncertainty band around the Standard Model expectation combines the
statistical uncertainty on the simulated event samples with
the systematic uncertainties on the jet energy scale,
$b$-tagging,
data-driven multijet background, and luminosity. The systematic
uncertainties are largely correlated from bin to bin. An example of the
distribution for a simulated signal is also shown (not
stacked); the signal point is chosen to be near the exclusion
limit of the analysis in Ref. \cite{Aad:2012rt,ATLAS:2012ag}.}
\label{fig:h2lCR}
\end{figure*}

\subsection{Reweighting of $W$+jets and $Z$+jets Simulated Samples}
\label{sec:reweight}

The samples of simulated $W$+jets and $Z$+jets events are reweighted as a
function of the generated \pt~of the vector boson.
A common set of corrections to the \pt~of the
vector boson, applied to both
$W$+jets and $Z$+jets samples, is found to improve the agreement
between data and simulation for a number of variables (\met,
$m_{\rm{eff}}^{\rm{inc}}$, and jet \pt).

The $\pt^{Z}$~distribution is measured in data by selecting a sample with two
oppositely-charged, same-flavor leptons with an invariant mass between
80 \GeV~and 100 \GeV, $\ge 3$ signal jets with \pt~$>$25 \GeV, and
$m_{\rm{eff}}^{\rm{inc}} > 400$~\GeV.   The
$\pt^{Z}$~distribution in five bins of reconstructed \pt~is compared
to the ALPGEN simulation in five bins
of generated \pt, with the first four bins ranging from 0 to
200 \GeV~and the last bin integrated above 200 \GeV; the ratio of
the two distributions is taken as the $\pt^{Z,\rm{gen}}$-dependent
weighting factor.  The simulation  employed
here uses the cross sections listed in Table \ref{tab:MC}.
Only the systematic uncertainty from the
jet energy scale is considered (in addition to statistical
uncertainties) when computing the uncertainty on the
weighting factors.

Figure \ref{fig:zptfit} (top) shows the $\pt^{Z}$~distribution before the
application of the reweighting factors and after the final fit to all
background control regions (described in Sec. \ref{sec:bkgfit}), which
includes the reweighting.
The bottom half of the figure shows the \met~ distribution in the
hard-lepton $W$+jets
control region (with the lower requirement on \met~set to 50 GeV and
the upper requirement removed).

\begin{figure*}[htbp]
\center
\includegraphics[width=0.49\textwidth]{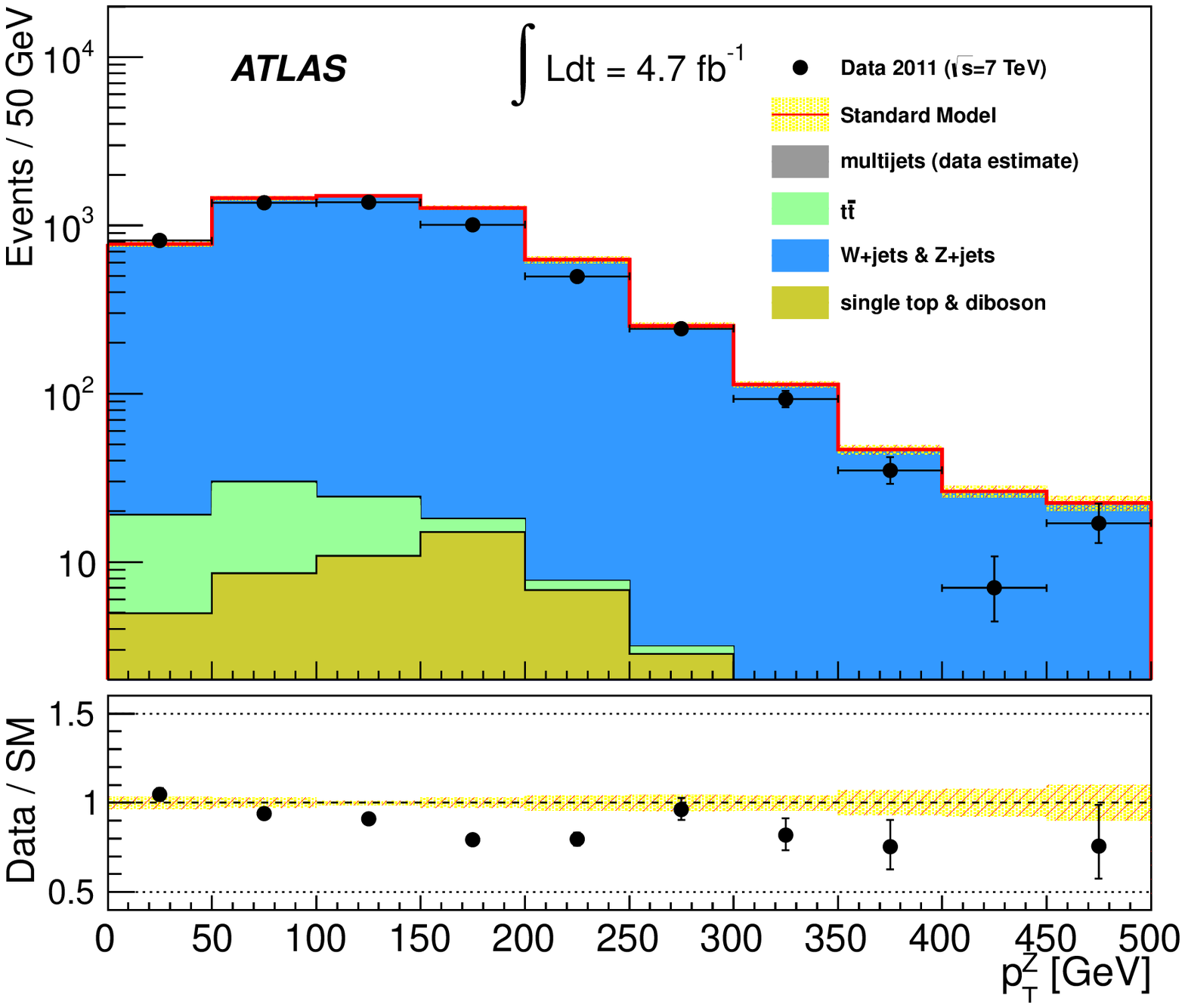}
\includegraphics[width=0.49\textwidth]{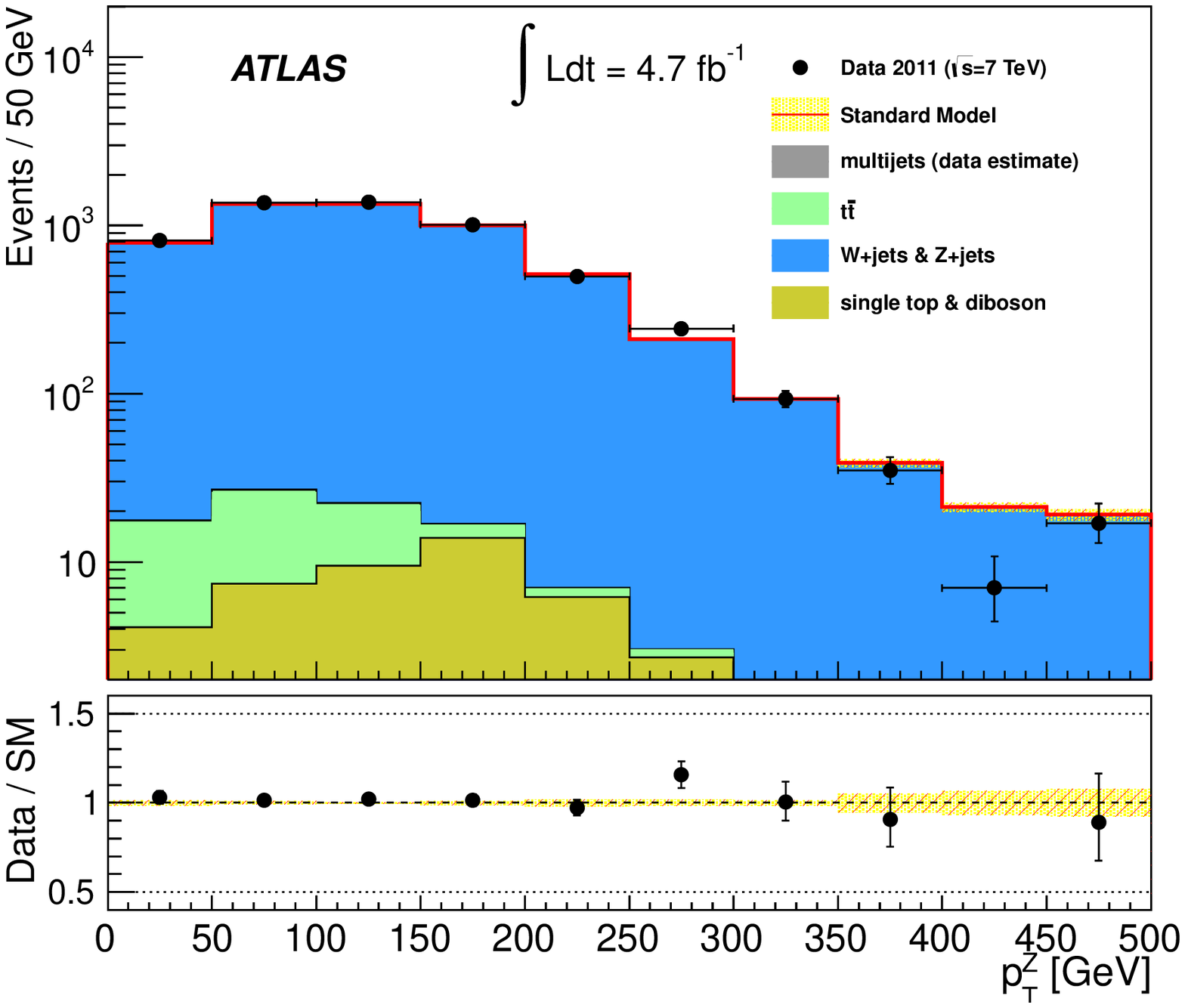}
\newline
\includegraphics[width=0.49\textwidth]{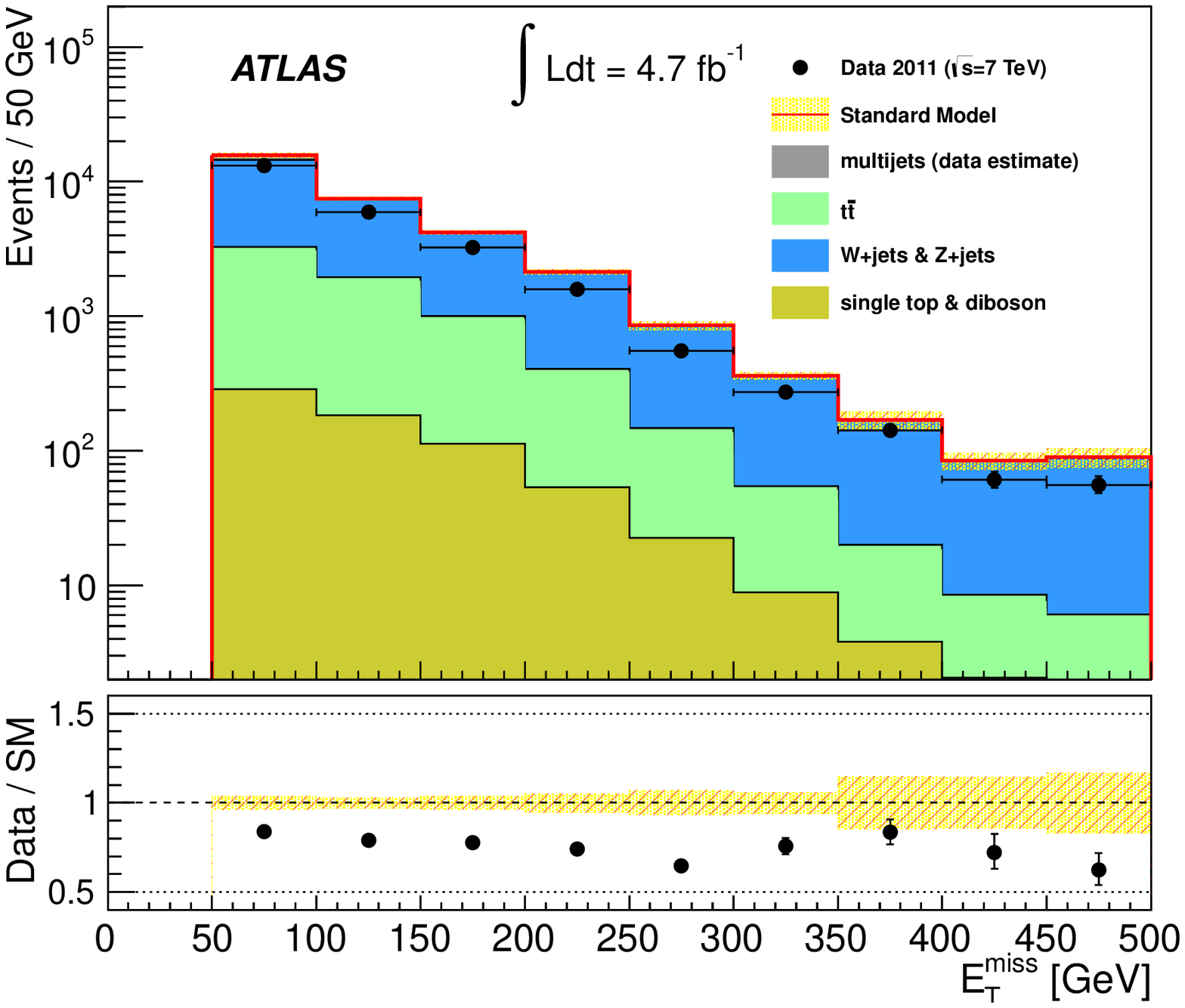}
\includegraphics[width=0.49\textwidth]{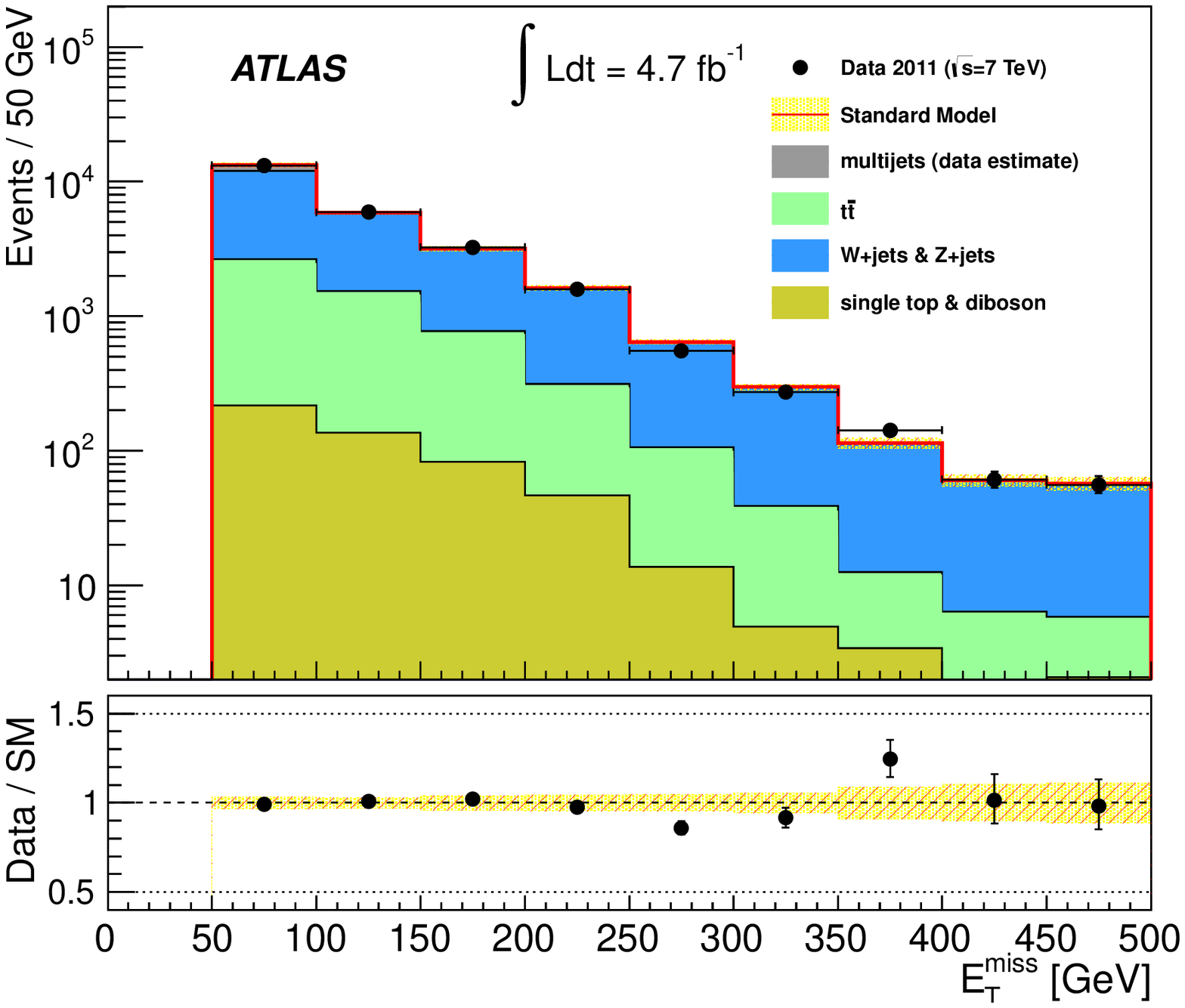}
\caption{
  Top: Distribution of the \pt~of the $Z$ boson in a region enhanced
  in $Z$+jets events ($ee$ and $\mu\mu$ final states combined) before
  (left) the application of any reweighting factors, and after
  (right) the final fit to all background control regions (described
  in Sec. \ref{sec:bkgfit}). 
  Bottom: Distribution of \met~in the hard-lepton $W$+jets control
  region (electron and muon channels combined, lower requirement on
  \met~set to 50 GeV and upper requirement removed) before (left) application of reweighting factors and after
(right) the final fit to all background control regions. Events in the
overflow bin have not been plotted.}
\label{fig:zptfit}
\end{figure*}

\subsection{Multijet Background}

Multijet events
become a background when a jet is misidentified as an isolated lepton
or when a real lepton appears as a decay product of hadrons in jets,
for example from $b$- or $c$-jets, and is sufficiently isolated.  In
the following, such lepton-like objects
are collectively referred to as misidentified leptons.
The multijet background in each signal region, and in the
$W$+jets and \ttbar~control regions, where it is more significant,
is estimated from the data following a matrix
method similar to that employed in Ref. \cite{ATLAS:2011ad}.  

The multijet background from all sources (but separated by lepton
flavor) is determined collectively.  In the single-lepton channels,
the multijet process is
enhanced in control samples with all the signal region criteria
applied but where the lepton isolation criteria are not imposed and
the shower shape requirements on electrons are relaxed.
Defining $N_{\mathrm{pass}}$ and $N_{\mathrm{fail}}$ as the number of events in such a loose
sample passing or failing the final lepton selection criteria,
and defining $N_{\mathrm{real}}$ and $N_{\mathrm{misid.}}$ as the number of real and
the number of misidentified leptons,
the following equations hold:
\begin{displaymath}
N_{\mathrm{pass}} = \epsilon_{\mathrm{real}} N_{\mathrm{real}} + 
\epsilon_{\mathrm{misid.}} N_{\mathrm{misid.}} ,
\end{displaymath}
\begin{displaymath}
N_{\mathrm{fail}} = (1 - \epsilon_{\mathrm{real}}) N_{\mathrm{real}} + 
( 1- \epsilon_{\mathrm{misid.}}) N_{\mathrm{misid.}} ,
\end{displaymath}
where $\epsilon_{\mathrm{real}}$ is the relative identification efficiency for
real leptons, and $\epsilon_{\mathrm{misid.}}$ is the misidentification
efficiency for misidentified leptons. Solving the equations leads to:
\begin{displaymath}
N_{\mathrm{misid.}}^{\mathrm{pass}} = \epsilon_{\mathrm{misid.}} N_{\mathrm{misid.}} =
\frac{N_{\mathrm{fail}} - (1/\epsilon_{\mathrm{real}} - 1) N_{\mathrm{pass}}}
{1/\epsilon_{\mathrm{misid.}} - 1/\epsilon_{\mathrm{real}}} .
\end{displaymath}
The efficiency $\epsilon_{\mathrm{real}}$ is measured from data samples of
$Z~\rightarrow~\ell\ell$ decays. 

The lepton misidentification efficiency is obtained as follows.
For electrons (muons) with \pt~$>$
25 (20)~\GeV\ $\epsilon_{\rm{misid.}}$ is
estimated with events containing at least one electron (muon)
satisfying the relaxed criteria, and at least one signal jet with
\pt~$>$ 30 (60) \GeV. In addition, for the electron case, \met~$<$ 30
\GeV~is required. For the muon case, the event is required to
contain exactly one muon with  $|d_{0}|/\sigma_{d_{0}} > 5$ where
$d_{0}$ and $\sigma_{d_{0}}$ are
the transverse impact parameter and its uncertainty, respectively,
measured with respect to the primary vertex.
For the soft-lepton channel, the sample for deriving
$\epsilon_{\rm{misid.}}$
consists of events containing a same-sign and same-flavor
lepton pair where the leptons satisfy the relaxed isolation criteria.
The selection of a lepton pair allows the low-\pt~region to be studied
with a large data sample.  The same-sign requirement reduces the
dominance of $b$-hadrons in the sample, providing a better mix of
the different mechanisms by which leptons can be misidentified.
One of the leptons is required to fail the signal
lepton criteria to further enhance the background;
the misidentification efficiency is measured with the
other lepton. An additional veto around the
$Z$~boson mass is applied.  In all channels, the electron misidentification
efficiency is evaluated separately for samples enhanced (depleted)
in heavy-flavor contributions by requiring (vetoing) a $b$-tagged jet
in the event.

For the multi-lepton channels, the misidentification probabilities as
determined above are applied to the number of events where two leptons
pass the loose selection criteria.  The contribution from processes
where one lepton is real and the other misidentified has been studied
in both simulation and data, using a generalization
of the above matrix method to two leptons.  Both methods give similar
results; the final estimate is taken from
the simulation.

\subsection {Other Backgrounds}

The backgrounds from single-top, diboson and \ttbar + vector boson production are
estimated almost purely from simulation, as is the $Z$+jets background
for the single-lepton channels. The
background from cosmic-ray muons overlapping a hard-scattering event
is estimated from a control
sample with large $z_{0}$, defined as the distance in
the $z$ direction with respect to the primary vertex, evaluated at the
point of closest approach of the muon to the primary vertex in the
transverse plane.  The extrapolated contribution to the signal region,
$|z_{0}| < 5$~mm, is found to be negligible.

\section{Systematic Uncertainties}
\label{sec:systematics}

Systematic uncertainties have an impact on the expected background and
signal event yields in the control- and signal regions.
These uncertainties are
treated as nuisance parameters in a profile likelihood fit described in
Sec. \ref{sec:bkgfit}.
The following systematic uncertainties on the reconstructed objects
are taken
into account.
The jet energy scale (JES)
uncertainty has been determined from a combination of test beam,
simulation and in-situ measurements from 2010 $pp$ collision data
\cite{Aad:2011he}.  Additional contributions 
from the higher luminosity and pile-up in 2011 are taken into account.
Uncertainties on the lepton identification, momentum/energy scale and
resolution are estimated from samples of $Z \rightarrow \ell^{+}\ell^{-}$,
$J/\psi \rightarrow \ell^{+}\ell^{-}$ and
$W^{\pm} \rightarrow \ell^{\pm}\nu$ decays in data \cite{Aad:2011mk,ATLAS-CONF-2011-021,ATLAS-CONF-2011-063}.  The
uncertainties on the jet and lepton energies are propagated to the
\met; an additional \met~ uncertainty arising from energy deposits not
associated with reconstructed objects is also included \cite{Aad:2012re}.
Uncertainties on the $b$-tagging efficiency are derived from dedicated
data samples 
\cite{ATLAS-CONF-2012-043, ATLAS-CONF-2011-143}, e.g. containing muons
associated with jets.  Uncertainties on the light-flavor
mis-tag rate are derived by examining tracks with negative impact parameter
\cite{ATLAS-CONF-2012-040} while charm mis-tag uncertainties are
obtained from data samples tagged by reconstructing $D^{\ast}$ mesons
\cite{ATLAS-CONF-2012-039}.

Uncertainties in the matrix method for the determination of the
multijet background include the statistical uncertainty
in the number of events available in the various control samples, the
difference in
misidentification efficiency for electrons from heavy- versus
light-flavored jets, the dependence of the misidentification
efficiency on the jet multiplicity, and the uncertainty in the
subtraction of other backgrounds from the samples used to estimate the
misidentification efficiency.

Uncertainties from the 
identification efficiency for 
jets associated with the primary vertex and from the overlay of
pile-up in simulated events are both found to be negligible.

Theoretical modeling uncertainties in the simulation include the
following contributions.
In previous versions of the analysis, renormalization and
factorization
scale uncertainties were estimated by varying the corresponding
parameters in the ALPGEN generator by a factor of two, up and down
from
their nominal settings. Since these variations affect mostly the
overall
normalization of the cross sections for the samples with different 
values of $N_{\rm{parton}}$, they are replaced here by a normalization of the
individual light-parton bins to the data (see Sec.
\ref{sec:bkgfit}).
Additional generator uncertainties arise from the 
parameter that describes the jet \pt~threshold used in the
matching ($p_{\rm{T,min}}$). This uncertainty is assessed
by changing its default value from 15 \GeV~to 30 \GeV; the difference
is assigned as both a positive and negative uncertainty.
Uncertainties arising from initial- and final-state radiation
are taken into account by the variation of the MLM matching
parameter in multi-leg generators as well as by studying dedicated
PYTHIA tunes with increased or decreased radiation \cite{Skands:2010ak}. 
Fragmentation/hadronization uncertainties are
estimated by comparing HERWIG with PYTHIA.
In order to vary the heavy-flavor fraction, the
cross sections for $Wb\overline{b}$+jets and $Wc\overline{c}$+jets in
Table \ref{tab:MC} are scaled by
$1.63\pm0.76$, while $Wc$+jets is scaled by $1.11\pm0.35$, based
on correction factors derived from data \cite{ATLAS:2012an}.  The
uncertainty on $Zb\overline{b}$+jets is taken to be  $\pm$100\%.  The
uncertainties on the cross sections for \ttbar+$W$ and \ttbar+$Z$ are
taken from the NLO calculations in Refs. \cite{Campbell:2012dh,Lazopoulos:2008de}.

The uncertainty in the signal cross section is taken from an envelope
of cross section predictions using different PDF sets (including the
$\alpha_{\rm S}$ uncertainty) and
factorization and renormalization scales, as described in
Ref. \cite{Kramer:2012bx}. 
 For the simplified models, uncertainties in the modeling of
initial-state radiation play a significant role for low 
gluino masses and for small mass differences in the decay cascade.
These uncertainties are estimated by varying generator tunes in
the simulation as well as by generator-level studies
of $\tilde{g}\tilde{g}$ and
production with an additional ISR jet generated
in the matrix element with MADGRAPH5.

The impact of these systematic uncertainties on the background yields
and signal estimates are evaluated via an overall fit, described in
Sec. \ref{sec:bkgfit} and \ref{sec:results}.

\section{Background Fit}
\label{sec:bkgfit}

The background in the signal region is estimated with a fit based on
the profile likelihood method \cite{Cowan:2010js}.
The inputs to the
fit are as follows:
\begin{enumerate}
  \item The observed numbers of events in the $W$+jets (or $Z$+jets in
    the multi-lepton channels)
    and \ttbar~control
    regions, and the numbers expected from simulation. These are
    separated into 7 jet-multiplicity bins, ranging from
    3 to 9 jets for the hard-lepton channel, 8 jet multiplicity bins
    ranging from 2 to 9 jets for the
    multi-lepton channels, and 6 bins ranging from 2 to 7 jets for the
    soft-lepton channel.
    This information is shown
    in the bottom half of Fig. \ref{fig:h1lCR}
    to \ref{fig:h2lCR}.
  \item Transfer factors (TF), derived from simulation, are
    multiplicative factors that propagate
    the event counts for $W$+jets, $Z$+jets and \ttbar~backgrounds
    from one control region to another, or from one control region to
    the signal region. Typical values of the TFs from the control
    to the signal region are $10^{-2}$ and $10^{-4}$ for the
    soft- and hard-lepton channels, respectively.
  \item The number of multijet background events
    in all control and signal regions, as
    derived from the data.
  \item Expectations from simulation for the number of events from 
    the minor backgrounds
    (single-top, diboson) in all
    control and signal regions.
\end{enumerate}
For each analysis channel (hard-lepton, soft-lepton, multi-lepton) the event
count in each bin of the control region 
is treated with a Poisson probability density function.
The statistical and systematic uncertainties on the
expected yields are included in the probability density function as
nuisance parameters, constrained to be Gaussian with a width
given by the size of the uncertainty.  Approximately 150 nuisance
parameters are included in the fit.  Correlations in the nuisance
parameters from bin to bin are taken into account where necessary.
The Poisson probability density
functions also include free parameters, for example to scale the expected
contributions from the major backgrounds; these are described in more
detail below.  A likelihood is formed as
the product of these probability density functions and the constraints
on the nuisance parameters.  Each lepton
flavor (in the multi-lepton channel, each combination of flavors
of the two leading leptons) is treated separately in the likelihood function.
The free parameters and nuisance
parameters are adjusted to maximize the likelihood.  An important
difference with respect to the analysis in Ref. \cite{ATLAS:2011ad} is the
increase in the number of measurements, allowing the 
fit to be constrained.
This has been used in this analysis to constrain the nuisance
parameters for the jet energy scale and the uncertainty
in the ALPGEN scale parameters from the shape information
provided in the control regions.  

The free parameters considered in the fit are as follows:
\begin{enumerate}
  \item \ttbar~background: Each \ttbar~sample, broken down by
    $N_{\rm{parton}}$ bin (from 0 to 3, with the last being inclusive), is
    scaled by a free
    parameter.  For each $N_{\rm{parton}}$ bin, a common parameter is used
    for semi-leptonic and dileptonic \ttbar~samples.  
  \item $W/Z$ background: Each $W$+jets and $Z$+jets sample, again
    broken down by $N_{\rm{parton}}$ bin from 2 to 5, is scaled
    by a free parameter. The $N_{\rm{parton}}=6$ bin for $W$ +
    light-flavored jets shares its fit parameter with
    $N_{\rm{parton}}=5$.  The vector boson plus heavy-flavor samples
    share the same relative normalization parameters as the
    light-flavor samples.  Only $N_{\rm{parton}}$ bins
    between two and five are allowed to float, as the lower
    multiplicity bins suffer from small numbers of events due to
    the jet and effective mass requirements.
    
\end{enumerate}
The backgrounds from multijets and the sub-dominant backgrounds from
single-top and diboson production are allowed to float in the fit
within their respective uncertainties.

Notable nuisance parameters in the fit are:
\begin{enumerate}
  \item The uncertainty in the
    ALPGEN MLM-matching parameter
    $p_{\rm {T,min}}$ manifests itself in the relative
    normalization of the ALPGEN $N_{\rm{parton}}$ samples and in the
    jet \pt~spectra within each sample.  The change
    in the event counts in the array of all control regions,
    resulting from this shift
    in the relative normalization, is mapped to one parameter for
    both $W$+jets and $Z$+jets and a separate parameter for \ttbar.
  \item The uncertainty in the normalization of the
    $N_{\rm{parton}}=0,1$ bins for $W$+jets and $Z$+jets, due to
    uncertainties in renormalization and factorization scales, is
    treated by one nuisance parameter.
  \item The overall normalization of the vector boson plus heavy
    flavor samples is assigned a nuisance parameter reflecting the
    uncertainty in the cross section.
  \item The uncertainty from the fit of the $\pt^{Z}$~distribution is
    treated by assigning one nuisance parameter for each bin in true
    \pt.  Four equal-width bins are used from 0 to 200 GeV, and one
    bin for \pt~$>$ 200 GeV.
  \item The uncertainty due to the jet energy scale is considered in
    three jet \pt~bins (25--40 \GeV, 40--100 \GeV~and $>$ 100 \GeV).  The
    resulting uncertainty in the event counts in the array of all control
    regions is mapped to one
    nuisance parameter for  each of the three jet \pt~bins.  The usage
    of three jet \pt~bins prevents the fit from artificially
    over-constraining the jet energy scale.
\end{enumerate}

\subsection{Background Fit Validation}

The background fit is cross-checked in a number of validation regions,
situated between the control and signal regions,
where the results of the background fit can be compared to
observation.  These validation regions are not used to constrain the fit.
For the single hard-lepton channels, one common set of
validation regions, which receives contributions from both 3- and
4-jet channels,
is defined as follows:
\begin{enumerate}
  \item The $W$+jets validation region is identical to the $W$+jets
    control region for the 3-jet channel
    except that the \met~ requirement is changed to 150~\GeV~$<$ \met
    $<$ 250 \GeV~(from [40, 150] \GeV). 
  \item Similarly, the \ttbar~validation region is identical to
    the \ttbar~control region for the 3-jet channel
    except for the change in the
    \met~requirement to 150~\GeV~$<$ \met $<$ 250 \GeV~(from [40, 150] \GeV).
  \item The high transverse mass validation region is defined
    by $m_{\rm{T}} > 100$~\GeV~and 40 \GeV~ $<$ \met $<$ 250 \GeV.
    This region tests the validity
    of the background yields from dileptonic \ttbar~events.
\end{enumerate}
For the soft-lepton channel, the validation region
is based on the sum of the $W$+jets and
\ttbar~control regions but with the
transverse mass selection changed to 80~\GeV~$< m_{\rm{T}}
<$~100~\GeV~(from [40, 80] \GeV). 

For the multi-lepton channels, two $Z$+jets validation  regions and
two \ttbar~validation regions are
defined:
\begin{enumerate}
  \item The 2-jet $Z$+jets validation region is similar to
    the $Z$+jets control region with $\ge$~2 jets, but  the
    leading two jets are required to have \pt~$>$120~\GeV~(instead of 
    80~\GeV~and 50~\GeV); the fourth
    leading jet (if present) is required to have \pt~$<$50 \GeV.
  \item The 4-jet $Z$+jets validation region  is similar
    to the control region with
    $\ge$~4 jets but the leading jet \pt~ requirement is tightened to
    \pt~$>$ 80~\GeV~(instead of 50~\GeV).
  \item The 2-jet \ttbar~validation region is similar to the
    \ttbar~control region with $\ge$~2 jets but the leading two jets
    are required to have \pt~$>$ 120~\GeV~(instead of 
    80~\GeV~and 50~\GeV); the fourth leading jet (if
    present) is required to have \pt~$<$~50 \GeV. The
    \met~requirement is changed to $100 \GeV < \met < 300$~\GeV.  
  \item The 4-jet \ttbar~validation region is similar to the
    \ttbar~control region with $\ge$~4 jets but the leading jet
    \pt~requirement is tightened to \pt~$>$ 80~\GeV~(instead of 50~\GeV).
    The \met~requirement is tightened to $80 \GeV < \met < 100$~\GeV.
\end{enumerate}
In both
$Z$+jets validation regions the \met~requirement is tightened to
$50 \GeV < \met < 100$~\GeV, and the number
of $b$-tagged jets is required to be zero in order to suppress
the \ttbar~contamination.  The selection requirements for the
validation regions are summarized in Tables \ref{tab:VR1} and
\ref{tab:VR2} for the single-lepton and multi-lepton channels,
respectively. 

\begin{table*}[ht!]
\begin{center}
\begin{tabular}{lcccc}
\hline\hline
 & \hspace*{27mm} & \hspace*{27mm} & \hspace*{27mm} & \hspace*{27mm} \\[-3mm]
 & \multicolumn{3}{c}{{\bf hard-lepton}} &   {\bf soft-lepton} \\[1mm]
 & {\bf $W$ VR} & {\bf \ttbar~VR} & {\bf High-$\rm{m}_{\rm{T}}$~VR} & \\
\hline\hline
$N_{\rm{jet}}$  & $\geq$ 3 & $\geq$ 3 & $\geq$ 3 & $\geq$ 2 \\
$p_{\rm{T}}^{\rm{jet}}$ (\GeV) & $>$ 80, 25, 25 & $>$ 80, 25, 25 & $>$~80, 25, 25 & $>$ 130,25 \\
$N_{\rm{jet}}$ ($b$-tagged) & 0 & $\geq$ 1 & --- & \\
\hline
\met\ (\GeV) & [150,250] & [150,250] & [40,250] & [180,250] \\
$m_{\rm{T}}$ (\GeV) & [40,80] & [40,80] & $>$~100 & [80,100] \\
$m_{\rm{eff}}^{\rm{inc}}$ (\GeV) & $>$ 500 & $>$ 500 & $>$ 500 & -- \\
\hline\hline
\end{tabular}
\caption{Overview of the
  selection criteria for the background validation
  regions (VR) for the single-lepton channels. Only the criteria that are different from
the signal selection criteria listed in Table \ref{tab:SR} are shown.}
\label{tab:VR1}
\end{center}
\end{table*}

\begin{table*}[ht!]
\begin{center}
\begin{tabular}{lcccc}
\hline\hline
 & \hspace*{27mm} & \hspace*{27mm} & \hspace*{27mm} & \hspace*{27mm} \\[-3mm]
 & \multicolumn{2}{c}{{\bf multi-lepton 2-jet}} &
\multicolumn{2}{c}{{\bf multi-lepton 4-jet}}  \\[1mm]
 & {\bf $Z$ VR} & {\bf \ttbar~VR} & {\bf $Z$ VR} & {\bf \ttbar~VR}\\
\hline\hline
$N_{\rm{jet}}$  & $\geq$ 2 & $\geq$ 2 & $\geq$ 4 & $\geq$ 4 \\
$p_{\rm{T}}^{\rm{jet}}$ (\GeV) & $>$ 120, 120 & $>$ 120, 120 & $>$~80,50,50,50 & $>$~80,50,50,50 \\
$N_{\rm{jet}}$ ($b$-tagged) & --- & $\geq$ 1 & --- & $\geq$ 1 \\
\hline
\met\ (\GeV) & [50,100] & [100,300] & [50,100] & [80,100]\\
$m_{\ell\ell}$ (\GeV) & [81,101] & $<$ 81 or $>$ 101 & [81,101] & $<$
81 or $>$ 101 \\ 
\hline\hline
\end{tabular}
\caption{Overview of the
  selection criteria for the background validation
  regions (VR) for the multi-lepton channels. Only the criteria that are different from
the signal selection criteria listed in Table \ref{tab:SR} are shown.  For
the 2-jet validation regions, the fourth leading jet (if present) is
required to have \pt~$<$~50 GeV.}
\label{tab:VR2}
\end{center}
\end{table*}

The results of the fit to the control regions, as well as the
comparison of observed versus predicted event counts in the validation
regions, are summarized in
Fig. \ref{fig:pulls}.  The difference between the observed and
predicted event counts is normalized by the total (statistical and
systematic) uncertainty on the prediction.  The
agreement between predicted and observed yields is good.

\begin{figure*}[htbp]
\center
\includegraphics[width=0.45\textwidth]{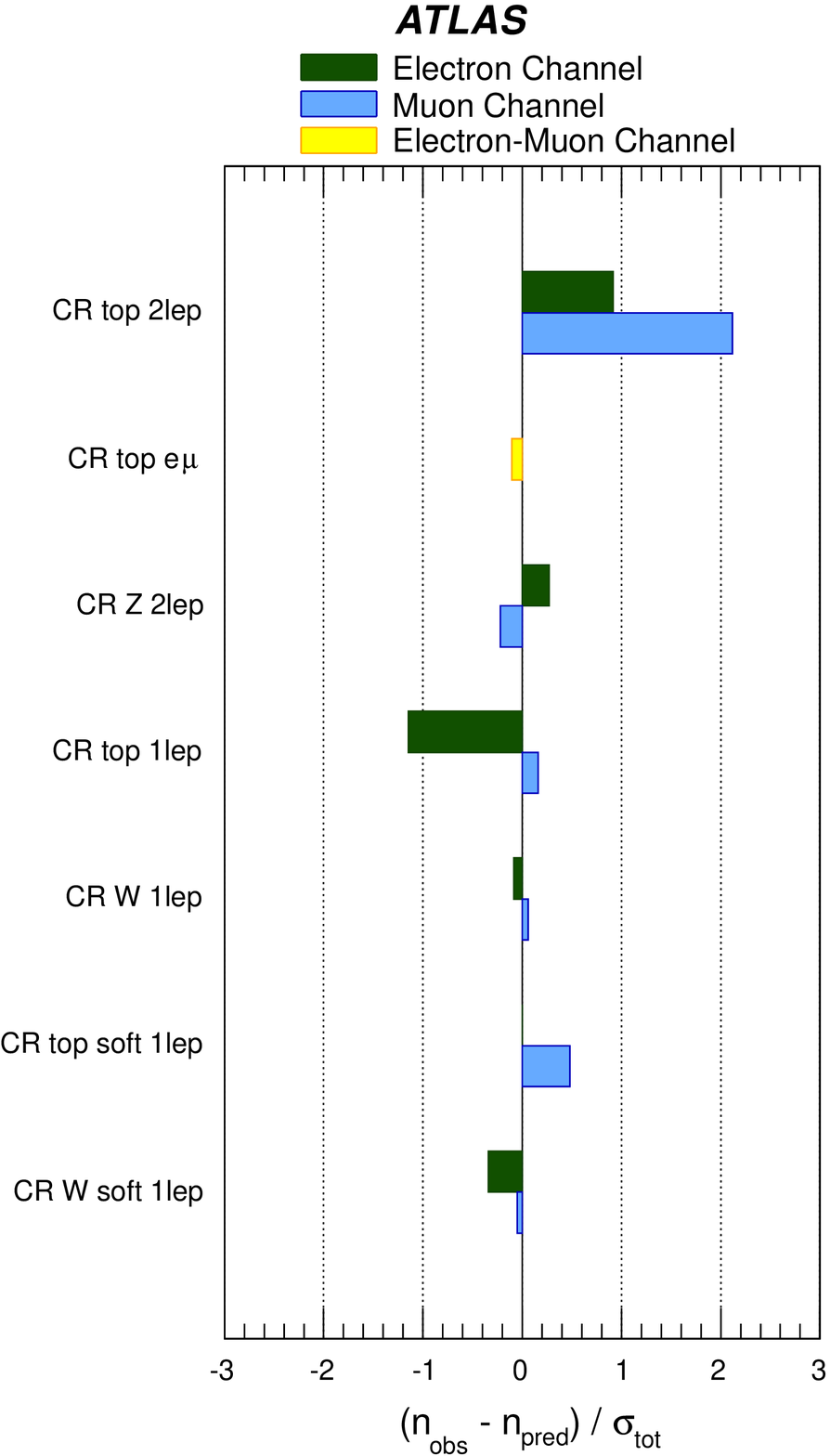}
\includegraphics[width=0.45\textwidth]{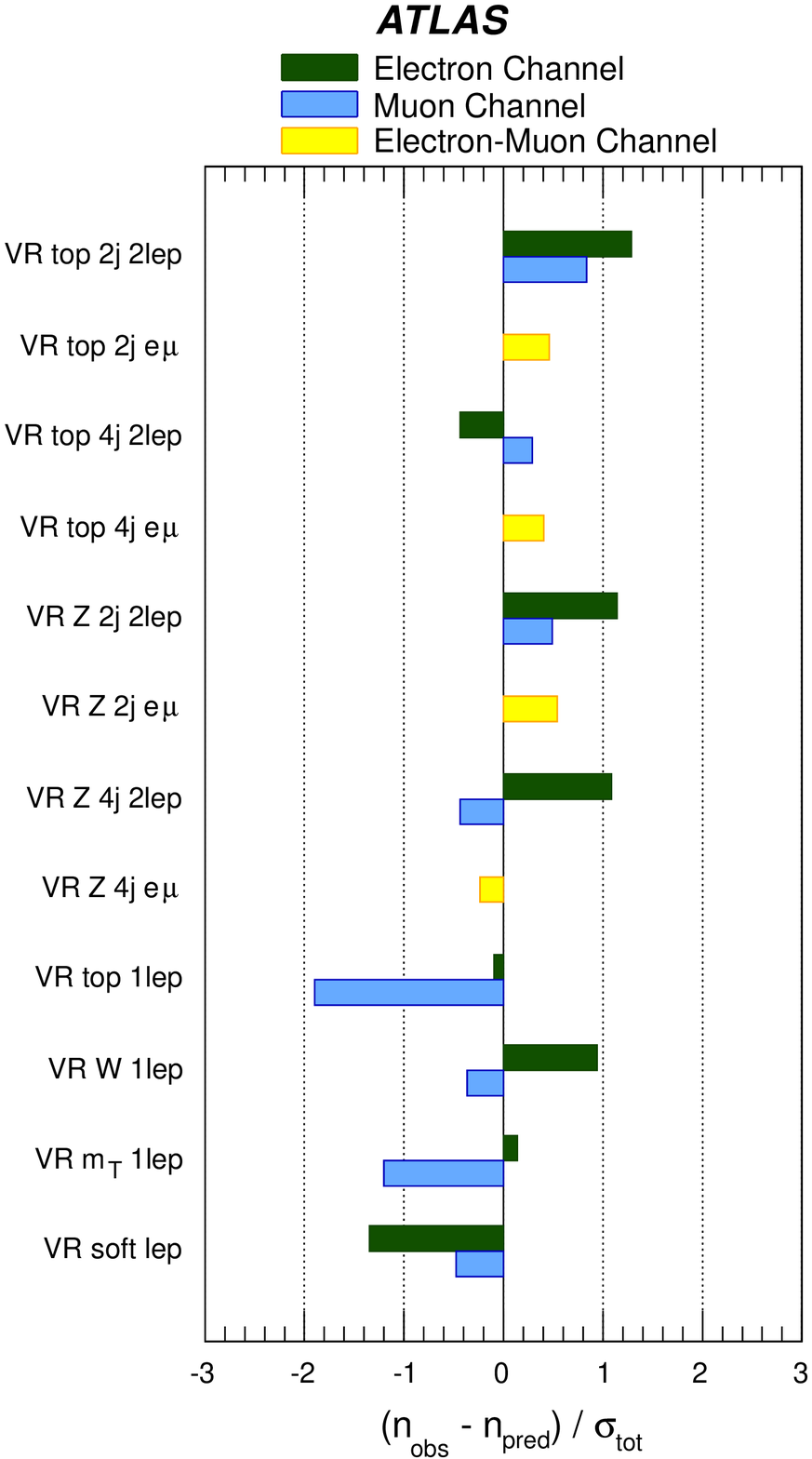}
\caption{
  Summary of the fit results in the control regions (left) and
  validation regions (right).  The difference between the observed
  and predicted number of events, divided by the  total (statistical and
  systematic) uncertainty on the prediction, is shown for each control
and validation region.}
\label{fig:pulls}
\end{figure*}

\section{Results and Interpretation}
\label{sec:results}

The predicted background in the signal regions and the observed numbers
of events are shown in Tables \ref{tab:DiscoveryFitOneLepton} and
\ref{tab:DiscoveryFitMultiLepton}.
The data are consistent with SM expectations in all signal regions.

The dominant background uncertainty
comes from the limited number of events in the background simulation
samples in the signal region.  Uncertainties on the jet energy scale
and the scale uncertainties for the \ttbar~background at high jet
multiplicity are also significant.  In the
soft-lepton channel, an important contribution comes from the
evaluation of the multijet background.

For the signal prediction,
the dominant uncertainties at the highest excluded SUSY
masses arise from the PDFs (30--40\%) and the
JES (10--20\%);
the former reflect the uncertainty in the gluon distribution at high
values of $x$.
In the simplified models with small mass differences typical
uncertainties from ISR variations are approximately 30\%.

\begin{table*}[htbp]
\begin{center}
\setlength{\tabcolsep}{0.0pc}
{\small
\begin{tabular*}{\textwidth}{@{\extracolsep{\fill}}lcccccc}
\noalign{\smallskip}\hline\hline\noalign{\smallskip}
{\bf Single-lepton} & \multicolumn{3}{c}{Electron} & \multicolumn{3}{c}{Muon}\\
Number of events & 3-jet & 4-jet & soft lepton & 3-jet & 4-jet & soft lepton  \\[-0.05cm]
\noalign{\smallskip}\hline\hline\noalign{\smallskip}
Observed  & $2$  & $4$  &  $11$ & $1$  & $2$  &  $14$                  \\
\noalign{\smallskip}\hline\noalign{\smallskip}
Fitted bkg  &
$2.3 \pm 0.9$ & $3.5 \pm 0.9$ & $14.0 \pm 3.3$ &  $2.6 \pm 0.8$ & $1.5 \pm 0.3$ & $19 \pm 5$         \\
\noalign{\smallskip}\hline\noalign{\smallskip}
Fitted top  &
$0.4 \pm 0.2$ & $2.3 \pm 0.6$ & $3.8 \pm 0.6$ &  $0.5 \pm 0.2$ & $1.3 \pm 0.3$ & $3.8 \pm 0.8$        \\
Fitted $W/Z$+jets  &
$1.5 \pm 0.6$ & $0.9 \pm 0.2$ & $5.8 \pm 1.0$ &  $2.0 \pm 0.6$ & $0.2 \pm 0.1$ & $11.4 \pm 2.3$        \\
Fitted other bkg   &
$0.0 \pm 0.0$ & $0.0^{+0.3}_{-0.0}$  & $0.6 \pm 0.1$ &  $0.1 \pm 0.1$ & $0.0 \pm 0.0$ & $0.2 \pm 0.1$         \\
Fitted multijet  &
$0.3\pm0.4$ & $0.3\pm0.4$ & $3.8\pm2.5$ &  $0.0\pm0.0$ & $0.0\pm0.0$ & $3.6\pm2.5$ \\
\noalign{\smallskip}\hline\noalign{\smallskip}
MC exp. SM  & $2.7$ & $5.3$ & $14.2$ & $2.8$ & $2.4$ & $18.0$           \\
\noalign{\smallskip}\hline\noalign{\smallskip}
MC exp. top  & $0.9$ & $3.1$ & $4.3$ & $0.6$ & $2.0$ & $3.8$         \\
MC exp. $W/Z$+jets   & $1.5$ & $1.3$ & $5.5$ & $2.0$ & $0.3$ & $10.5$          \\
MC exp. other bkg  & $0.0$ & $0.5$ & $0.5$ & $0.2$ & $0.1$ & $0.1$       \\
Data-driven multijet  & 0.3 & 0.3 & 3.8 & 0.0 & 0.0 & 3.6 \\
\noalign{\smallskip}\hline\hline\noalign{\smallskip}\noalign{\smallskip}
\end{tabular*}
}
\end{center}
\caption{ The observed numbers of events in the single-lepton signal
  regions, and the background expectations from the fit. 
  The inputs to the fit are
  also shown; these consist of the data-driven multijet background
  estimate and the  
nominal expectations from simulation (MC), normalized to theoretical
cross sections. 
  The errors shown are the statistical plus systematic
  uncertainties.}
\label{tab:DiscoveryFitOneLepton}
\end{table*}

\begin{table*}[htbp]
\begin{center}
\setlength{\tabcolsep}{0.0pc}
{\small
\begin{tabular*}{\textwidth}{@{\extracolsep{\fill}}lcccccc}
\noalign{\smallskip}\hline\hline\noalign{\smallskip}
{\bf Multi-lepton} & \multicolumn{3}{c}{2-jets} &
\multicolumn{3}{c}{4-jets}\\
Number of events & $ee$ & $\mu\mu$ & $e\mu$ & $ee$ & $\mu\mu$ & $e\mu$  \\[-0.05cm]
\noalign{\smallskip}\hline\hline\noalign{\smallskip}
Observed  & $0$  & $0$  &  $1$ & $8$  & $12$  &  $18$                  \\
\noalign{\smallskip}\hline\noalign{\smallskip}
Fitted bkg  &
$0.3 \pm 0.2$ & $0.4 \pm 0.2$ & $ 0.7 \pm 0.2$ & $9.1 \pm 1.5$ & $11.7 \pm 1.7$ & $21 \pm 3$         \\
\noalign{\smallskip}\hline\noalign{\smallskip}
Fitted top  &
$0.1 \pm 0.1$ & $0.2 \pm 0.1$ & $0.6 \pm 0.2$ &  $9.1 \pm1.4$ & $11.1 \pm 1.7$ & $20 \pm 3$        \\
Fitted $W/Z$+jets  &
$0.1 \pm 0.1$ & $0.1 \pm 0.0$ & $0.0 \pm 0.0$ & $0.0 \pm 0.0$ & $0.2 \pm 0.1$ & $0.4 \pm 0.1$        \\
Fitted other bkg   &
$0.1 \pm 0.1$ & $0.1 \pm 0.0$ & $0.1 \pm 0.0$ & $0.0 \pm 0.0$ & $0.4 \pm 0.1$ & $0.6 \pm 0.1$         \\
Fitted multijet  &
$0.0\pm0.0$ & $0.0\pm0.0$ & $0.0\pm0.0$ & $0.0\pm0.2$ & $0.0\pm0.0$ & $0.0\pm0.0$ \\
\noalign{\smallskip}\hline\noalign{\smallskip}
MC exp. SM  & $0.3$ & $0.5$ & $0.9$ & $11.4$ & $14.7$ & $27.1$           \\
\noalign{\smallskip}\hline\noalign{\smallskip}
MC exp. top  & $0.2$ & $0.3$ & $0.7$ & $11.1$ & $13.9$ & $26.0$         \\
MC exp. $W/Z$+jets   & $0.1$ & $0.1$ & $0.1$ & $0.1$ & $0.3$ & $0.4$     \\
MC exp. other bkg  & $0.1$ & $0.1$ & $0.1$ & $0.2$ & $0.5$ & $0.7$      \\
Data-driven multijet  & 0.0 & 0.0 & 0.0 & 0.0 & 0.0 & 0.0 \\
\noalign{\smallskip}\hline\hline\noalign{\smallskip}\noalign{\smallskip}
\end{tabular*}
}
\end{center}
\caption{ The observed numbers of events in the multi-lepton signal
  regions, and the background expectations from the fit. 
  The inputs to the fit are
  also shown; these consist of the data-driven multijet background
  estimate and the  
nominal expectations from simulation (MC), normalized to theoretical
cross sections. 
  The errors shown are the statistical plus systematic uncertainties.}
\label{tab:DiscoveryFitMultiLepton}
\end{table*}

\begin{table*}
\begin{center}
\setlength{\tabcolsep}{0.0pc}
\begin{tabular*}{\textwidth}{@{\extracolsep{\fill}}lcccc}
\noalign{\smallskip}\hline\hline\noalign{\smallskip}
{\bf Signal channel}                        & $\langle\epsilon{\rm \sigma}\rangle_{\rm obs}^{95}$[fb]  &  $S_{\rm obs}^{95}$  & $S_{\rm exp}^{95}$ & $CL_{B}$ \\
\noalign{\smallskip}\hline\hline\noalign{\smallskip}
hard electron, 3-jet   & $0.94$  & 4.4  & ${4.3}^{+2.0}_{-0.8}$ & 0.54  \\[0.07cm]
hard muon, 3-jet       & $0.75$         & 3.6  & ${4.2}^{+2.0}_{-0.7}$ & 0.27  \\[0.07cm]
hard electron, 4-jet   & $1.22$  & 5.8  & ${5.3}^{+2.6}_{-1.3}$ & 0.63  \\[0.07cm]
hard muon, 4-jet       & $0.95$  & 4.5  & ${3.8}^{+1.3}_{-0.7}$ & 0.75  \\[0.07cm]
soft electron          & $1.82$         & 8.6  & $10.4^{+4.2}_{-3.1}$  & 0.28  \\[0.07cm]
soft muon              & $1.92$         & 9.0  & $12.5^{+5.4}_{-3.8}$  & 0.21  \\[0.07cm]
multi-lepton, $ee$, 2-jet      & $0.71$ & 3.3  & $3.5 \pm 0.1$         & 0.48 \\[0.07cm]
multi-lepton, $\mu\mu$, 2-jet  & $0.76$ & 3.6  & $3.5 \pm 0.1$         & 0.46 \\[0.07cm]
multi-lepton, $e\mu$, 2-jet    & $0.83$ & 3.9  & $3.6^{+0.6}_{-0.2}$   & 0.85 \\[0.07cm]
multi-lepton, $ee$, 4-jet      & $1.53$ & 7.2 & $7.7^{+3.2}_{-2.1}$    & 0.39 \\[0.07cm]
multi-lepton, $\mu\mu$, 4-jet  & $1.93$ & 9.1 & $8.8^{+3.3}_{-3.0}$    & 0.55 \\[0.07cm]
multi-lepton, $e\mu$, 4-jet    & $2.14$ & 10.1& $11.5^{+4.8}_{-3.5}$   & 0.28 \\
\noalign{\smallskip}\hline\hline\noalign{\smallskip}
\end{tabular*}
\end{center}
\caption{
  Left to right: 95\% CL upper limits on the visible cross section
  ($\langle\epsilon\sigma\rangle_{\rm obs}^{95}$) in the various
  signal regions, and on the number of
  signal events ($S_{\rm obs}^{95}$ ).  The third column
  ($S_{\rm exp}^{95}$) shows the 95\% CL upper limit on the number of
  signal events, given the expected number (and $\pm 1\sigma$ 
  uncertainty on the expectation) of background events.
  The last column
  indicates the $CL_B$ value, i.e. the observed confidence level for
  the background-only hypothesis.  }
\label{table.results.exclxsec.pval}
\end{table*}

Model-independent
limits on the visible cross section (i.e. the cross section evaluated
inside a given signal region) are derived by including the number of
events observed in that region as an input to the fit and deriving an
additional parameter, representing the non-SM signal strength
(constrained to be non-negative), as the output of the fit.
Potential signal contamination in the control regions is ignored.
Limits on the number of non-SM events in the signal region,
derived using the $CL_{s}$ \cite{Read:2002hq} prescription, are divided by the
integrated luminosity to obtain the constraints on the visible cross
section.  The limits at 95\% confidence level (CL) are shown in Table
\ref{table.results.exclxsec.pval}.

\begin{figure*}[htb]
\center
\includegraphics[width=0.46\textwidth]{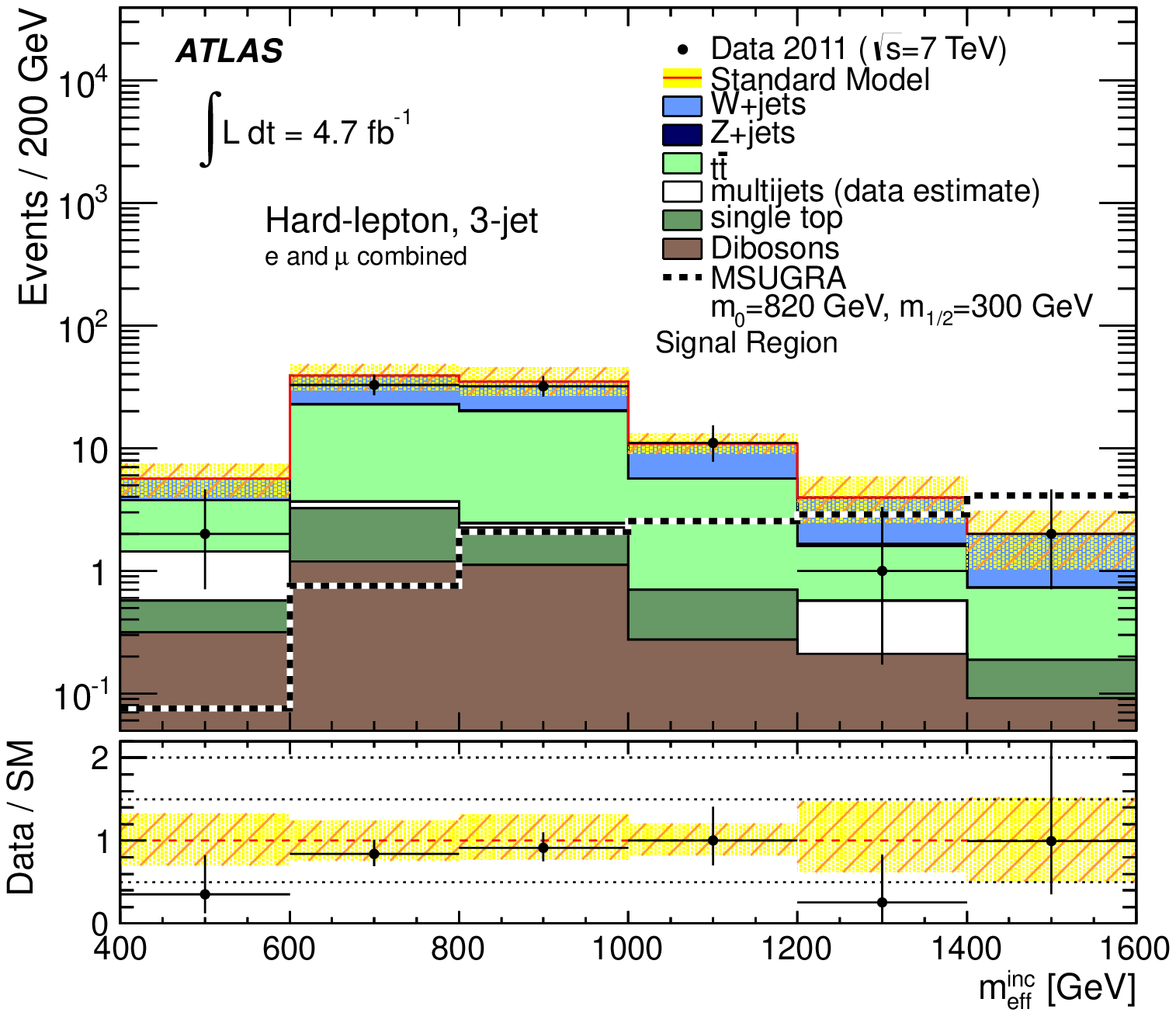}
\includegraphics[width=0.46\textwidth]{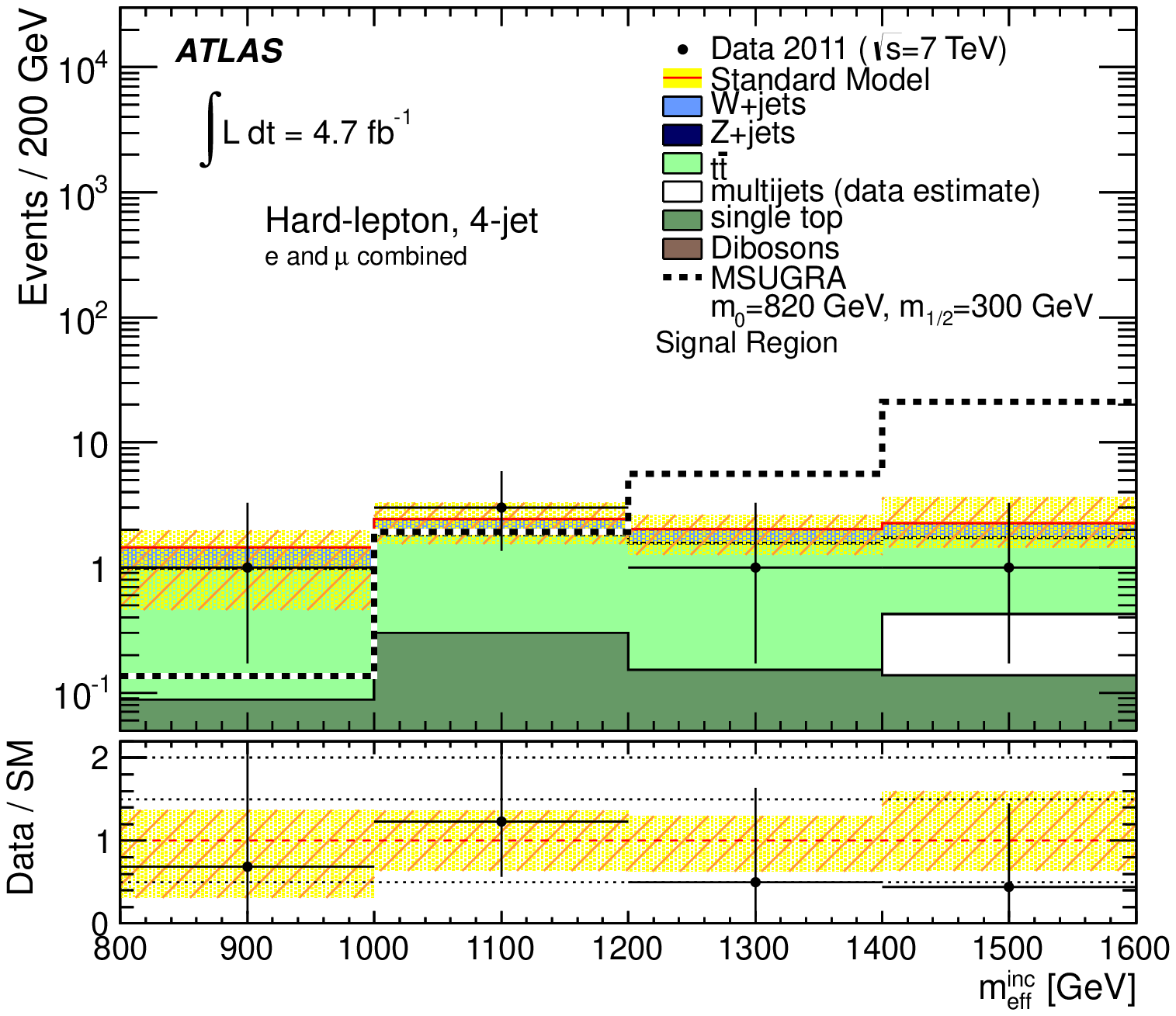}
\newline
\includegraphics[width=0.46\textwidth]{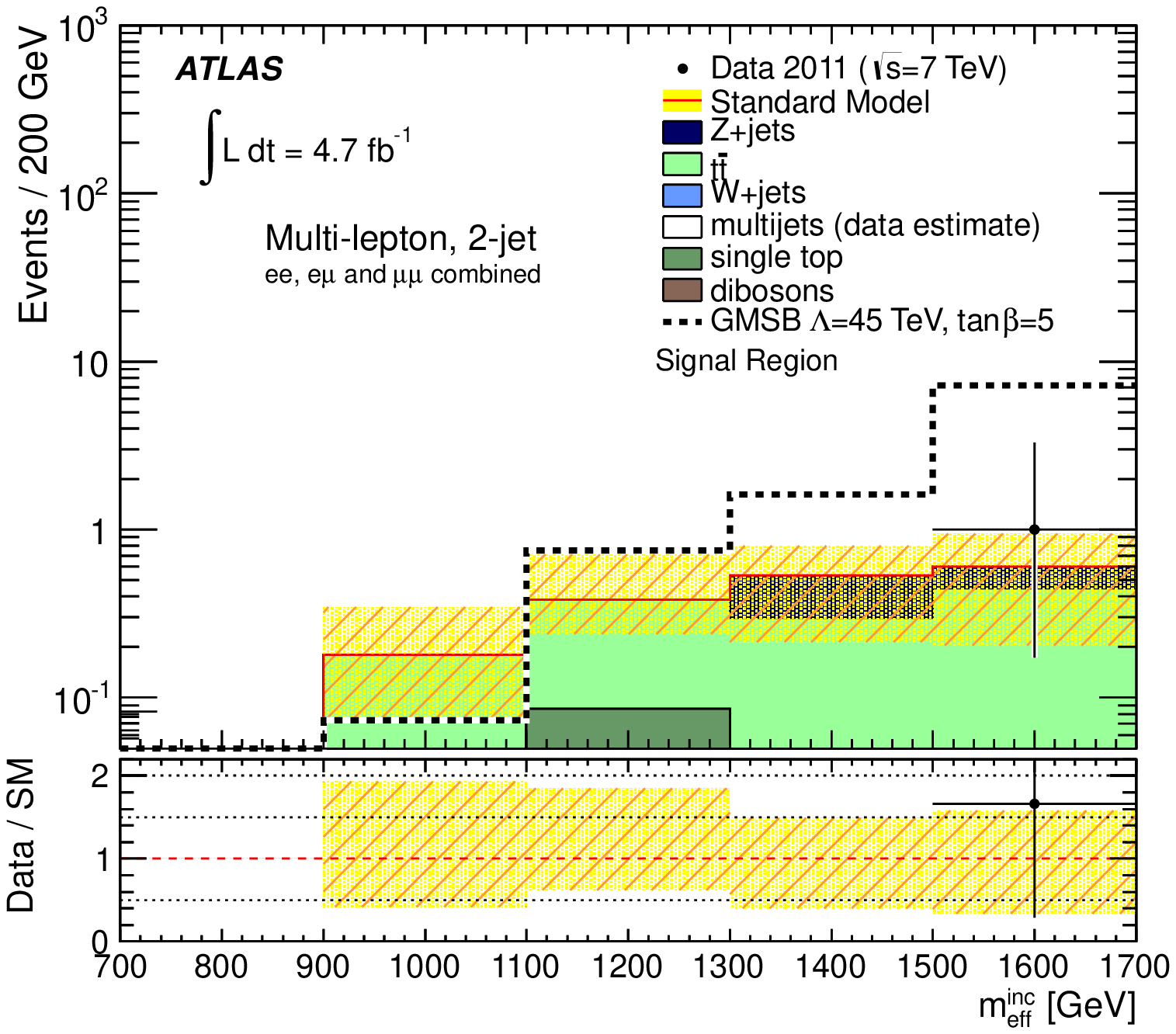}
\includegraphics[width=0.46\textwidth]{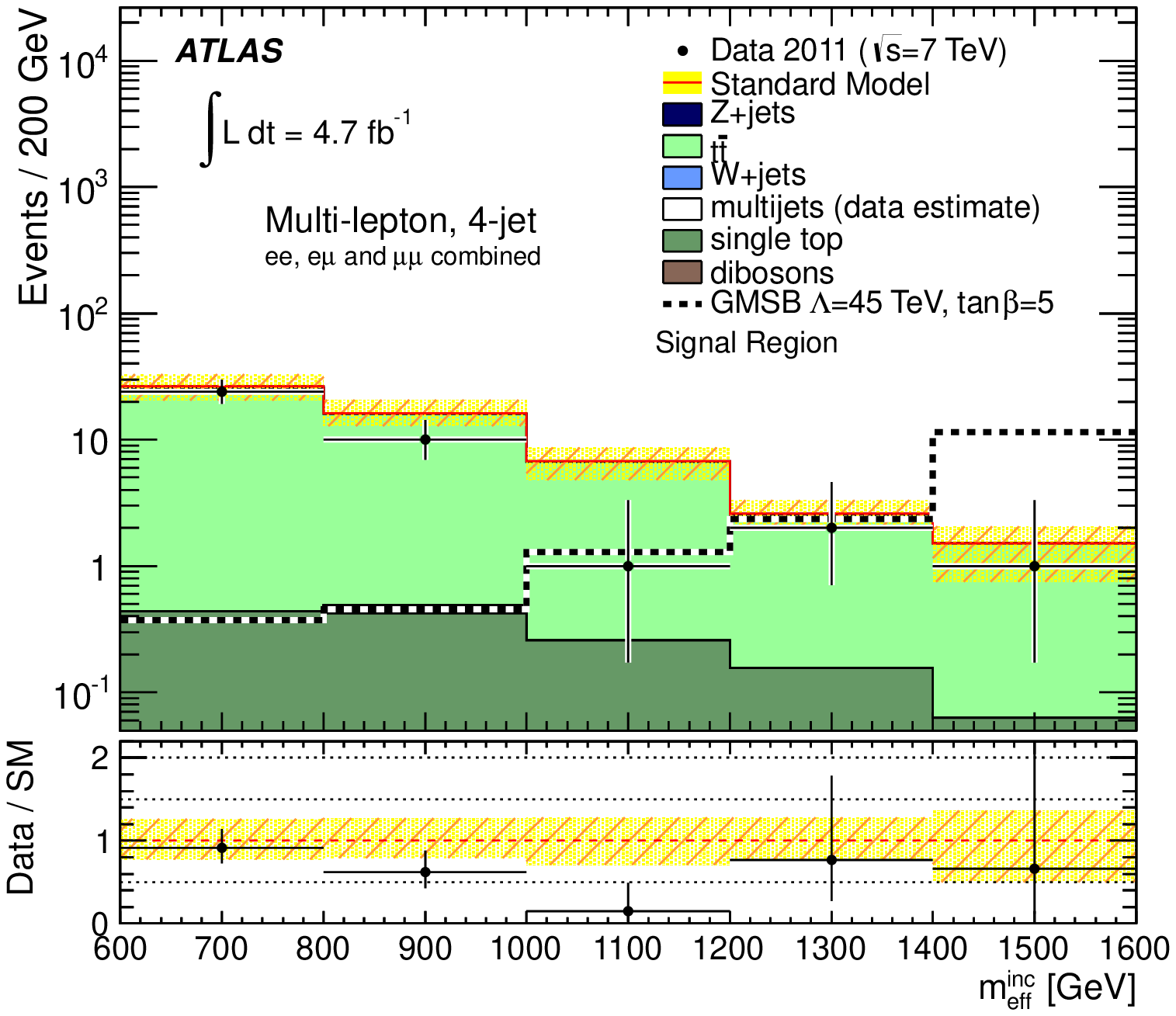}
\newline
\includegraphics[width=0.46\textwidth]{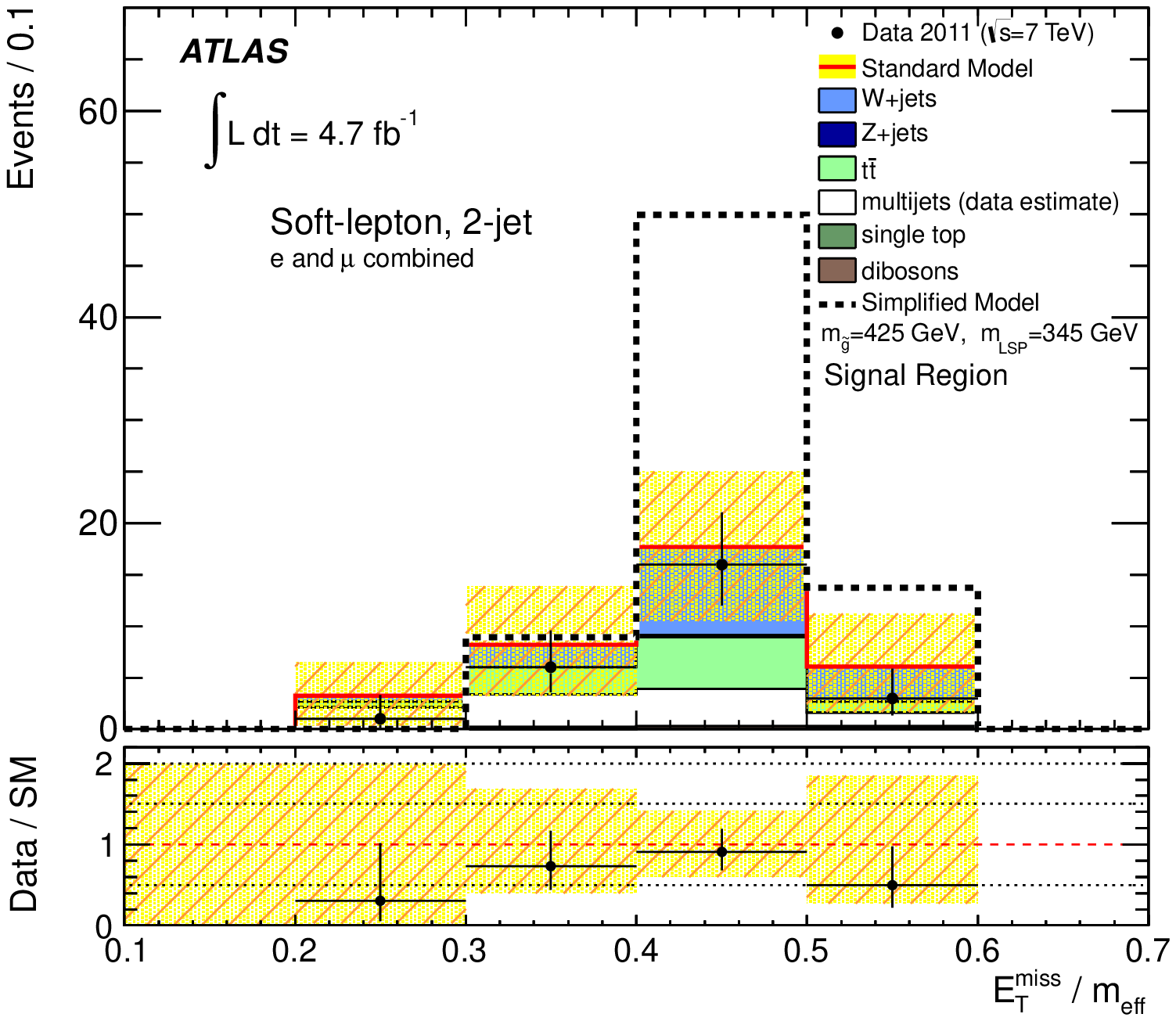}

\caption{
  Top and middle: Distribution of $m_{\rm{eff}}^{\rm{inc}}$ in the
  signal regions after all selection requirements except for that
  on the inclusive effective mass.  Top left: hard-lepton, 3-jet selection. Top
  right: hard-lepton, 4-jet selection. Middle left: multi-lepton, 2-jet
  selection.
  Middle right: multi-lepton, 4-jet selection.
  The last $m_{\rm{eff}}^{\rm{inc}}$ bin 
  includes all overflows.  The lowest
  $m_{\rm{eff}}^{\rm{inc}}$ bins are affected by the minimum
  \pt~requirements on jets and \met.
  Bottom: The
  \met/$m_{\rm{eff}}$  distribution in the soft-lepton signal
  region after all selection requirements except for that on
  \met/$m_{\rm{eff}}$.
  In all plots the different lepton flavors have been combined for
  ease of presentation.  
  The ``Data/SM'' plots show the ratio between
data and the total Standard Model expectation.  The Standard Model
expectation shown here is the input to the final fit, and is
derived from simulation only, normalized
to the theoretical cross sections. 
The uncertainty band around the Standard Model expectation combines the
statistical uncertainty on the simulated event samples with
the systematic uncertainties on the jet energy scale,
$b$-tagging,
data-driven multijet background, and luminosity. The systematic
uncertainties are largely correlated from bin to bin.  An example of the
distribution for a simulated signal is also shown (not
stacked); the signal point is chosen to be near the exclusion
limit of the analysis in Ref. \cite{ATLAS:2011ad}.}
\label{fig:fitressr}
\end{figure*}

For excluding specific models of new physics, 
the fit in the signal regions proceeds in the same way except that in
this case 
the signal contamination in control regions is treated by providing 
transfer factors from the signal regions to the control regions as
further input to the fit.
In addition, the likelihood fit makes use of the 
$m_{\rm{eff}}^{\rm{inc}}$ shape information (\met/$m_{\rm{eff}}$~for
the soft-lepton channel) in the signal region as a
further discriminant. Examples of these distributions
are shown in Fig.  \ref{fig:fitressr} (the figure shows
the distributions summed over lepton flavors, while the fit treats
each lepton flavor channel independently).
The likelihood is extended to include
bin-by-bin $m_{\rm{eff}}^{\rm{inc}}$ or \met/$m_{\rm{eff}}$~information
by dividing
the signal region into several bins of
$m_{\rm{eff}}^{\rm{inc}}$ or \met/$m_{\rm{eff}}$.  

The ten statistically independent
hard-lepton and
multi-lepton channels are combined
to set limits in the MSUGRA/CMSSM model.
For the minimal GMSB model, only the multi-lepton channels are used.
The soft-lepton channels are used together with the hard-lepton
and multi-lepton channels
to set limits in the one- and two-step simplified models.

\begin{figure*}
\begin{center}
\includegraphics[width=0.7\textwidth]{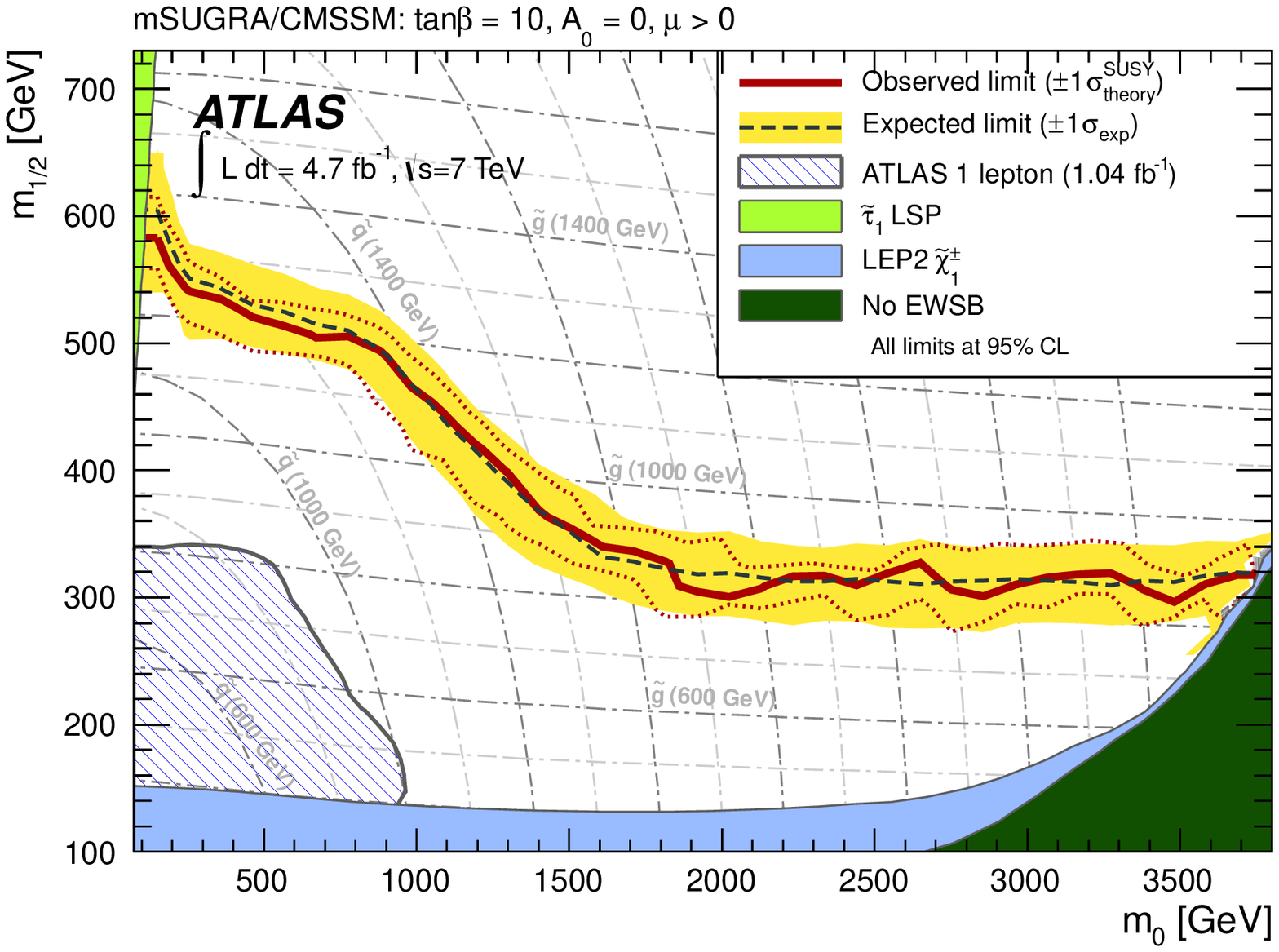}
\caption{
Expected and observed 95\% CL exclusion limits in the MSUGRA/CMSSM
model with $\tan\beta=10, A_{0}=0$ and the sign of $\mu$ taken to be
positive.
The results are
obtained by combining ten signal regions from the hard
single-lepton and multi-lepton channels. The band around the median
expected limit shows the $\pm 1\sigma$
variations, including all uncertainties except
theoretical uncertainties on the signal. The dotted lines around the
observed limit indicate the sensitivity to $\pm 1\sigma$ variations
on these theoretical uncertainties.
The dashed grid shows contours of
constant squark (curved lines) and gluino (nearly horizontal lines)
masses.
The previous limit from ATLAS \cite{ATLAS:2011ad} and the results
from the LEP experiments \cite{LEPSUSYWG} are also shown. }
\label{fig:msugra.exclusion}
\end{center}
\end{figure*}

The limit in the plane of $m_{1/2}$ versus $m_{0}$ in the MSUGRA/CMSSM
model is shown in Fig. \ref{fig:msugra.exclusion}.
The band around the expected limit
includes all uncertainties except theoretical uncertainties on the
signal prediction while the band on the observed
limit indicates the sensitivity to the theoretical uncertainties on the
signal. 
A large improvement in exclusion
coverage over the previous analysis \cite{ATLAS:2011ad} can be seen.
The simultaneous fit to 
the ten signal regions and the inclusion
of the shapes of the $m_{\rm{eff}}^{\rm{inc}}$
distributions  increase the expected reach 
in $m_{1/2}$ and $m_{0}$ by about 100 \GeV, approximately uniformly across
the plane.
Along the line of equal
masses between squarks and gluinos in the MSUGRA/CMSSM model,
masses below approximately 1200
\GeV~are excluded at 95\% CL.

\begin{figure*}
\begin{center}
\includegraphics[width=0.7\textwidth]{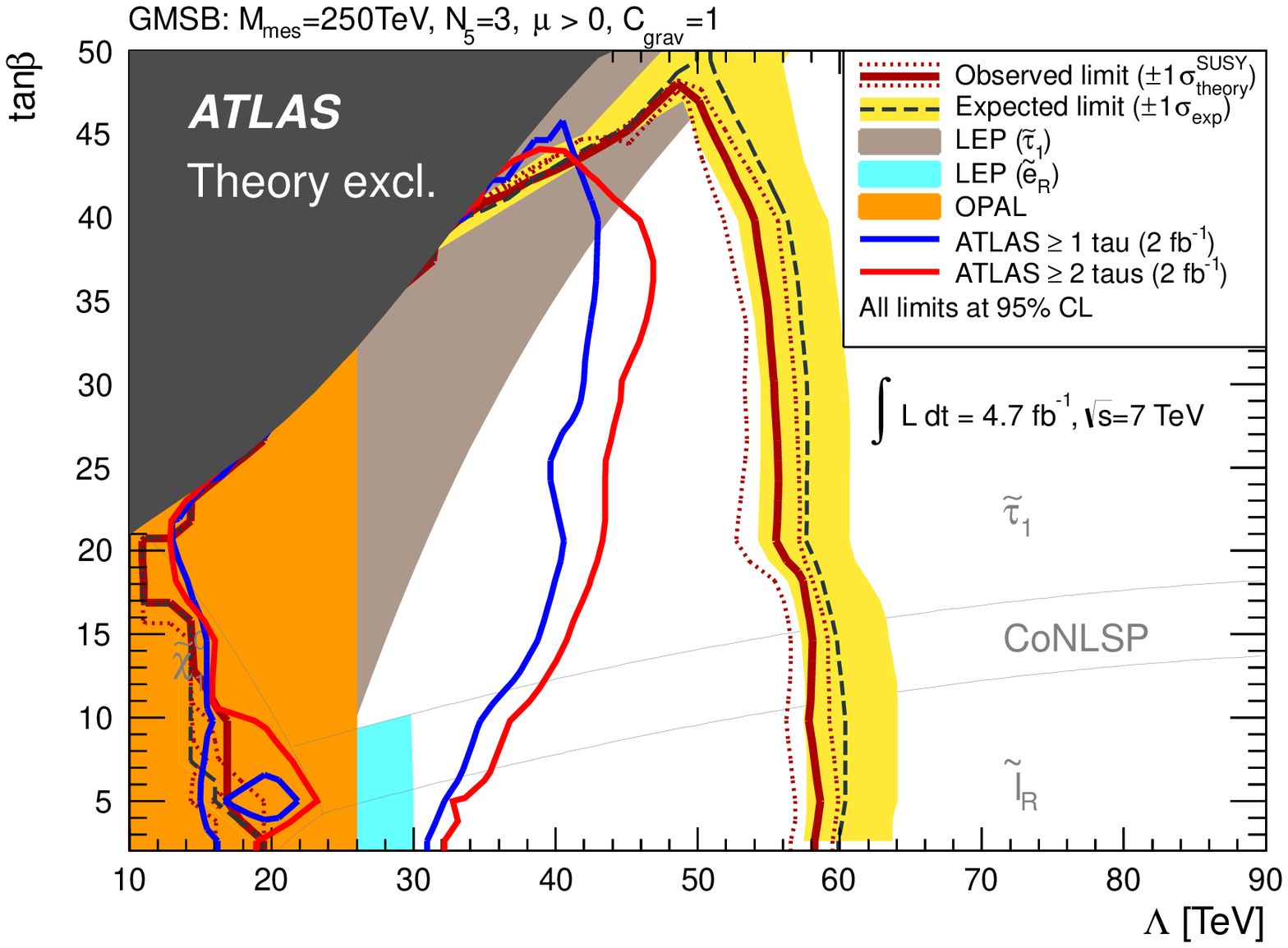}
\caption{
Expected and observed 95\% CL exclusion limits in the minimal GMSB
model, combining six signal regions from the multi-lepton channels.
The band around the median expected limit shows the $\pm 1\sigma$
variations, including all uncertainties except
theoretical uncertainties on the signal. The dotted lines around the
observed limit indicate the sensitivity to $\pm 1\sigma$ variations
on these theoretical uncertainties.
 The different next-to-lightest-SUSY
particle (NLSP) regions are indicated.  The coNLSP region denotes the
region where $\tilde{\tau}_{1}$ and $\tilde{\ell}_{R}$ are nearly mass
degenerate.  Previous OPAL and ATLAS limits 
in this model can be found in Refs. \cite{Abbiendi:2005gc} and
\cite{Aad:2012rt,ATLAS:2012ag}, respectively.  Limits derived from the
LEP slepton mass limits \cite{LEPSUSYWG2} are also shown.
}
\label{fig:mgmsb.exclusion}
\end{center}
\end{figure*}

For the minimal GMSB model, the limit in the plane of $\tan\beta$
versus $\Lambda$ is shown in Fig. \ref{fig:mgmsb.exclusion}.  The
exclusion reach is dominated by the dilepton plus two jets channel.
Values of $\Lambda$ below about 50 \TeV~are excluded at 95\% CL for
$\tan\beta < 45$, improving on previous constraints.

Exclusion limits in the one-step simplified models are shown in
Fig. \ref{fig.results.1step}.   The figures also show the
cross section excluded at 95\% CL.
The exclusion limits in the two-step simplified models are shown in
Fig. \ref{fig.results.2step.gg} for gluino pair production and
Fig. \ref{fig.results.2step.ss} for squark pair production.
Simplified models with varying
chargino mass and two-step simplified models are considered here for the
first time in leptonic SUSY searches.
For both one- and two-step models, for the case of low
LSP masses, gluinos with masses
below approximately 900--1000 \GeV~and squarks with masses below
approximately 500--600 \GeV~are excluded.  Squark limits are
considerably weaker, primarily due to the lower
  production cross section.  Furthermore in the one-step model,
gluinos with
mass below 550 \GeV~are excluded for essentially all values of the LSP
mass if the latter is more than 30 \GeV~smaller than the gluino mass.
Care has to be taken when interpreting
the simplified model limit in the context of a MSSM
scenario, where the mass of the sneutrino is lighter than
the mass of the left-handed slepton, as this can lead to 
modification of the lepton momenta.

\begin{figure*}[htbp]
\center
\includegraphics[width=0.49\textwidth]{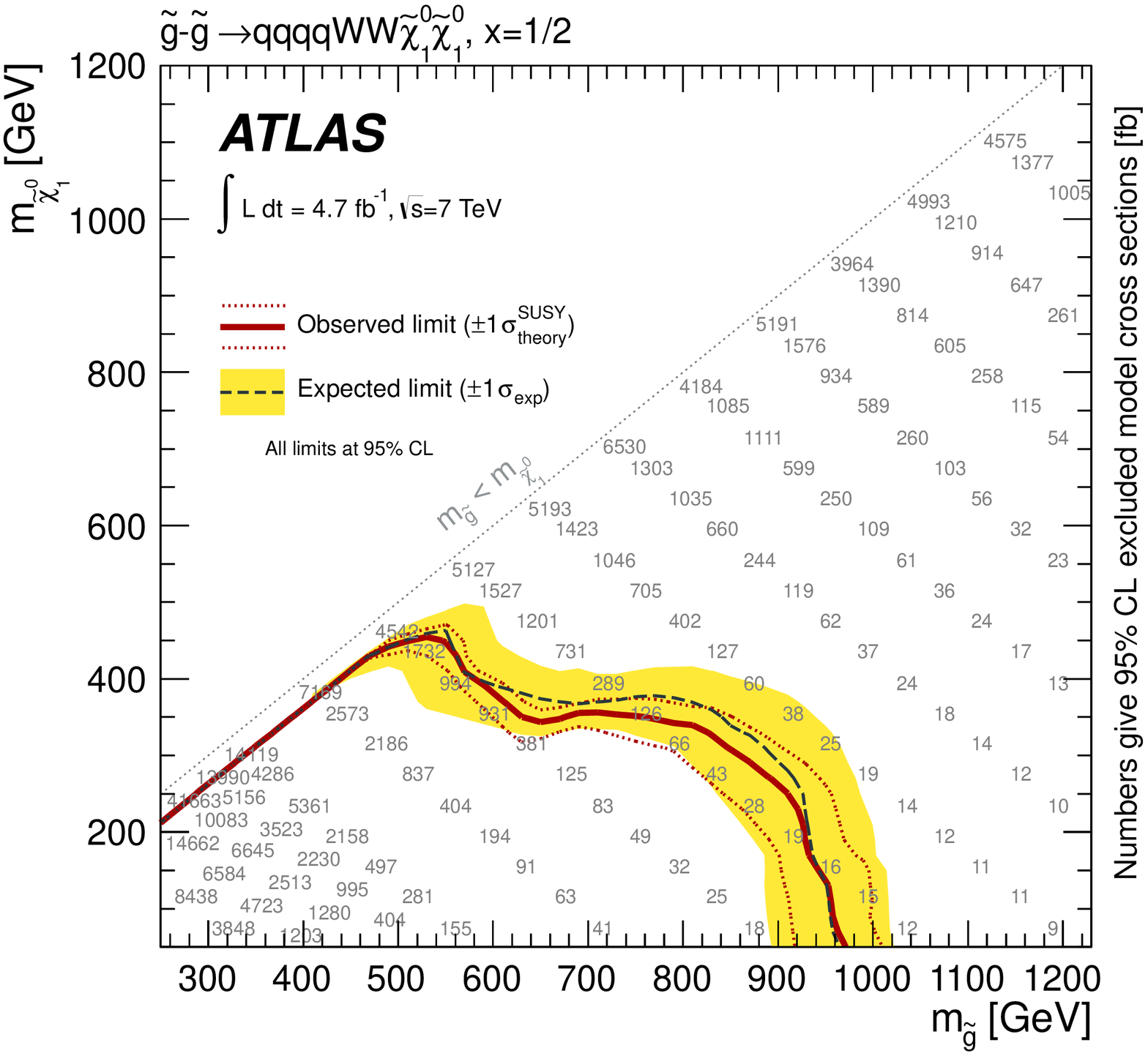}
\includegraphics[width=0.49\textwidth]{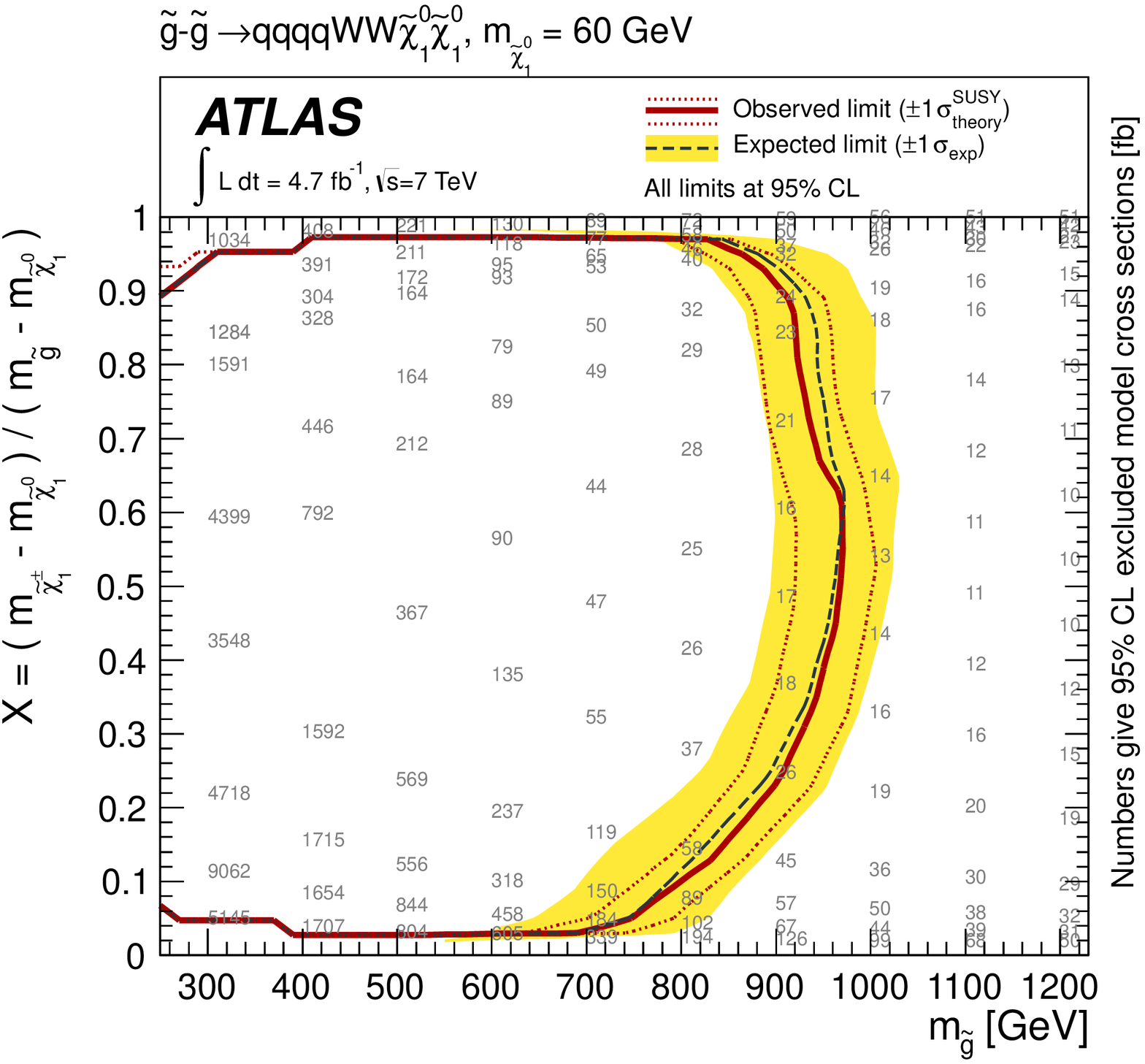}
\newline
\includegraphics[width=0.49\textwidth]{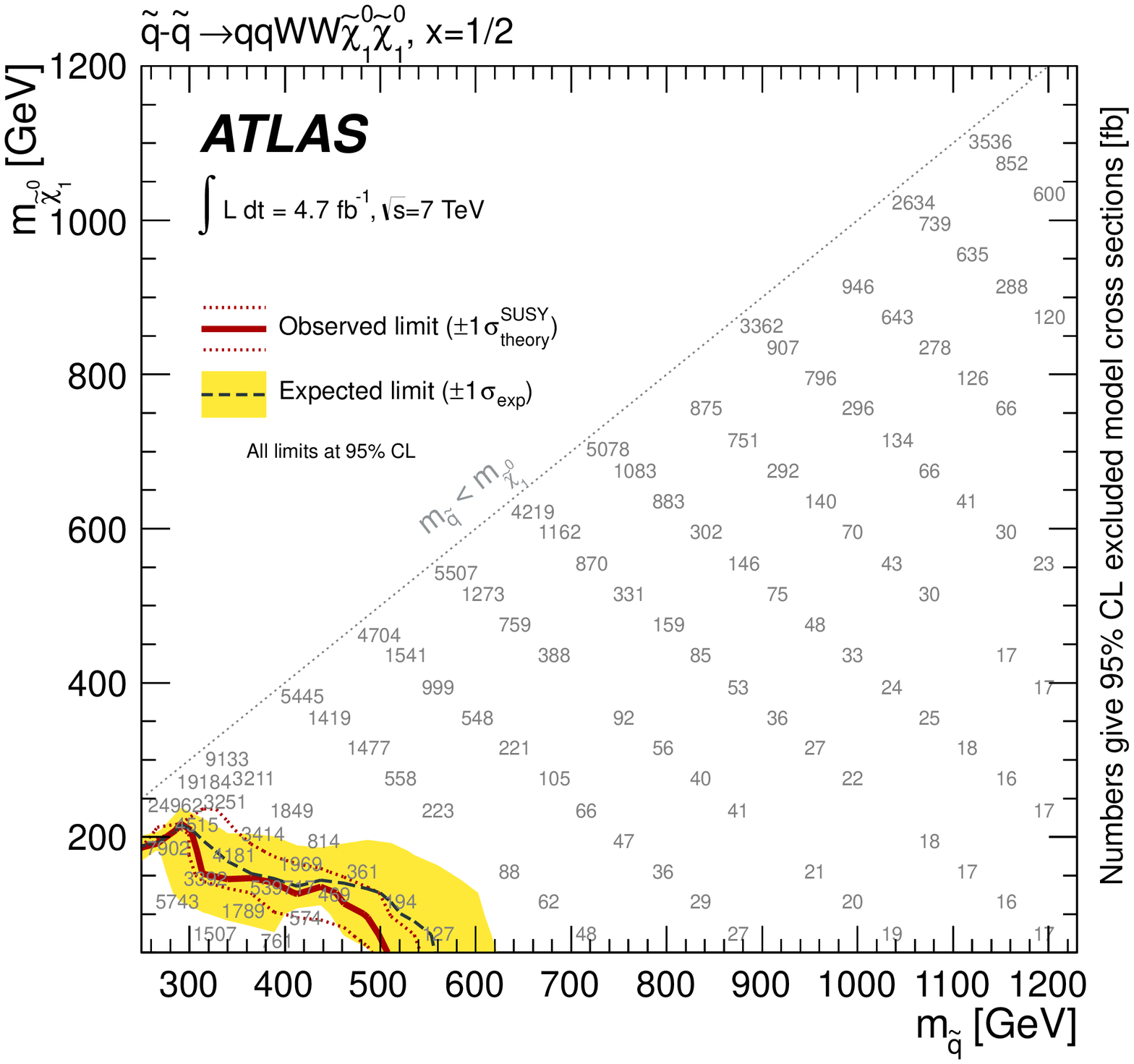}
\includegraphics[width=0.49\textwidth]{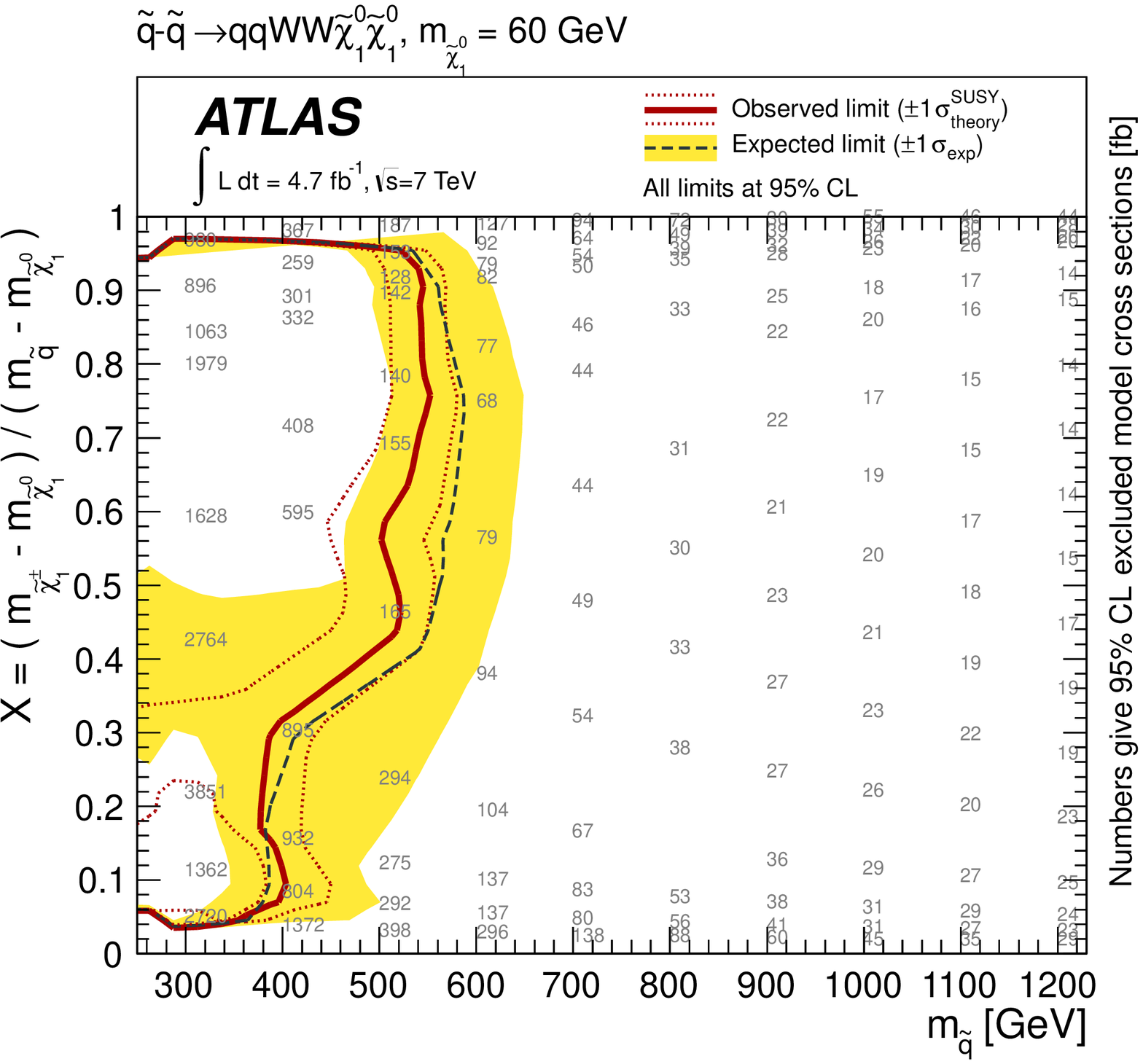}

\caption{Excluded regions at 95\% confidence level  in the parameter
  space of 
  one-step simplified models.  Top row: gluino pair production with
  $\tilde{g} \rightarrow
q\overline{q}'\tilde{\chi}_{1}^{\pm} \rightarrow
q\overline{q}'W^{\pm}\tilde{\chi}_{1}^{0}$.    Bottom
row: squark pair production with
$\tilde{q}_{L} \rightarrow
q' \tilde{\chi}_{1}^{\pm} \rightarrow q'W^{\pm}\tilde{\chi}_{1}^{0}$.
In the left column,  the
chargino mass is set to be halfway
between gluino (top) or squark (bottom) and LSP masses.   In the right
column, the LSP mass is
fixed at 60 \GeV~and the masses of the chargino and gluino (top) or
squark (bottom) are varied.
  The  band around the median expected limit shows the $\pm 1\sigma$
variations, including all uncertainties except
theoretical uncertainties on the signal. The dotted lines around the
observed limit indicate the sensitivity to $\pm 1\sigma$ variations
on these theoretical uncertainties.
  The plots are from the combination of the hard and soft
  single-lepton channels.
The numbers indicate the excluded cross section in fb.
A smaller excluded cross section implies a more stringent
limit. }
\label{fig.results.1step}
\end{figure*}

\begin{figure*}[htbp]
\center
\includegraphics[width=0.49\textwidth]{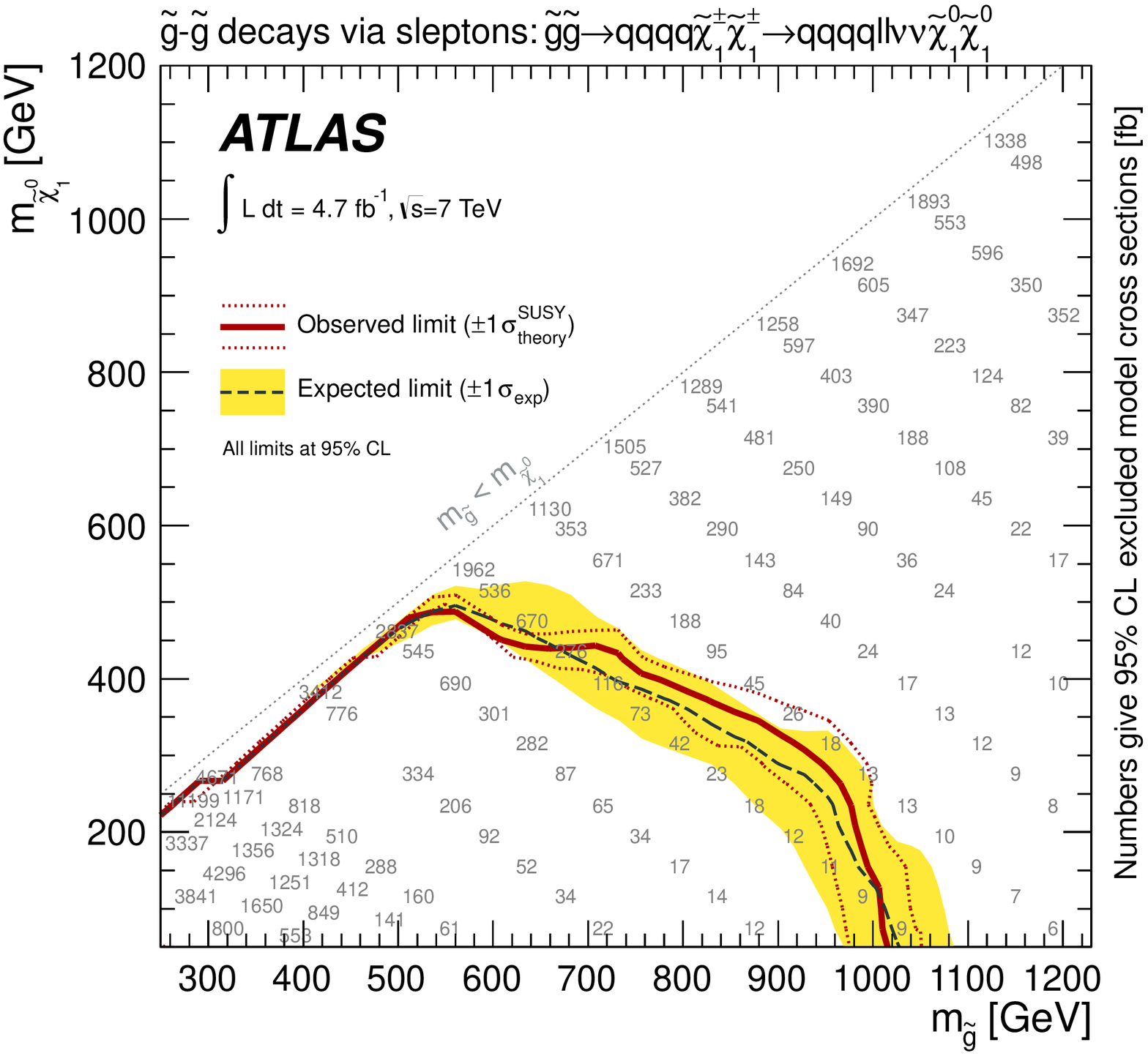}
\includegraphics[width=0.49\textwidth]{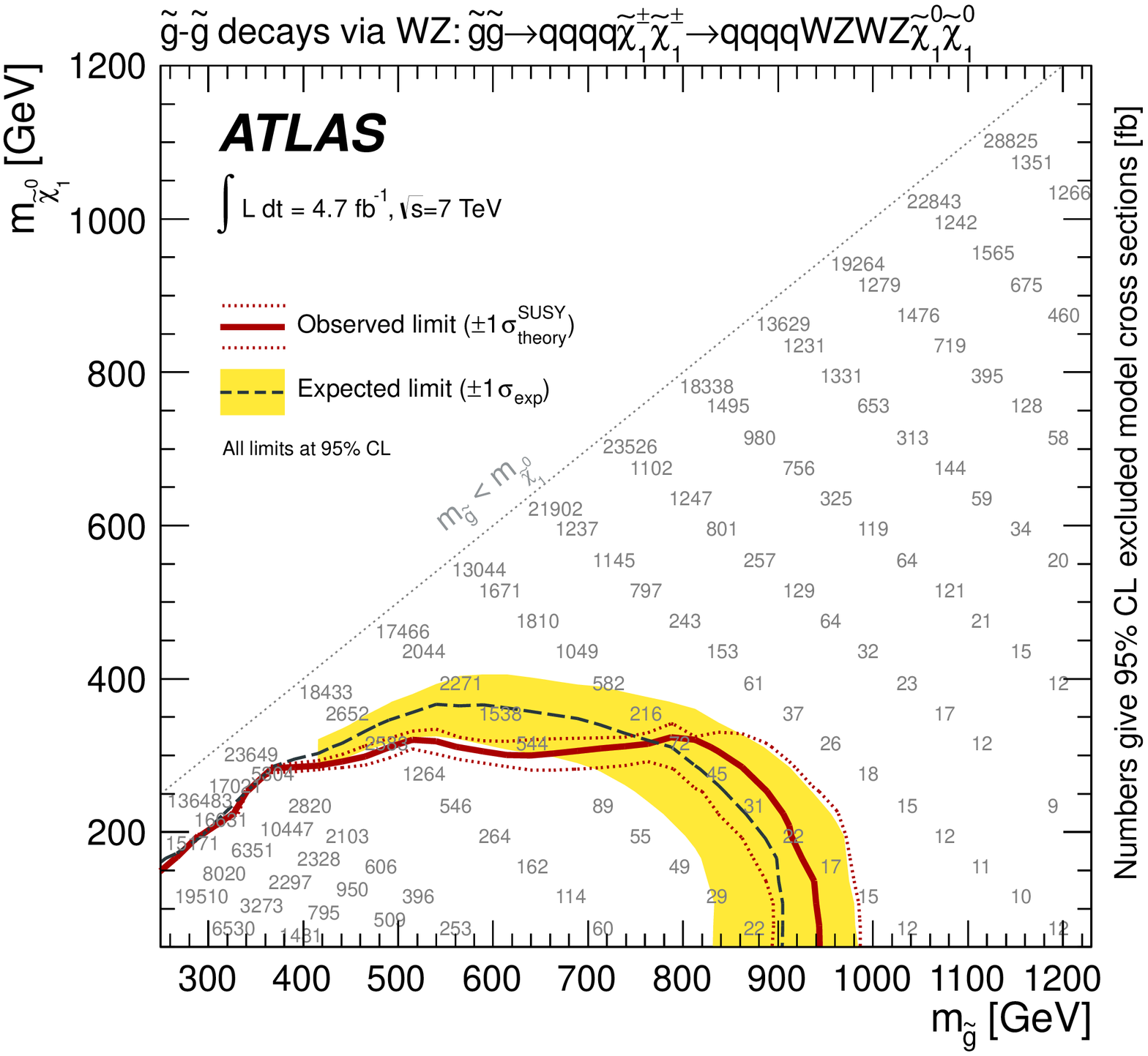}

\caption{Excluded regions at 95\% confidence level  in the parameter
  space of 
  two-step simplified models with gluino pair production.
  Left: both gluinos decay via 
  $\tilde{g} \rightarrow
q\overline{q}'\tilde{\chi}_{1}^{\pm} \rightarrow q\overline{q}'\ell^{\pm}
\tilde{\nu}_{L} \rightarrow q\overline{q}'\ell^{\pm}\nu\tilde{\chi}_{1}^{0}$ or  $\tilde{g} \rightarrow
q\overline{q}'\tilde{\chi}_{1}^{\pm} \rightarrow q\overline{q}'\nu
\tilde{\ell}_{L}^{\pm} \rightarrow q\overline{q}'\nu
\ell^{\pm}\tilde{\chi}_{1}^{0}$.    Right: both gluinos decay via 
$\tilde{g} \rightarrow
q\overline{q}'\tilde{\chi}_{1}^{\pm} \rightarrow q\overline{q}' W^{(\ast)\pm} \tilde{\chi}_{2}^{0} \rightarrow
W^{(\ast)\pm} Z^{(\ast)} \tilde{\chi}_{1}^{0}$.
  The  band around the median expected limit shows the $\pm 1\sigma$
variations, including all uncertainties except
theoretical uncertainties on the signal. The dotted lines around the
observed limit indicate the sensitivity to $\pm 1\sigma$ variations
on these theoretical uncertainties.
  The plots are dominated by the multi-lepton channels.
The numbers indicate the excluded cross section in fb.
A smaller excluded cross section implies a more stringent
limit. }
\label{fig.results.2step.gg}
\end{figure*}

\begin{figure*}[htbp]
\center
\includegraphics[width=0.49\textwidth]{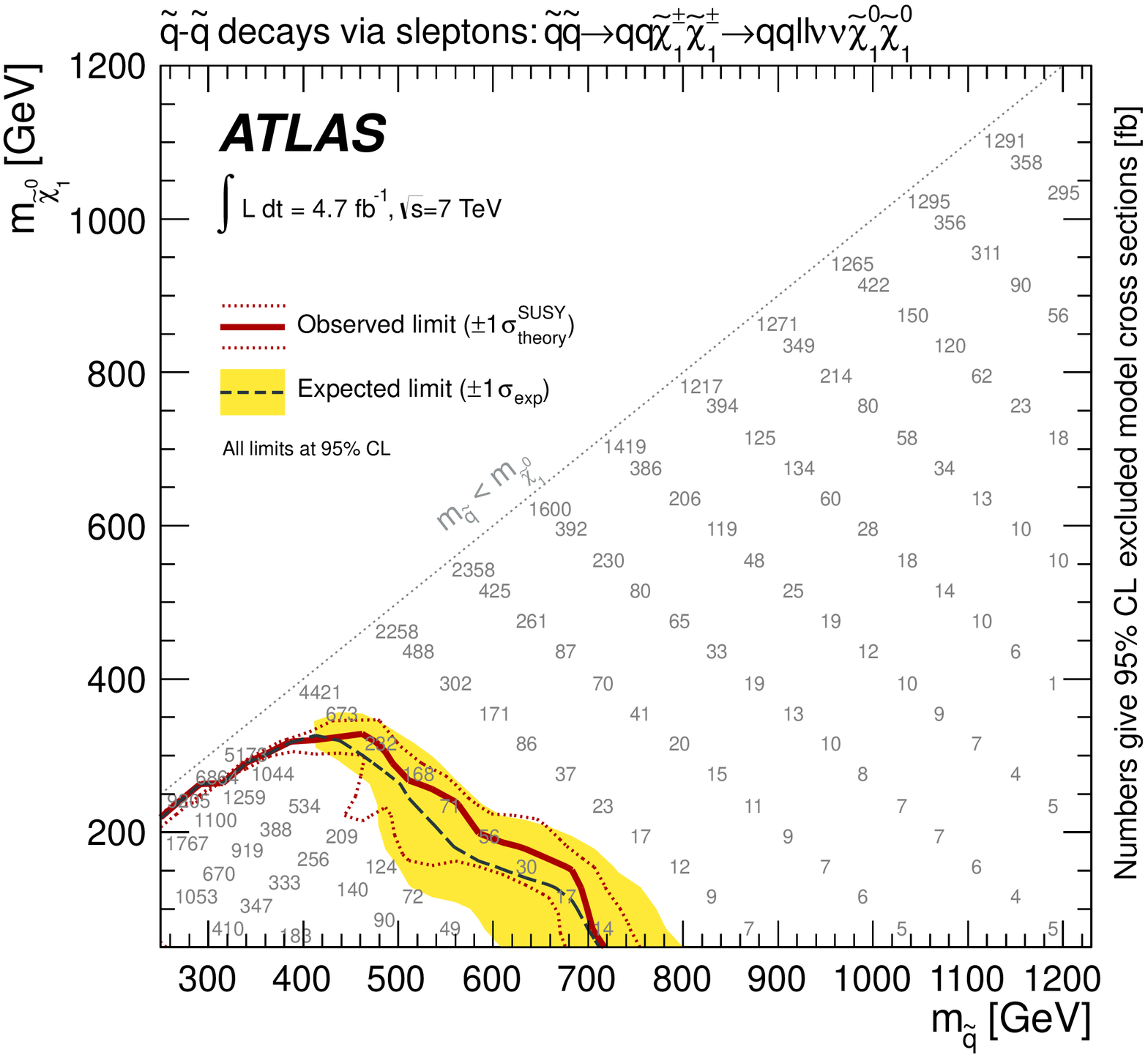}
\includegraphics[width=0.49\textwidth]{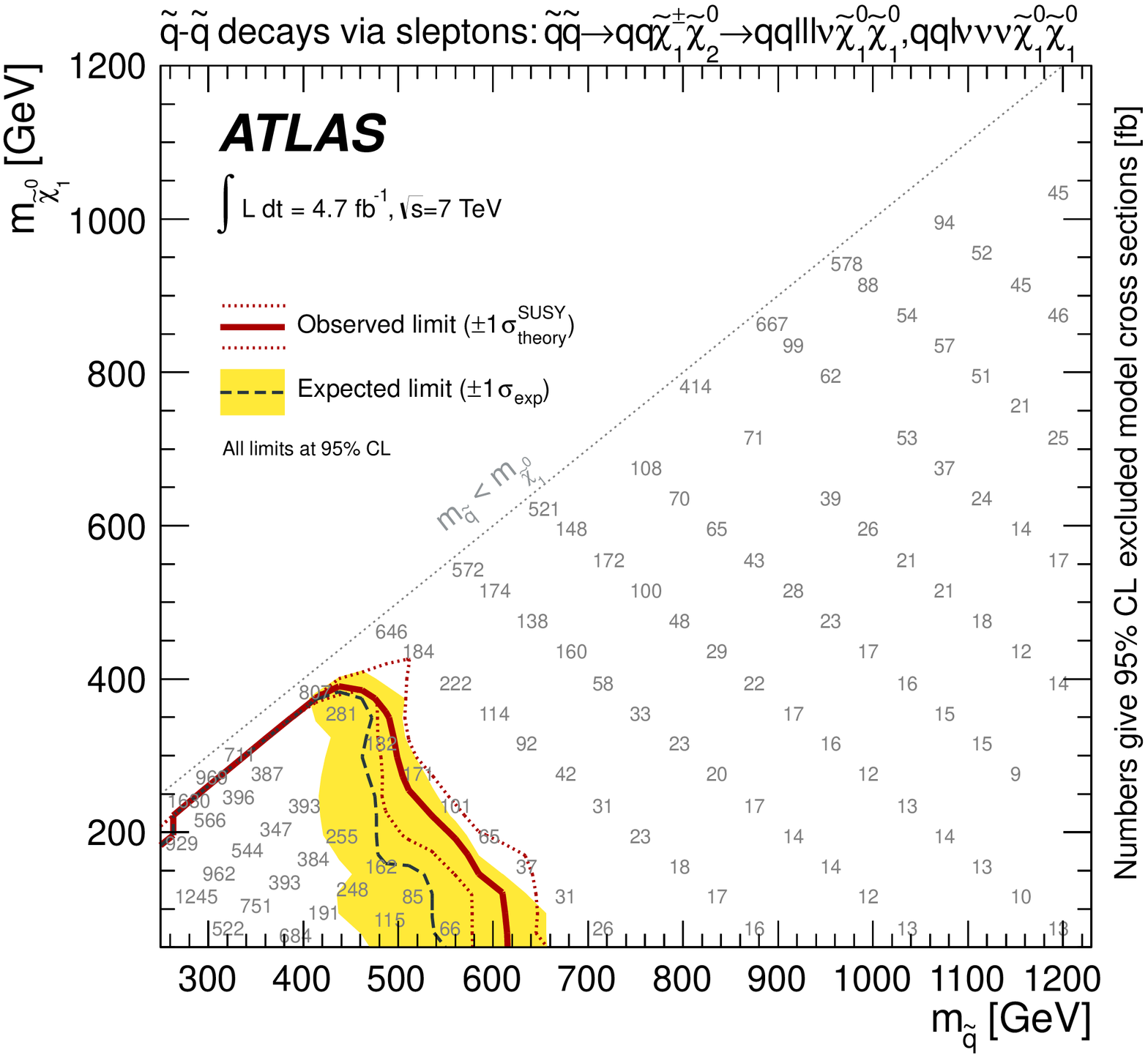}
\newline
\includegraphics[width=0.49\textwidth]{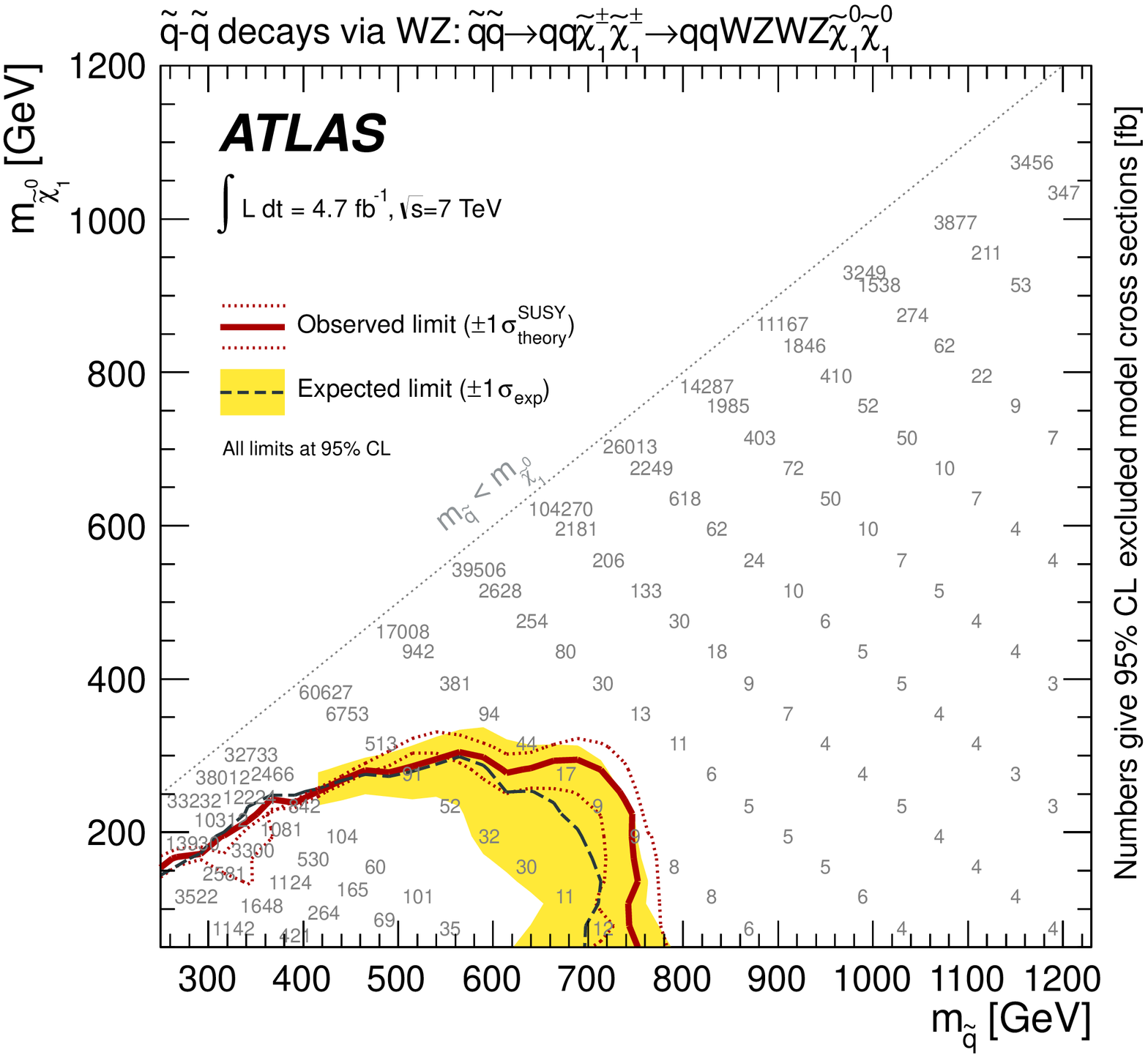}

\caption{Excluded regions at 95\% confidence level in the parameter
  space of 
  two-step simplified models with squark pair production.
  Top left: both squarks decay
  via $\tilde{q}_{L} \rightarrow
q'\tilde{\chi}_{1}^{\pm} \rightarrow q'\ell^{\pm}\tilde{\nu}_{L}
\rightarrow q'\ell^{\pm}\nu\tilde{\chi}_{1}^{0}$ or
$\tilde{q}_{L} \rightarrow
q'\tilde{\chi}_{1}^{\pm} \rightarrow q'\tilde{\ell}^{\pm}\nu
\rightarrow q'\ell^{\pm}\nu\tilde{\chi}_{1}^{0}$.
Top right: 
one squark decays via $\tilde{q}_{L} \rightarrow
q' \tilde{\chi}_{1}^{\pm} \rightarrow q'\ell^{\pm}\tilde{\nu}_{L}
\rightarrow q'\ell^{\pm}\nu\tilde{\chi}_{1}^{0}$ or
$\tilde{q}_{L} \rightarrow
q' \tilde{\chi}_{1}^{\pm} \rightarrow q'\tilde{\ell}^{\pm}\nu
\rightarrow q'\ell^{\pm}\nu\tilde{\chi}_{1}^{0}$
and the other squark decays
via $\tilde{q}_{L} \rightarrow q \tilde{\chi}_{2}^{0} \rightarrow q
\ell^{\pm}\tilde{\ell}_{L}^{\mp} \rightarrow
q\ell^{\pm}\ell^{\mp}\tilde{\chi}_{1}^{0}$ or 
$\tilde{q}_{L} \rightarrow q \tilde{\chi}_{2}^{0} \rightarrow q
\nu\tilde{\nu}_{L} \rightarrow
q\nu\nu\tilde{\chi}_{1}^{0}$ .
Bottom row: both 
squarks decay via $\tilde{q}_{L} \rightarrow
q' \tilde{\chi}_{1}^{\pm} \rightarrow W^{(\ast)\pm} \tilde{\chi}_{2}^{0} \rightarrow
W^{(\ast)\pm} Z^{(\ast)} \tilde{\chi}_{1}^{0}$.
  The  band around the median expected limit shows the $\pm 1\sigma$
variations, including all uncertainties except
theoretical uncertainties on the signal. The dotted lines around the
observed limit indicate the sensitivity to $\pm 1\sigma$ variations
on these theoretical uncertainties.
  The plots are dominated by the multi-lepton channels.
The numbers indicate the excluded cross section in fb.
A smaller excluded cross section implies a more stringent
limit. }
\label{fig.results.2step.ss}
\end{figure*}

\FloatBarrier
\section{Conclusion}

A new search with the ATLAS detector
for SUSY in final states containing jets, one or more isolated leptons
(electron or muon) and \met~has been presented. 
Data from the full 2011 data-taking period,
corresponding to an integrated
luminosity of 4.7 \ifb, have been analyzed.
Single- and multi-lepton channels are treated in one analysis.
A 
signal region with a soft lepton and soft jets has been introduced to
increase the sensitivity to SUSY decay spectra involving small mass differences
(``compressed SUSY''),
where the sensitivity is improved by a factor of 10--30 compared to the
hard-lepton channel.
A simultaneous fit is performed to
the event yield in
multiple signal and control regions and to the shapes of distributions in
those regions.

Observations are in good agreement with SM expectations and constraints
have been set on the visible cross section for new physics
processes.  Exclusion limits have also been extended for the
MSUGRA/CMSSM and minimal GMSB models as well as for a number of
simplified models.  In
MSUGRA/CMSSM, squark and gluino masses below approximately
1200~\GeV~are
excluded at 95\% CL (for equal squark and gluino masses).  In minimal
GMSB, values of $\Lambda$ below about 50~\TeV~are excluded for
$\tan\beta < 45$.

In one-step simplified models (with the chargino mass halfway between
the masses of the gluino/squark and LSP) gluinos are excluded for masses below
approximately 900~\GeV~for low values of the LSP mass.
Gluinos with mass below 550~\GeV~are excluded
for essentially all values of the LSP mass if the latter is
more than 30~\GeV~smaller than the mass of the gluino.
    In the one-step
simplified model with a fixed LSP mass and varying chargino and gluino
(squark) masses, gluinos below approximately 950~\GeV~are excluded for
a wide range of chargino masses; squarks are excluded below 500~\GeV,
albeit for a narrower range of chargino masses.

A variety of two-step simplified models have been considered.  Limits
on gluino masses range from about 900~\GeV~to 1000~\GeV, while squark mass
limits range from about 500~\GeV~to 600~\GeV, all for low LSP masses.

These results improve significantly on previous constraints.

\begin{acknowledgments}
We thank CERN for the very successful operation of the LHC, as well as the
support staff from our institutions without whom ATLAS could not be
operated efficiently.

We acknowledge the support of ANPCyT, Argentina; YerPhI, Armenia; ARC,
Australia; BMWF, Austria; ANAS, Azerbaijan; SSTC, Belarus; CNPq and FAPESP,
Brazil; NSERC, NRC and CFI, Canada; CERN; CONICYT, Chile; CAS, MOST and NSFC,
China; COLCIENCIAS, Colombia; MSMT CR, MPO CR and VSC CR, Czech Republic;
DNRF, DNSRC and Lundbeck Foundation, Denmark; EPLANET and ERC, European Union;
IN2P3-CNRS, CEA-DSM/IRFU, France; GNSF, Georgia; BMBF, DFG, HGF, MPG and AvH
Foundation, Germany; GSRT, Greece; ISF, MINERVA, GIF, DIP and Benoziyo Center,
Israel; INFN, Italy; MEXT and JSPS, Japan; CNRST, Morocco; FOM and NWO,
Netherlands; RCN, Norway; MNiSW, Poland; GRICES and FCT, Portugal; MERYS
(MECTS), Romania; MES of Russia and ROSATOM, Russian Federation; JINR; MSTD,
Serbia; MSSR, Slovakia; ARRS and MVZT, Slovenia; DST/NRF, South Africa;
MICINN, Spain; SRC and Wallenberg Foundation, Sweden; SER, SNSF and Cantons of
Bern and Geneva, Switzerland; NSC, Taiwan; TAEK, Turkey; STFC, the Royal
Society and Leverhulme Trust, United Kingdom; DOE and NSF, United States of
America.

The crucial computing support from all WLCG partners is acknowledged
gratefully, in particular from CERN and the ATLAS Tier-1 facilities at
TRIUMF (Canada), NDGF (Denmark, Norway, Sweden), CC-IN2P3 (France),
KIT/GridKA (Germany), INFN-CNAF (Italy), NL-T1 (Netherlands), PIC (Spain),
ASGC (Taiwan), RAL (UK) and BNL (USA) and in the Tier-2 facilities
worldwide.

\end{acknowledgments}

\bibliographystyle{atlasBibStyleWoTitle}
\bibliography{paper}

\onecolumngrid
\clearpage
\begin{flushleft}
{\Large The ATLAS Collaboration}

\bigskip

G.~Aad$^{\rm 48}$,
T.~Abajyan$^{\rm 21}$,
B.~Abbott$^{\rm 111}$,
J.~Abdallah$^{\rm 12}$,
S.~Abdel~Khalek$^{\rm 115}$,
A.A.~Abdelalim$^{\rm 49}$,
O.~Abdinov$^{\rm 11}$,
R.~Aben$^{\rm 105}$,
B.~Abi$^{\rm 112}$,
M.~Abolins$^{\rm 88}$
% ...
\bigskip

% ...
$^{11}$ Institute of Physics, Azerbaijan Academy of Sciences, Baku, Azerbaijan\\
$^{12}$ Institut de F\'{i}sica d'Altes Energies and Departament de F\'{i}sica de la Universitat Aut\`{o}noma de Barcelona and ICREA, Barcelona, Spain\\
% ...
$^{21}$ Physikalisches Institut, University of Bonn, Bonn, Germany\\
% ...
$^{48}$ Fakult\"{a}t f\"{u}r Mathematik und Physik, Albert-Ludwigs-Universit\"{a}t, Freiburg, Germany\\
$^{49}$ Section de Physique, Universit\'{e} de Gen\`{e}ve, Geneva, Switzerland\\
% ...
$^{88}$ Department of Physics and Astronomy, Michigan State University, East Lansing MI, United States of America\\
% ...
$^{105}$ Nikhef National Institute for Subatomic Physics and University of Amsterdam, Amsterdam, Netherlands\\
% ...
$^{111}$ Homer L. Dodge Department of Physics and Astronomy, University of Oklahoma, Norman OK, United States of America\\
$^{112}$ Department of Physics, Oklahoma State University, Stillwater OK, United States of America\\
% ...
$^{115}$ LAL, Universit\'{e} Paris-Sud and CNRS/IN2P3, Orsay, France\\
% ...
\end{flushleft}


\end{document}